\documentclass[fleqn,usenatbib]{mnras}

\usepackage{newtxtext,newtxmath}
\usepackage[dvipsnames, table]{xcolor}

\usepackage[T1]{fontenc}

\DeclareRobustCommand{\VAN}[3]{#2}
\let\VANthebibliography\thebibliography
\def\thebibliography{\DeclareRobustCommand{\VAN}[3]{##3}\VANthebibliography}

\usepackage{graphicx}	% Including figure files
\usepackage{amsmath}	% Advanced maths commands
\usepackage[normalem]{ulem} % for \sout

\newcommand{\hMpc}{\,h\mathrm{Mpc}^{-1}}
\newcommand{\sB}[1]{\textcolor{black}{#1}}

\title[The web-halo model of structure formation]{Web-Halo Model (WHM): Accurate non-linear matter power spectrum predictions without free parameters}

\author[S. Brieden et al.]{
S. Brieden,$^{1,\sB{2}}$\thanks{E-mail: \sB{brieden@physik.rwth-aachen.de}}
F. Beutler,$^{1}$\thanks{\sB{E-mail: florian.beutler@ed.ac.uk}}
M. Pellejero-Ibañez,$^{1}$
\\
$^{1}$Institute for Astronomy, University of Edinburgh, Blackford Hill, Edinburgh EH9 3HJ, UK \\
\sB{$^{2}$Institute for Theoretical Particle Physics and Cosmology (TTK), RWTH Aachen University, Sommerfeldstr. 16, D-52056 Aachen, Germany} \\
}

\date{Accepted 2026 February 9. Received 2026 January 27; in original form 2025 August 22}

\pubyear{\sB{2026}}

\begin{document}
\label{firstpage}
\pagerange{\pageref{firstpage}--\pageref{lastpage}}
\maketitle

% Abstract of the paper
\begin{abstract}
We present a parameter-free variant of the halo model that significantly improves the precision of matter clustering predictions, particularly in the challenging 1-halo to 2-halo transition regime, where standard halo models often fail. Unlike \texttt{HMcode-2020}, which relies on 12 phenomenological parameters, our approach achieves comparable or superior accuracy without any free fitting parameters. This new \emph{web-halo model} (WHM) extends the traditional halo model by incorporating structures that have collapsed along two dimensions (filaments) and one dimension (sheets), in addition to haloes, and combines these with 1-loop Lagrangian Perturbation Theory (1$\ell$-LPT) in a consistent framework. We show that WHM matches N-body simulation power spectra within the precision of state-of-the-art emulators at the 2-halo to 1-halo transition regime at all redshifts. Specifically, the WHM achieves better than 2\% accuracy up to scales of $k = 0.4\, h\,\mathrm{Mpc}^{-1}$, $0.7\, h\,\mathrm{Mpc}^{-1}$, and $1.3\, h\,\mathrm{Mpc}^{-1}$ at redshifts $z = 0.0$, $0.8$, and $1.5$, respectively, for both the \texttt{baccoemu} and \texttt{EuclidEmu2} emulators, across their full $w_0w_a\mathrm{CDM} + \sum m_\nu$ cosmological parameter space. This marks a substantial improvement over 1$\ell$-LPT and \texttt{HMcode-2020}, the latter of which performs similarly at low redshift but deteriorates at higher redshifts despite its 12 tuned parameters. We publicly release WHM as \texttt{WHMcode}, integrated into existing \texttt{HMcode} implementations for \texttt{CAMB} and \texttt{CLASS}.

\end{abstract}

% Select between one and six entries from the list of approved keywords.
% Don't make up new ones.

%%%%%%%%%%%%%%%%%%%%%%%%%%%%%%%%%%%%%%%%%%%%%%%%%%

%%%%%%%%%%%%%%%%% BODY OF PAPER %%%%%%%%%%%%%%%%%%

\section{Introduction}
\label{sec:intro} 
%This is a simple template for authors to write new MNRAS papers.
%See \texttt{mnras\_sample.tex} for a more complex example, and \texttt{mnras\_guide.tex}
%for a full user guide.

Extracting the full scientific potential of current and future large-scale structure (LSS) surveys requires highly accurate predictions of non-linear matter clustering. Central to this effort is the non-linear matter power spectrum $P_\mathrm{}^\mathrm{NL}(k)$, which underpins theoretical models of virtually all LSS observables. In particular, weak gravitational lensing, a primary probe of the matter distribution, directly measures a projection of $P_\mathrm{}^\mathrm{NL}(k)$, making its accurate modelling critical for lensing surveys such as the Kilo Degree Survey (KiDS, \citealt{Wright_2025}) the Dark Energy Survey (\citealt{DES:2025xii}), the Hyper Suprime-Cam (HSC, \citealt{Dalal:2023olq}), Euclid (\citealt{euclid2025arXiv250315302E,euclid2025A&A...697A...1E}), and Rubin (\citealt{lsst2022ApJS..258....1B}). 
Beyond weak lensing there are many other LSS observations of biased tracers of the matter density field, such as the $21\,$cm-line (\citealt{pritchard2012RPPh...75h6901P}), the Lyman-$\alpha$ forest (\citealt{DESI:2025zpo,DESI:2024lzq,Ravoux:2025uik}), Sunyaev Zel'dovich photons (\citealt{1972CoASP...4..173S,2021ApJ...908..199R}), and galaxies
(\citealt{desidr1fs2024arXiv241112021D,desidr1fscosmo2024arXiv241112022D,desidr2bao2025arXiv250314738D,DESIDR1cosmo2025JCAP...02..021A,desidr1bao2025JCAP...04..012A}) which critically rely on accurate models of $P_\mathrm{}^\mathrm{NL}(k)$.

There are many different approaches to predict $P_\mathrm{}^\mathrm{NL}(k)$ using perturbative, phenomenological or simulation-based strategies. On relatively large scales (small $k$), the matter power spectrum can be modelled with Perturbation Theory (PT, \citealt{BERNARDEAU_2001}). However, it fails on smaller scales, where density fluctuations grow up to order $\mathcal{O}(1)$, such that the perturbative series diverges. Within the Effective Field Theory of large-scale structure (EFTofLSS, \citealt{Baumann2012JCAP...07..051B,Senatore2020JCAP...05..005D,Ivanov2020JCAP...05..042I}) framework, these PT divergences are compensated by effective `counterterms', that cannot be predicted \textit{a priori} and need to be marginalized over. These additional parameters not only lead to volume effects complicating the data analysis (\citealt{Carrilho_2022_priors}), but also have a restricted regime of validity up to mildly non-linear scales only. For precise predictions of $P_\mathrm{}^\mathrm{NL}(k)$ on smaller scales (high $k$), full N-body simulations (\citealt{Springel_2005}) calculating the gravitational force on each matter particle in infinitesimal timesteps are needed. However, these come with a significant computational cost and are unfeasible to perform at each step in the cosmological parameter inference. Instead, they can be run for a predefined grid of cosmological models that is used as a training set for a Neural Network to interpolate between grid points. This technique is used to built so-called emulators such as \texttt{baccoemu} \citet{Angulo_2021}, \texttt{EuclidEmu2} \citet{Euclid:2020rfv}, \texttt{CosmicEmu} \citet{Heitmann_2014}, and \texttt{MiraTitan} \citet{MoranMiraTitan-2023}. Their disadvantage is their restriction to the cosmological model and parameter boundaries used for training. 

The halo model framework (\citealt{Peacock_2000,Seljak_2000,Ma_2000}, reviewed in \citealt{Cooray_2002,Asgari_2023}) attempts to combine the best of both the PT and N-body simulation worlds: it recovers PT on large scales and extends PT towards smaller scales via the notion that gravitational collapse does not lead to singularities (leading to divergencies in PT) but rather to virialised objects, called haloes. Their abundance can be estimated using the excursion set formalism \citet{Cooray_2002} based on spherical (or ellipsoidal) collapse theory. Their internal structure can be extracted from N-body simulations \citet{NFW_1997}. Then it is assumed that all \textit{inter-halo} clustering can be described by PT as a `2-halo' term, whereas the beyond PT behaviour is captured via the \textit{intra-halo} or `1-halo' term. However, this split turns out to be too simplistic when compared with N-body simulation results. While the halo model is able to recover the large- and small-scale limits, it fails severely on intermediate scales, the 2-halo to 1-halo transition regime, by up to 30-40\% \citet{Mead_2021}. At the same time, as first noted by \citet{Mead_2016b}, the \textit{relative} response of halo model calculations to extended cosmological models such as varying dark energy or modified gravity is quite accurate compared to simulations. The \texttt{ReAct} \citet{Cataneo_2019a,Cataneo_2019b,Bose_2020,Bose_2021} formalism takes advantage of this by multiplying precise $\Lambda$CDM emulator predictions with the `halo model reaction' to obtain good non-standard cosmology predictions of $P_\mathrm{}^\mathrm{NL}(k)$. This showcases the predictive power of the halo model even in the era of precision cosmology despite its simplistic nature. 

On the other hand, there have been many attempts to ameliorate the \textit{absolute} uncertainty of the halo model, either from the PT perspective (using the Zel'dovich approximation (ZA, \citet{1970A&A.....5...84Z}) \citet{Mohammed_2014,Seljak_2015,Sullivan_2021} or the EFTofLSS \citet{Philcox_2020}) or from the N-body simulation perspective (phenomenological fits such as \texttt{Halofit} \citet{Smith_2003,Takahashi_2012,Bird_2014,Smith_2019}, \texttt{HMcode} \citet{Mead_2015,Mead_2016a,Mead_2021}, or the inclusion of non-linear halo bias \citet{MeadVerde_2021}). But all these improvements come at the cost of additional fitting parameters that are purely phenomenological and not derived from first principles. This is a concern, given that the purpose of these augmented halo models is to extrapolate beyond the range of the emulators they were fitted to. But if the fitting parameters and their functional forms have been tuned to the emulators, it is impossible to demonstrate that they are still valid beyond the emulated range. Therefore, any analytic model can only compete with emulators if it is as accurate and at the same time parameter-free. While the modelling of the low-$k$ (validity of PT) and high-$k$ (dominance of the 1-halo term) regions is well under control, the transition region from the 2-halo to the 1-halo term on scales $0.3 \hMpc < k < 2 \hMpc$ is notoriously hard to model. At the same time, these scales correspond to the sweet-spot where LSS observations are least affected by cosmic variance (dominating at low-$k$) and shot noise (dominating at high-$k$, not to mention baryonic feedback uncertainties) and hence have the potential to deliver most of the constraining power. Thus, it is important to improve our theoretical understanding of this region in order to optimally extract information from LSS data. 

A natural path to explore is combining the halo model with the notion that they are embedded within the `cosmic web' structure (e.g., \citealt{IckevdW1987,Bond_1996CosmicWeb}). Changing halo properties based on their cosmic web environment has already been studied by, e.g., \citet{GilMarin2011MNRAS}, \citet{Alonso:2014zfa} and \citet{Voivodic2020JCAP}. In this work, we instead treat haloes as substructure of other collapsed structures, i.e., filaments and sheets, analogous to the treatment of halo substructure in \citet{Sheth_2003}. We do not include matter in voids, since uncollapsed matter is naturally included in PT. By incorporating filaments and sheets into the halo model language, we arrive at its
physical extension, the \textit{web-halo model} (WHM). We demonstrate its excellent accuracy compared to various emulators without any free parameters across all redshifts and scales, unaffected by other uncertainties such as baryonic feedback. The model acknowledges that the ellipsoidal collapse calculation within excursion set theory not only provides halo properties (such as the halo mass function from \citet{Sheth_2001} routinely used within the standard halo model) but also the formation history of filaments and sheets as described by \citet{Shen_2006}. Incorporating them naturally provides a smoother 2-halo to 1-halo transition, in agreement with N-body simulation results. 

In short, the reasoning behind our WHM is as follows. We assume PT is valid up to `shell-crossing', where PT diverges. As an idealised scenario, consider a homogeneous sphere collapsing isotropically into a black hole at its center, such that radial shells do not cross until entering the black hole. In this case, the exact collapse evolution can be described perturbatively up to order $N$, except for the central singularity, where the series diverges even for $N\rightarrow\infty$. Of course, rather than forming black holes, collisionless cold dark matter further propagates through the centre (meaning that inner shells start crossing outer shells and PT fails) various times and instead form stable haloes at the end of this `virialisation' process. As described before, the halo model considers the knowledge of these `end-products' to regularise the PT divergence at shell-crossing. In reality, collapse is not spherical, but ellipsoidal. Hence, collapse occurs along each of the three spatial dimensions, separately. Matter shells collapse along a preferred axis (dimension 1), forming (2D) sheets, then along the next preferred axis (dimension 2), forming (1D) filaments, before finally collapsing along the remaining axis (dimension 3) into (point-like, 0D) haloes. Shell crossing already occurs at the first step of collapse, the creation of sheets. Equivalently to the reasoning of the (spherical) halo model, for our (ellipsoidal) WHM, we consider a 1-sheet (1-filament, and 1-halo) term regularising the (PT, PT+sheet, and PT+sheet+filament) divergence at the shell crossing associated with each step of collapse. In other words, we recognise that even at $N=\infty$ order PT fails to describe the formation of sheets (also see, e.g., \citealt{Bharadwaj_1996,McQuinn_2016}) and use a 1-sheet term representing the beyond $N=\infty$ order of PT, the so-called `sheet-order'. The 1-filament and 1-halo term correspond to further regularisations associated with further collapsed dimensions representing the `filament-order' and `halo-order', respectively. We hence identify the reason for the inaccurate 2-halo to 1-halo transition regime in the standard halo model as its inability to describe the stages of collapse before forming haloes, also not covered by PT.

The WHM framework and its differences with respect to the standard halo model are described in section \ref{sec:whm}, while the detailed model ingredients are specified in section \ref{sec:ing}. The non-linear matter power spectrum results are presented and compared to various emulator predictions in section \ref{sec:res}. Finally, we provide a prospect for further improving the agreement in the 1-halo term using N-body simulations in section \ref{sec:1h} before concluding with section \ref{sec:concl}. 

\section{Web-halo model}
\label{sec:whm} 

In this section, we present the WHM step by step. We introduce our notation in section \ref{sec:whm-definitions} and briefly recapitulate the standard halo model formulation pioneered by \citet{Peacock_2000,Seljak_2000,Ma_2000} in section \ref{sec:whm-halomodel}. We summarise its shortcomings and potential solutions in section \ref{sec:whm-solutions}, paving the way for section \ref{sec:whm-webhalomodel}, where we finally provide its extension to the WHM.

\subsection{Definitions}
\label{sec:whm-definitions} 

Throughout this work, we use interchangeably either the scale-invariant power spectrum $\Delta^2(k,z)$ or the standard power spectrum $P(k,z)$, which are related to each other as
\begin{equation}
\left(\Delta^\mathrm{X}\right)^2(k,z)=4\pi\left(\frac{k}{2\pi}\right)^3P^\mathrm{X}(k,z)\ .
\label{eq:whm-def-power}
\end{equation}
The superscript `X' refers to the method of computation, i.e. `PT', `L', `HM', `WHM', `1h', `2h', `NL', for the general PT, linear PT, halo model, web-halo model, 1-halo, 2-halo, or fully non-linear prediction, respectively. 
If the superscript is not specified (to keep the notation concise), we refer to linear PT by default. 

Moreover, we find it useful to define the variance of the linear field, filtered on a comoving scale $R$, given as
\begin{equation}
\sigma^2(R,z)=\int_0^{\infty}\Delta^2(k,z)\, W_\mathrm{t}^2(kR)\;\mathrm{d}\ln{k}\ ,
\label{eq:whm-def-variance}
\end{equation}
where $W_\mathrm{t}(x)$ is the Fourier transform of the real-space top-hat filter
\begin{equation}
W_\mathrm{t}(x)=\frac{3}{x^3}(\sin{x}-x\cos{x})\ .
\label{eq:whm-def-tophat}
\end{equation}

Note that most of the quantities introduced here depend on wavenumber $k$ and redshift $z$. However, in what follows, we may omit the redshift dependence in case it is not relevant, to make the notation less cluttered.

\subsection{Brief overview of the standard halo model formalism}
\label{sec:whm-halomodel}

The standard halo model prediction $P_\mathrm{}^\mathrm{HM}$ for the non-linear matter power spectrum is given by the sum of a two-halo term $P_\mathrm{}^\mathrm{2h}$ describing inter-halo clustering and a one-halo term $P_\mathrm{}^{\mathrm{1h}}$  describing intra-halo clustering of matter,
\begin{equation}
    P_\mathrm{}^\mathrm{HM}(k) = P_\mathrm{}^\mathrm{2h}(k) + P_\mathrm{}^\mathrm{1h}(k)\ .
    \label{eq:whm-hm-powersum}
\end{equation}

In this paper, we retain this basic 1-halo + 2-halo superposition, but we modify both of these terms from their usual forms. Firstly, in Section \ref{sec:whm-solutions-he}, we follow previous suggestions for dealing with the effects of mass conservation
on the 1-halo term. Then, in Section \ref{sec:whm-webhalomodel}, we introduce a novel recipe for computing the 2-halo term, which goes beyond PT and explicitly accounts for the fact that haloes are substructure of the cosmic web.

Back to the standard halo model, assuming that all sources of non-linearities are isolated to intra-halo scales, i.e., the halo-halo power spectrum can be written as
\begin{align}\label{eq:whm-hm-linearhalobias}
    P_\mathrm{hh}(k,M_1,M_2) = b_h(M_1)b_h(M_2)P^\mathrm{L}(k)~,
\end{align}
the terms in equation \eqref{eq:whm-hm-powersum} are given by
\begin{align}
P_\mathrm{}^\mathrm{2h}(k) &\equiv
 \int_0^\infty \! \int_0^\infty \!  \, W_\mathrm{h}(k,\nu_1)\, f_\mathrm{h}(\nu_1) \; P_\mathrm{hh}(k, \nu_1, \nu_2) \nonumber \\
& \qquad \qquad ~~ \cdot W_\mathrm{h}(k,\nu_2)\, f_\mathrm{h}(\nu_2)\;\mathrm{d}\nu_1 \mathrm{d}\nu_2
\label{eq:whm-hm-twohalo-full} \\
&\approx P^\mathrm{L}_\mathrm{}(k)
\left[\int_0^\infty b_\mathrm{h}(\nu)\, W_\mathrm{h}(k,\nu)\, f_\mathrm{h}(\nu)\;\mathrm{d}\nu\right]^2\ ,
\label{eq:whm-hm-twohalo} \\
P_\mathrm{}^\mathrm{1h}(k) &\equiv\frac{1}{\bar{\rho}}\int_0^\infty M(\nu)\, W_\mathrm{h}^2(k,\nu)\, f_\mathrm{h}(\nu)\;\mathrm{d}\nu\ ,
\label{eq:whm-hm-onehalo}
\end{align}
with halo mass $M(\nu)$ which is related to the peak-height $\nu$ via
\begin{equation}
\nu=\frac{\delta_\mathrm{crit}}{\sigma_\mathrm{cc}(R(M))} \ , \quad \mathrm{with} \qquad R(M)= \left(\frac{3M}{4\pi \bar{\rho} }\right)^{1/3} \ , 
\label{eq:whm-hm-peakheight}
\end{equation}
where $\delta_\mathrm{crit}$ is the spherical collapse linear theory threshold overdensity and $\sigma_\mathrm{cc}(R(M))$ is defined in equation~(\ref{eq:whm-def-variance}). Note that, when considering cosmologies including massive neutrinos, 
\citet{Castorina_2014JCAP} showed (also see \citealt{Massara_2014}) that halo model calculations are more universal when using the cold matter field (`cc', not including massive neutrinos) instead of the total matter field. The filter scale $R(M)$ corresponds to the radius of a homogeneous sphere with constant density given by the mean background density $\bar{\rho}$ and mass $M$. Hence, the variables $\nu, M,R$ follow a deterministic relationship, and in what follows, we use them interchangeably within function arguments.

Next, $b_\mathrm{h}(\nu)$ is the linear scale-independent halo bias and $f_\mathrm{h}(\nu)$ is the normalised halo mass function, corresponding to the probability density for finding a halo of a given mass.
In case of describing the matter field, these functions must fulfil the conditions
\begin{equation}
\int_0^\infty f_\mathrm{h} (\nu)\;\mathrm{d}\nu = 1, \quad \mathrm{and} \qquad  \int_0^\infty b_\mathrm{h}(\nu) f_\mathrm{h} (\nu)\;\mathrm{d}\nu = 1\ , 
\label{eq:whm-hm-conditions}
\end{equation}
meaning that, at all times, all matter must be associated with haloes and, on average, matter must be unbiased with respect to itself. 

Finally, the two-halo and one-halo terms depend on the spherical Fourier transform of the normalised halo density profile $\rho_\mathrm{h}(M,r)/M$, i.e., the halo window function
\begin{equation}
W_\mathrm{h}(k, M)=\frac{1}{M}\int_0^{R}4\pi r^2\frac{\sin(kr)}{kr}\rho_\mathrm{h}(r, M)\;\mathrm{d}r\ .
\label{eq:whm-hm-window}
\end{equation}

Since haloes are virialised objects, their density profile is normally set to zero for radii larger than their virial radius 
$r_\mathrm{v}$, i.e., $\rho_\mathrm{h}(M,r) \propto \Theta(r_\mathrm{v}-r)$ with $\Theta(x)$ being the Heaviside step function. Hence, all halo mass is assumed to reside within its virial radius $r_v$, which is related to the initial radius $R$ of the total collapsed region via the virial overdensity 
\begin{equation}
\Delta_\mathrm{v} = \left(\frac{R}{r_\mathrm{v}}\right)^3 . 
\label{eq:whm-hm-virialoverdensity}
\end{equation}

We have now described all the ingredients that appear within the integrals of equations \eqref{eq:whm-hm-twohalo} and \eqref{eq:whm-hm-onehalo}. However, there are many options for their functional forms. For an in-depth comparison between various choices for mass functions, the mass dependence of density profiles, and the cosmology dependence of $\delta_\mathrm{crit}$ and $\Delta_\mathrm{v}$, we refer to \citet{Mead_2021}. We adopt their baseline ingredient choices as our baseline presented in section ~\ref{sec:ing}.   

\subsection{Shortcomings of the halo model and potential solutions}
\label{sec:whm-solutions}

The strengths of the halo model clearly rely on its simplicity, low computational cost, and physical intuition.  However, it has not been intended to accurately describe the clustering of matter in the first place. Its initial intention has been to compute rough estimates of halo occupation densities for the analysis of galaxy clustering. Therefore, it is no surprise that the halo model comes with several shortcomings regarding the calculation of non-linear matter power spectra. Here, we provide a concise overview of attempts in the literature to address these issues and connect them to our own potential solutions, foreshadowing the WHM formalism introduced in section~\ref{sec:whm-webhalomodel}.

\subsubsection{2-halo to 1-halo transition regime}
\label{sec:whm-solutions-2-1}

A fundamental problem of the halo model is its inability to accurately model the 2-halo to 1-halo transition regime. \citet{Mead_2021} showed that even for the most optimal choice of ingredients, both the mean deviation between halo model predictions and emulator results and their scatter reach a maximum in that regime of scales, around $k = 0.5h$/Mpc. In particular, the mean deviation and scattering at that scale are of order $20\%$ and $5\%$, respectively. 

Many attempts to address this issue can be found in the literature. \citet{Smith_2003} introduced a `quasilinear' term to equation~(\ref{eq:whm-hm-powersum}), \citet{Mohammed_2014} use a polynomial while \citet{Seljak_2015} and \citet{Sullivan_2021} use a Pade expansion of the 1-halo term, respectively. \citet{Philcox_2020} add a Pade-resummed counterterm to the two-halo term, and all \texttt{HMcode} versions (\citealt{Mead_2015,Mead_2016a,Mead_2021}) introduce a "transition smoothing" parameter that depends on the effective spectral tilt of the linear power spectrum. However, all these attempts are of phenomenological nature and do not derive from first principles. 

The purpose of this work is to provide a parameter-free treatment of the transition regime, which is exactly the regime of intermediate scales where the impact of collapsed structures beyond haloes, i.e., filaments and sheets, is expected to be important. The corresponding formalism is introduced in section \ref{sec:whm-webhalomodel}.

\subsubsection{Non-linear halo bias}
\label{sec:whm-solutions-nhb}

The elegance of the halo model stands (and falls) with its ability to separate linear and non-linear structure formation via the 2-halo and 1-halo split of equation \eqref{eq:whm-hm-powersum}. Ideally, all sources of non-linearities should be confined to intra-halo scales, and the 2-point clustering of halo centres should be described by linear PT assuming linear, scale-independent halo bias. In that case, equation \eqref{eq:whm-hm-linearhalobias} holds, such that the double integral in equation \eqref{eq:whm-hm-twohalo-full} simplifies to equation \eqref{eq:whm-hm-twohalo}.

Unfortunately, comparisons with higher-order PT and full N-body simulations demonstrate that this approximation is far from the truth \citep{Scocc2001ApJ,Desjacques:2016bnm}. This was explicitly demonstrated by \citet{MeadVerde_2021} who built an emulator for $P^\mathrm{NL}_\mathrm{hh}(k, \nu_1, \nu_2)$, i.e., the non-linear halo bias, inserted it in equation \eqref{eq:whm-hm-twohalo-full}, and recovered accurate halo model predictions of the non-linear matter power spectrum, even in the 2-halo to 1-halo transition regime. The fact that the accuracy of the halo model ansatz of equation \eqref{eq:whm-hm-twohalo-full} relies on emulated non-linear halo power spectra as input, challenges the halo model's usability as a prediction scheme for non-linear power spectra in the first place.

One explanation provided by \citet{MeadVerde_2021} for the necessity of modelling the non-linear halo bias is the contribution of filamentary, sheet, and void structure, i.e., the cosmic web, on intermediate scales. The WHM, which we will introduce in section~\ref{sec:whm-webhalomodel}, aims to explicitly incorporate these structures in the form of additional 1-sheet and 1-filament terms, bridging the transition from the 2-halo (or rather 2-sheet) term to the 1-halo term.
We isolate the perturbative scales within the 2-sheet term, and describe the non-linear scales beyond PT via the 1-sheet, 1-filament, and 1-halo terms. This restores the original halo model idea of separating perturbative and non-perturbative scales by separating the physics.

\subsubsection{Halo exclusion and stochasticity}
\label{sec:whm-solutions-he}

Another fundamental problem is the halo model's incapacity to consider 'halo exclusion', i.e., the fact that haloes cannot overlap in space. Despite many attempts, see \cite{Asgari_2023} and references therein, a first-principle analytic treatment does not yet exist (apart from \citealt{ShethLemson_1999} for Poissonian initial conditions).
One manifestation of this problem is that on very large scales, i.e., in the limit $k \rightarrow 0$ and $W_\mathrm{h}(k \rightarrow 0, \nu) \rightarrow 1$, the 1-halo term in equation \eqref{eq:whm-hm-onehalo} converges towards a constant shot-noise term $P_\mathrm{}^\mathrm{1h}(k\rightarrow0) = \left<M\right>/\bar{\rho}$, with $\left<M\right>$ the mean mass per halo. Eventually, this shot-noise term dominates over the two-halo term on large scales, a correct behaviour for discrete tracers such as haloes and galaxies, but not observed for the matter power spectrum in simulations.

Correctly modelling halo exclusion resulting in sub-Poissonian noise on large scales could solve this issue, as shown by \citet{Smith_2007} and \citet{Baldauf_2013}. The latter introduced an effective halo exclusion scale $R_\mathrm{excl}$ as a free parameter below which the halo correlation function becomes $-1$. This approach is good enough for discrete tracers of matter occupying a narrow halo mass range, but does not work when describing the matter power spectrum itself, since matter is distributed across a huge dynamic range of halo masses and radii.

But even if we were able to model halo exclusion correctly, we would still be in the paradoxical situation that when describing the continuous, deterministic matter fluid (left hand side of equation \ref{eq:whm-hm-powersum}), we rely on discrete, stochastic tracers (right hand side) as a detour for describing it. In other words, halo clustering is stochastic in nature, while matter clustering is not, as it needs to fulfil conservation laws. This paradox is solved \textit{a posteriori} via the integral constraints in equation \eqref{eq:whm-hm-conditions}, but does not compensate for the unphysical noise term in the low $k$ limit of the 1-halo term.

\subsubsection{Haloes as ``substructure'' of the perturbative field}
\label{sec:whm-solutions-sub}

In this work, we do not model halo exclusion explicitly, but instead remove large-scale noise from the 1-halo term by transforming the halo window via 
\begin{equation}
    \widehat{W}_\mathrm{h}^2(k, R) =  W_\mathrm{h}^2(k, R) - W_\mathrm{PT}^2(k, R)\ ,
    \label{eq:whm-s-compensation}
\end{equation}
where $W_\mathrm{PT}$ is the window the halo would have before shell crossing, i.e., following the particle trajectories at given order in PT. Here, haloes can be interpreted as substructure of the ``primordial'' haloes before their collapse, described purely by PT. In analogy to \citet{Sheth_2003} who described substructure within haloes, we arrive at equation \eqref{eq:whm-s-compensation} assuming that all mass collapsed into haloes. This approach is also similar to the `window compensation' scheme introduced by \citet{Baldauf_2016} to estimate the `noiseless' 1-halo term. For example, at zeroth order in PT, the window $W_\mathrm{PT}(k, R) = W_\mathrm{t}(kR)$ corresponds to a homogeneous spherical top-hat. As a consequence, the 1-halo term
\begin{equation}
    \widehat{P}_\mathrm{}^\mathrm{1h}(k) =  P_\mathrm{}^\mathrm{1h}(k, W_\mathrm{h}^2\rightarrow \widehat{W}_\mathrm{h}^2) = P_\mathrm{}^\mathrm{1h}(k) - P_\mathrm{}^\mathrm{1t}(k) \xrightarrow{k\rightarrow 0} 0 \ ,
    \label{eq:whm-s-phcomp}
\end{equation}
is subtracted by the equivalent pre-collapse 1-halo (or 1-top-hat) term. As such, it tends towards zero on large scales, where the halo and top-hat window both converge to one. Note that \citet{Valageas_2011a} follow a similar procedure, where they identify the subtraction of $W_\mathrm{t}^2(kR)$ as a counterterm. In contrast, we interpret equation \eqref{eq:whm-s-compensation} as mass conservation of halo substructure within the perturbative field as parent structure (partitioning of mass) rather than halo exclusion (partitioning of volume).

\subsubsection{Large scale limit}
\label{sec:whm-solutions-lsl}

While equation \eqref{eq:whm-s-phcomp} ensures the correct large-scale limit of the 1-halo term, applying a similar window compensation to the 2-halo term in eq.~\eqref{eq:whm-hm-twohalo} would result in zero power on large scales, i.e., prohibit recovery of the PT prediction.
\citet{Cooray_2002} already noted this issue with their halo profile compensation scheme. \citet{Chen_2020} solved this by adding the PT prediction to the compensated 1-halo and 2-halo terms. 
Following a similar but slightly different logic, we simply set
\begin{equation}
P_\mathrm{}^\mathrm{2h}(k)=P^\mathrm{PT}_\mathrm{}(k)\ ,
\label{eq:whm-s-P2hPlin}
\end{equation}
i.e., assume that the 2-halo term (or later on the 2-sheet term) can be described perturbatively and neglect the integral term in equation \eqref{eq:whm-hm-twohalo}. In fact, \citet{Mead_2021} have shown that this term is subdominant with respect to the 1-halo term. Furthermore, thanks to equation \eqref{eq:whm-s-P2hPlin}, we avoid the detour of modelling the halo power spectrum with its underlying caveats mentioned in sections \ref{sec:whm-solutions-nhb} and \ref{sec:whm-solutions-he} and focus on the matter-matter clustering statistics, which can be conveniently described with PT. 

In summary, and in preparation for section \ref{sec:whm-webhalomodel}, we formulate our modified halo model prediction as a sum of the PT term and the 1-halo term with stochasticity removed (via the substructure treatment),
\begin{equation}
P^\mathrm{HM}_\mathrm{}(k) = P^\mathrm{PT}_\mathrm{}(k) +  \widehat{P}_\mathrm{}^\mathrm{1h}(k)~,
    \label{eq:whm-s-powersum}
\end{equation}
which recovers PT at large scales and confines all non-linear, beyond perturbative, scales to the 1-halo term. As said, the latter assumption is not accurate, but paves the way for including the cosmic web.

\subsection{Extension towards the web-halo model}
\label{sec:whm-webhalomodel}

Here, we present the formalism to incorporate collapsed structures beyond haloes, i.e. filaments and sheets. It primarily builds on the notion of triaxial collapse, as described by \citet{Shen_2006}. Their ellipsoidal excursion set theory, reviewed in more detail in section \ref{sec:ing}, considers that collapse is not spherical, but occurs along three dimensions. An overdense sphere, due to tidal fields or angular momentum, first collapses along one pronounced dimension into a planar structure (a 2D sheet), before collapsing along the second dimension into filament-like, 1D  structures and finally along the third dimension into (virialised) `points' (0D). This is true from the smallest scales like the solar system (gas cloud collapses into a plane, forms rings around the center, and finally planets orbiting around the Sun) to the largest scales like the formation of the cosmic web.\footnote{Note that the collapse of a gas cloud into a star takes a blink of an eye compared to cosmic time-scales, whereas the collapse of all matter into the central nodes of the cosmic web is still ongoing.}

We combine the ellipsoidal collapse picture with the reasoning of sections \ref{sec:whm-halomodel} and \ref{sec:whm-solutions} as follows. We assume that the PT prediction for the matter power spectrum $P^\mathrm{PT}_\mathrm{}(k)$ is correct until some scale before shell crossing, i.e., the formation of sheets.\footnote{For PT to be strictly correct up to shell crossing, it would need to be evaluated up to infinite order. Therefore, when cutting the perturbative series at order N, the 1-sheet term can be seen as an extension beyond order infinity.} We introduce a 1-sheet term $\widehat{P}^\mathrm{1s}_\mathrm{}$ accounting for the sheet profile. These two terms fully describe matter clustering at the `sheet-order' and can be interpreted as the 2-filament $P^\mathrm{2f}_\mathrm{}$ term. We add a 1-filament term $\widehat{P}^\mathrm{1f}_\mathrm{}$ describing the field at the `filament order' that can be interpreted as the 2-halo $P^\mathrm{2h}_\mathrm{}$ term. Finally, we add the 1-halo term $\widehat{P}^\mathrm{1h}_\mathrm{}$ resulting in the full `halo order' expression

\begin{equation}
P^\mathrm{WHM}_\mathrm{}(k) = \underbrace{\underbrace{\underbrace{\underbrace{P^\mathrm{PT}_\mathrm{}(k)}_{P^\mathrm{2s}_\mathrm{}: ~\mathrm{PT~order}} \!\!\!+\, \widehat{P}^\mathrm{1s}_\mathrm{}(k)}_{P^\mathrm{2f}_\mathrm{}:~\mathrm{Sheet~order}} \,+\, \widehat{P}^\mathrm{1f}_\mathrm{}(k)}_{P^\mathrm{2h}_\mathrm{}:~ \mathrm{Filament~order}} \,+\, \widehat{P}^\mathrm{1h}_\mathrm{}(k)}_{\mathrm{P^\mathrm{NL}_\mathrm{}:~ Halo~order}}~,
\label{eq:whm-whm-powersum}
\end{equation}
with
\begin{align}
\widehat{P}^\mathrm{1s}_\mathrm{}(k) &=\frac{1}{\bar{\rho}} \int_0^\infty M(\nu)\, \left[W_\mathrm{s}^2(k, \nu) - W_\mathrm{PT}^2(k, \nu)\right]\, f_\mathrm{s}(\nu)\;\mathrm{d}\nu\ \label{eq:whm-whm-1s} \\
\widehat{P}^\mathrm{1f}_\mathrm{}(k) &=\frac{1}{\bar{\rho}} \int_0^\infty M(\nu)\, \left[W_\mathrm{f}^2(k, \nu) - W_\mathrm{s}^2(k, \nu)\right]\, f_\mathrm{f}(\nu)\;\mathrm{d}\nu\  \label{eq:whm-whm-1f} \\
\widehat{P}^\mathrm{1h}_\mathrm{}(k) &=\frac{1}{\bar{\rho}} \int_0^\infty M(\nu)\, \left[W_\mathrm{h}^2(k, \nu) - W_\mathrm{f}^2(k, \nu)\right]\, f_\mathrm{h}(\nu)\;\mathrm{d}\nu ~.
\label{eq:whm-whm-1h}
\end{align}
Here, we generalize the halo model window compensation of equation \eqref{eq:whm-s-compensation} towards the web-halo model by subtracting the window of the `lower order', when going from order N in PT to the sheet order (equivalent to beyond order infinity, i.e., shell crossing), filament order, and finally the halo order, respectively. Consequently, the 1-halo term $\widehat{P}^\mathrm{1h}_\mathrm{}$ in equation \eqref{eq:whm-whm-powersum} differs from the one in \eqref{eq:whm-s-powersum}. \sB{The newly added symbols $W_\mathrm{s}, W_\mathrm{f}, f_\mathrm{s}$ and $f_\mathrm{f}$ correspond to the sheet (subscript s) and filament (subscript f) window functions and mass functions and will be introduced in section \ref{sec:ing}.}

Note that the integrands in equations (\ref{eq:whm-whm-1s}-\ref{eq:whm-whm-1h}) can be summed before carrying out the integration, such that the additional computational cost for evaluating the filament and sheet terms is minimal.

\begin{figure*}
    \centering
    \includegraphics[width=\textwidth]{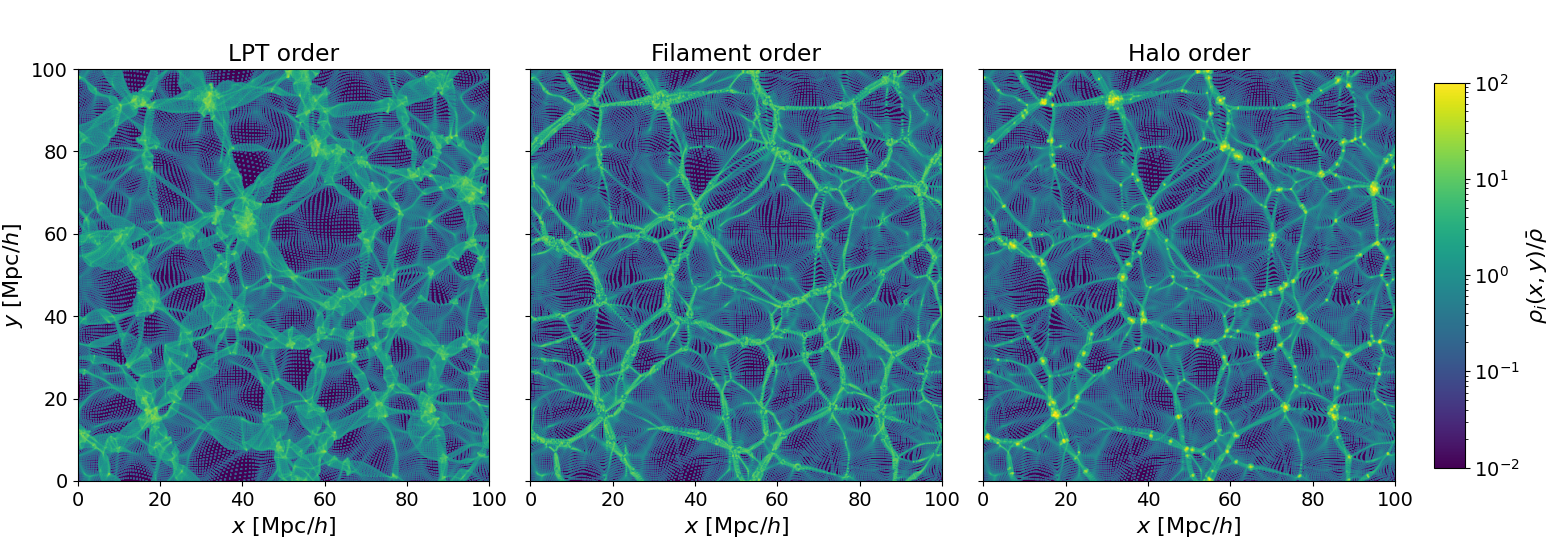}
    \caption{We show the resulting $z=0$ density fields of a toy 2D N-body simulation illustrating the logic behind the WHM workhorse equations \eqref{eq:whm-whm-powersum}-\eqref{eq:whm-whm-1h}. The left panel depicts the field assuming ballistic particle trajectories (a single timestep, ZA). The resulting power spectrum would be described by tree-level LPT. The middle panel displays the fully evolved field, but with haloes removed and their mass redistributed into filaments. Its power spectrum would be described by adding a 1-filament term compensated by the LPT filament profiles to the `LPT order', yielding the `filament order' (note that there is no sheet order in 2D). Finally, the right panel shows the exact fully evolved field including haloes. Within our WHM, its power spectrum would be described by the `halo order' that adds to the filament order a 1-halo term compensated by the power these haloes had contributed when they were filaments.}
    \label{fig:whm-LFH}
\end{figure*}

We stress that equation \eqref{eq:whm-whm-powersum} represents an extension not only of the standard halo model but also of PT. It describes non-linear clustering beyond the perturbative regime by adding the sheet, filament, and halo orders corresponding to the different stages of collapse indicated above. The sheet order describes the field in case we would smooth the matter field such that the substructure in sheets, i.e. their filaments and haloes, is removed. The filament order resolves sheets and filaments, but the filament substructure is smoothed out. At the halo order, we arrive at the most basic ingredient of the matter field, resolving it at all cosmological scales (neglecting halo substructure). 

To intuitively illustrate this picture, we show in figure \ref{fig:whm-LFH} the result of a 2D N-body simulation\footnote{At \href{https://github.com/mpelleje/2D\_Nbody\_sim\_python}{https://github.com/mpelleje/2D\_Nbody\_sim\_python} the 2D simulation code is publicly available as a fork of the original code in \href{https://github.com/jstuecker/sheet-unfolding}{https://github.com/jstuecker/sheet-unfolding}.}
employed here for purely conceptual reasons.
The rightmost panel shows the full density field corresponding to the `halo order' described above. Next,  we redistribute the mass attributed to haloes\footnote{We find haloes using the friends-of-friends (FoF) algorithm \texttt{pyfof}, \href{https://github.com/simongibbons/pyfof}{https://github.com/simongibbons/pyfof}.} randomly into filaments.\footnote{We find filaments by requiring one of the two eigenvalues $\lambda_{1,2}$ of the 2D tidal tensor to be larger, the other one smaller than the threshold $\lambda_\mathrm{c}=1$. In principle, particles within haloes should be assigned to their parent filaments, but for simplicity (given the purpose of visualisation only), we assign all halo particles randomly to all filaments as described in the text.} The corresponding `filament order' density field is shown in the middle panel. Compared to the right panel, haloes are removed and filaments appear denser due to the halo-to-filament redistribution, ensuring mass conservation. Finally, the left panel shows the density field using the ZA, i.e., tree-level Lagrangian PT. While it predicts the large scale density fluctuations and filament positions quite well, the underlying assumption that particles evolve `ballistically' (no change in direction across evolution) leads to `puffed' filaments (or sheets in 3D). In section \ref{sec:ing-dp-pt} we present our ansatz for modelling the corresponding PT window functions $W_\mathrm{PT}(k,\nu)$ entering the compensation in equation \eqref{eq:whm-whm-1s}. Figure \ref{fig:whm-LFH} also helps to understand the purpose of the window function compensation mentioned. Since in each panel the total mass is the same, when going from `filament order' to `halo order', we cannot simply add mass via the standard 1-halo term. Instead, we also need to subtract the power arising from haloes when their mass was in filaments (recall the reduced filament density of the right compared to the middle panel). Likewise when going from `sheet order' to `filament order', and `PT order' to `sheet order'.     

This hierarchical structure illustrated above is also the reason why no cross-terms (like a sheet-filament or a filament-halo term) appear in equation \eqref{eq:whm-whm-powersum}. Note that the excursion set picture by \citet{Shen_2006} does \textit{not} assume matter would be \textit{either} in sheets, filaments \textit{or} haloes, in which case such cross-terms would be necessary. Instead, haloes reside within parent filaments, and filaments within parent sheets. Hence, any mass within a halo is necessarily \textit{also} attributed to a filament \textit{and} a sheet. And since \textit{all} mass is attributed to haloes, \textit{all} mass is also attributed to filaments and sheets, i.e., $f_\mathrm{s}(\nu)$ and $f_\mathrm{f}(\nu)$ fulfill the same integral constraint \eqref{eq:whm-hm-conditions} as $f_\mathrm{h}(\nu)$.\footnote{\sB{Note that within the substructure picture this ensures mass conservation and does \textit{not} mean there would be three times as much mass as expected.}} Consequently, there is no `void order', as the presence of voids is resolution dependent and, if any, uncollapsed mass in voids is incorporated at the PT order.

Our view of haloes as substructures of filaments, which are substructure of sheets, shares similarities to the treatment of halo substructure by \citet{Sheth_2003}. They investigate the presence of dark matter clumps within the smooth 1-halo component, giving rise to additional smooth-clump and clump-clump terms within the standard 1-halo term. In contrast to their more general approach, we work under the assumptions that 
\begin{itemize}
    \item[(i)] all sheet and filament substructures have collapsed into haloes, such that no (smooth-clump) cross terms arise, and
    \item[(ii)] every overdense sphere of mass $M$ collapses into one sheet, one filament, and finally one halo, each of the same mass $M$.
\end{itemize}
Assumption i) is always true up to the resolution limit, i.e., we can describe a smooth filament as a set of haloes with mass equal to the mass resolution. Assumption ii) can be motivated by the conditional mass function of \citet{Shen_2006}, finding that most mass within a sheet is indeed found in a filament with slightly smaller mass and a subsequent halo with slightly smaller mass. Independent of that, in appendix \ref{app:window} we show that assumption ii) still provides a good representation of reality even if a single filament collapses into 8 haloes with equally distributed mass. This justifies our simplified choices that allow us to write equations \eqref{eq:whm-whm-powersum}-\eqref{eq:whm-whm-1h} as separate integrals. A fully general prescription equivalent to \citet{Sheth_2003} would instead require nested integrals. We leave this topic to future work.

\section{Ingredient choices}
\label{sec:ing}

Before we can evaluate equations (\ref{eq:whm-whm-powersum}-\ref{eq:whm-whm-1h}), we have to choose their ingredients. I.e., we need to select the PT input (section \ref{sec:ing-pt}), the parameters $\delta_\mathrm{crit}$ and $\Delta_\mathrm{v}$ derived from spherical collapse (section \ref{sec:ing-sc}), mass functions $f(\nu)$ (section \ref{sec:ing-mf}), and window functions $W(k, \nu)$ (section \ref{sec:ing-dp}), for sheets, filaments, and haloes. While the latter have been studied extensively in the literature, the investigation of sheets and filaments is rather sparse. Already, the step of identifying sheets and filaments from N-body simulation output is highly non-trivial. Several methods exist, and their results differ greatly, as shown in the comparison project of \citet{Libeskind_2018}. Also, these methods depend on hyperparameters like the linking length much more sensitively than for halo identification, where their impact is also not negligible, see \citet{Knebe_2011}.
Therefore, while our choices (consistent with those of \citealt{Mead_2021}) for halo mass functions and window functions largely agree with simulation outputs, for sheets and filaments we can only follow well-motivated theoretical reasoning. In particular, we use sheet and filament mass functions (section \ref{sec:ing-mf}) and derive corresponding window functions (section \ref{sec:ing-dp}) based on the work of \citet{Shen_2006}. Based on excursion set theory, their results are independent of N-body simulation-based methods that still exhibit large uncertainties on hyperparameters described before. The excursion set is also closer to cosmic web dynamics than spatial statistics. I.e., a web environment classifier finds filaments as bridges between haloes, whereas in the excursion set, filaments embed one halo at their centre. 

\subsection{Perturbation Theory}
\label{sec:ing-pt}

We combine the WHM with two types of perturbation theory (PT), linear PT (section \ref{sec:ing-pt-lin}) and one-loop Lagrangian PT (section \ref{sec:ing-pt-lag}). 

\subsubsection{Linear PT} \label{sec:ing-pt-lin}
Linear PT enters the halo model and web-halo model at various levels. First and foremost, linear PT is needed to evaluate the filtered matter density variance $\sigma(R)$ in equation \eqref{eq:whm-def-variance} and the peak height $\nu$ via equation \eqref{eq:whm-hm-peakheight} (also see section \ref{sec:ing-sc}). Therefore, it is natural to also use linear PT to predict the matter power spectrum $P_\mathrm{}^\mathrm{PT}(k)$ in equation \eqref{eq:whm-whm-powersum}. For this we use the linear theory Boltzmann codes \texttt{CAMB}\footnote{\href{https://camb.info/}{https://camb.info/}} and \texttt{CLASS}\footnote{\href{http://class-code.net/}{http://class-code.net/}}. However, as we confirm later, the linear PT prediction is known to lack accuracy at quasi-linear scales, where the 1-halo, 1-filament and 1-sheet terms are still subdominant.  

\subsubsection{One-loop Lagrangian PT} \label{sec:ing-pt-lag}
Therefore, by default, we use the one-loop Lagrangian PT (1$\ell$-LPT) prediction for $P_\mathrm{}^\mathrm{PT}(k)$ in equation \eqref{eq:whm-whm-powersum} following \citet{Chen_LPT_2020,Chen_velo_2021}. Unlike one-loop Eulerian PT (1$\ell$-EPT), where the perturbed quantity is the overdensity field $\delta(\mathbf{x})$, 1$\ell$-LPT operates on the displacement field $\boldsymbol{\Psi}(\mathbf{q}) = \mathbf{x}(\mathbf{q})-\mathbf{q}$ between initial (Lagrangian) and final (Eulerian) particle positions $\mathbf{q}$ and $\mathbf{x}$, respectively. Its regime of validity extends up to the linear displacement field variance
\begin{align} \label{eq:ing-pt-sigmav}
    \sigma_\mathrm{v,cc}^2 = \frac{1}{6\pi^2}\int_0^\infty \! P_\mathrm{cc}^\mathrm{L}(k)\,\mathrm{d}k~.
\end{align}

There are a few fundamental reasons for choosing 1$\ell$-LPT instead of 1$\ell$-EPT within our web-halo model framework. First, the excursion set approach to derive mass functions (section \ref{sec:ing-mf}) also operates in Lagrangian space, and the associated ellipsoidal collapse model by \citet{Shen_2006} is inspired by the LPT predictions of the tidal field of \citet{Bond_1996}. Thus, LPT naturally predicts the formation of sheets, whereas EPT does not account for the formation of highly asymmetric, web-like structures. Second, studying EPT and LPT in one spatial dimension only, \citet{McQuinn_2016} showed that the EPT perturbative expansion converges towards the zero-order LPT prediction, which is exact up to shell crossing. While this result cannot be extrapolated to three spatial dimensions, the assumption that LPT's validity goes deeper towards the shell crossing regime than EPT is supported by its prediction of web-like structures. Related to that, the LPT perturbative expansion is better behaved than EPT, in the sense that it does not overpredict the full non-linear power spectrum at any order, unlike EPT. Therefore, 1$\ell$-EPT (which overpredicts $P_\mathrm{}^\mathrm{NL}(k)$) cannot be combined with the halo model (which is purely additive) without considering an additional (negative) counterterm introducing a free parameter as in \citet{Philcox_2020}.
Finally, by construction, LPT accounts for the suppression of BAO wiggles via infrared (IR) resummation \citet{Chen_LPT_2020}. This procedure does not naturally emerge within EPT frameworks, where it can only be added \textit{ad-hoc} via phenomenological dewiggling techniques, see \citet{Blas_2011}. 

To obtain our 1$\ell$-LPT power spectrum predictions, we use the \texttt{`lpt\_rsd\_fftw'} module within the \texttt{velocileptors}\footnote{\href{https://github.com/sfschen/velocileptors}{https://github.com/sfschen/velocileptors}} code by \citet{Chen_velo_2021}. While the code is designed to provide accurate power spectra on perturbative scales (up to $k\approx0.25\,h\mathrm{Mpc}^{-1}$ at $z=0$), for the web-halo model we need sensible estimates of the PT term deep into the non-linear regime ($k\approx2\,h\mathrm{Mpc}^{-1}$). Obtaining such predictions with \texttt{velocileptors} is challenging, as in that regime they show non-negligible variations depending on some hyperparameters. The most significant variations arise from the choice of loop integral cutoff scale $k_\mathrm{cutoff}$ and IR resummation scale $k_\mathrm{IR}$.
For this work, we chose
\begin{align} \label{eq:ing-pt-kcutoffkIR}
    k_\mathrm{cutoff} = k_\mathrm{IR} = \sqrt{5} \cdot r_\mathrm{nl}^{-1}
\end{align}
with the non-linear scale $r_\mathrm{nl}$ defined as
\begin{align} \label{eq:ing-pt-rnl}
    \sigma_\mathrm{cc}(r_\mathrm{nl}) = \delta_\mathrm{crit}~,
\end{align}
where, as in \citet{Alonso:2014zfa}, the factor $\sqrt{5}$ \sB{comes from} match\sB{ing the leading order Taylor expansions of} the Gaussian filter used within \texttt{velocileptors} to the \sB{one of the} tophat filter used in the halo model approach.\footnote{We checked that the results are unaltered if modifying the \texttt{velocileptors} source code, replacing the Gaussian filter by a tophat. The scaling with $\sqrt{5}$ is hence preferred, to avoid the need for the user to modify the source. Also, we note that employing a sharp real-space cut (tophat) and hence an oscillating $k$-cut rather than a smooth $k$-cut (Gaussian) could negatively interfere with the strictly Gaussian $k$-cut within \texttt{velocileptor}'s IR resummation machinery.} We find that this choice of cutoff scale optimally reduces the variance of 1$\ell$-LPT predictions with respect to N-body simulation results. For a detailed justification, we refer to appendix \ref{app:1l-LPT}, including an additional convergence test not mentioned here. At the same time, equation \eqref{eq:ing-pt-kcutoffkIR} ensures that all scales entering the one-loop calculation are resummed, matching the spirit of LPT. 

\subsection{Spherical Collapse}
\label{sec:ing-sc}

The ellipsoidal collapse model described later is based heavily on two basic ingredients obtained from spherical collapse calculations. The linear theory collapse threshold overdensity $\delta_\mathrm{crit}\approx1.686$ enters the peak height variable $\nu$ in equation \eqref{eq:whm-hm-peakheight} important to determine the mass functions in section \ref{sec:ing-mf} and the halo mass-concentration relation in section \ref{sec:ing-dp-h}. The virial halo overdensity $\Delta_\mathrm{v} \approx 200$ introduced in equation \eqref{eq:whm-hm-virialoverdensity} plays an important role in determining the density profiles and window functions in section \ref{sec:ing-dp}. 

In Einstein-de Sitter universes, the numerical values for $\delta_\mathrm{crit}$ and $\Delta_\mathrm{v}$ 
can be derived straight-forwardly from the spherical collapse theory \citep{Gunn_1972,Cooray_2002}  yielding $\delta_\mathrm{crit}\approx1.68$ and $\Delta_\mathrm{v} \approx 200$. In general, cosmologies with a cosmological constant $\Lambda$ or Dark Energy (DE) component, more complex differential equations arise. In this work, we include the cosmology and redshift dependence of $\delta_\mathrm{crit}$ and $\Delta_\mathrm{v}$ using the fitting functions to the numerical solution of Appendix A in \citet{Mead_2021}, initially derived in \citet{Mead_2016b} with slight modifications to account for massive neutrino cosmologies. \sB{As can be seen in figure A3 therein, the (rather weak) cosmology dependence and redshift dependence is fully captured by $\Omega_\mathrm{m}(z)$.}

\subsection{Mass functions}
\label{sec:ing-mf}

Among the most important ingredients of the halo model and its generalisation, the web-halo model, are the mass functions $f(\nu)$. They provide the probability that the mass of an object is within the range $\mathrm{d}\nu(M)$, where the relation between mass $M$ and peak-height $\nu$ is given by equation~(\ref{eq:whm-hm-peakheight}).

The prediction of such functions from theory dates back to the work by \citet{Press_1974}, who derived the halo mass function 

\begin{equation}
f_\mathrm{h}^{\mathrm{PS}}(\nu) = \frac{2}{\sqrt{2\pi}} e^{-\frac{\nu^2}{2}} 
\label{eq:ing-mf-ps}
\end{equation}
from spherical collapse calculations. This result was reproduced by \citet{Bond_1991} within the context of excursion set theory. In this framework, the probability of forming a halo (i.e., a spherically collapsed region) is represented by the probability of a random walk, whose stepsize is given by the variance $\sigma(R)$ of the density field filtered on scale $R$ as defined in equation~(\ref{eq:whm-def-variance}), crossing the critical overdensity, or "barrier", $\delta_\mathrm{crit}$.  While equation~(\ref{eq:ing-mf-ps}) delivered an astonishingly good fit to numerical simulations for intermediate masses, later on \citet{Sheth_1999} adjusted equation~(\ref{eq:ing-mf-ps}) to
\begin{equation}
f_\mathrm{h}^{\mathrm{ST}}(\nu) = A \left(1+ \frac{1}{(a\nu^2)^p} \right) e^{-\frac{a\nu^2}{2} }\ , 
\label{eq:ing-mf-st}
\end{equation}
with the purpose of fitting the high and small mass limits using two phenomenological fitting parameters $a=0.707$ and $p=0.3$, respectively, and $A=0.21616$ to satisfy the mass constraint of equation~(\ref{eq:whm-hm-conditions}). In a follow-up work, \citet{Sheth_2001} \textit{a posteriori} provided a physical motivation for the functional form of equation~(\ref{eq:ing-mf-st}): ellipsoidal collapse and the moving barrier. 

In a nutshell, in contrast to the spherical collapse framework, where collapse occurs along all three dimensions simultaneously once the constant barrier $\delta_\mathrm{crit}$ is crossed, in their work, \citet{Sheth_2001} include the possibility of ellipsoidal collapse along one, two and then three dimensions. This is achieved via a "moving barrier" where the collapse thresholds along different dimensions become more degenerate with smaller masses. By imposing the threshold for collapse along three dimensions to increase with increasing $\sigma(M)$ and hence decreasing $M$, they motivated the functional form of equation~(\ref{eq:ing-mf-st}).

Similarly, it is possible to derive mass functions corresponding to the barrier crossings of two- and one-dimensional collapse as well. In particular, \citet{Shen_2006} introduced the generalised mass function for a moving barrier of type
\begin{equation}
    \delta_\mathrm{crit}^\mathrm{e}(\sigma) = \delta_\mathrm{crit} \left(1+\frac{\beta}{(a\nu^2)^{\alpha}}\right)\ , 
    \label{eq:ing-mf-mb}
\end{equation}
corresponding to ellipsoidal collapse (with $\beta=0$ recovering spherical collapse), given by
\begin{equation}
    f^{\mathrm{g}}(\nu) = A e^{-\frac{a\nu^2}{2} \left(1+\frac{\beta}{(a\nu^2)^{\alpha}}\right)^2 } \left(1+ \frac{\beta}{(a\nu^2)^{\alpha}} \sum_{N=0}^5 \frac{(-1)^N \alpha!}{(\alpha-N)! N!} \right) \ . 
    \label{eq:ing-mf-gen}
\end{equation}
Here, we use $\alpha!/(\alpha-N)!=\alpha(\alpha-1)\dots(\alpha-N+1)$ as shorthand notation. In addition, we include the same parameter $a=0.707$ from equation~(\ref{eq:ing-mf-st}) in equations~(\ref{eq:ing-mf-mb}) and (\ref{eq:ing-mf-gen}), although it does not appear in the original source. We choose the same parameters as \citet{Shen_2006} for sheets, filaments and haloes, respectively,
\begin{equation}
\begin{aligned} \label{eq:ing-mf-numbers}
    &&\mathrm{Sheets:}\quad  &\alpha_\mathrm{s} = 0.55, &&\beta_\mathrm{s} = -0.56, &A_\mathrm{s} = 0.631\ , \\
    &&\mathrm{Filaments:}\quad &\alpha_\mathrm{f} = 0.28,  &&\beta_\mathrm{f} = -0.012, &A_\mathrm{f} = 0.672\ , \\
    &&\mathrm{haloes:}\quad &\alpha_\mathrm{h} = \sB{0.61}, &&\beta_\mathrm{h} = \sB{0.45}, &A_\mathrm{h} = 0.639 \ ,
\end{aligned}
\end{equation}
where we have slightly adjusted the $A$ parameters to satisfy the mass normalisation constraint of equation~(\ref{eq:whm-hm-conditions})~. It is important to reiterate the logic of the excursion set theory that \textit{all} the mass is attributed to sheets, filaments \textit{and} haloes at the same time. Again, this seems different from the standard picture (in the context of web environment classifications of N-body simulations), where mass is attributed to \textit{either} sheets, filaments \textit{or} haloes. However, both views are not contradictory: The normalisation constraint of equation~(\ref{eq:whm-hm-conditions}) holds for \textit{all} masses from zero to infinity, while N-body simulations and structure identifiers are always limited in mass resolution. 

\begin{figure}
    \centering
    \includegraphics[width=\columnwidth]{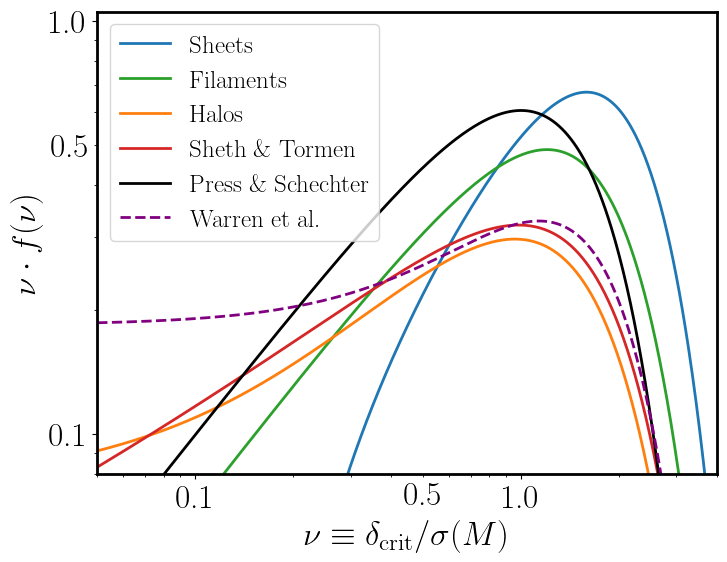}
    \caption{This shows the mass functions for sheets, filaments and haloes in equations \eqref{eq:ing-mf-gen} and \eqref{eq:ing-mf-numbers} following \citet{Shen_2006}, but with $a=0.707$ and hence readjusted normalisation $A$. These are compared to the halo mass functions corresponding to spherical collapse from \citet{Press_1974}, our baseline, ellipsoidal collapse \citet{Sheth_1999} halo mass function, and the fit to FoF haloes from \citet{Warren2006ApJ}.}
    \label{fig:ing-hm}
\end{figure}

We show the corresponding mass functions of equation~(\ref{eq:ing-mf-gen}) in comparison to the halo mass functions defined in equations~(\ref{eq:ing-mf-ps}) and (\ref{eq:ing-mf-st}) in figure~\ref{fig:ing-hm}. Interestingly, the general \citet{Shen_2006} halo mass function (orange line) matches the original one by \citet{Sheth_1999} (red line) reasonably well.\footnote{The unphysical increase in low masses is due to the series in equation \eqref{eq:ing-mf-gen} not having converged at the 5th order for haloes. We checked that higher orders converge towards equation \eqref{eq:ing-mf-st}, and that truncating at order 5 as in \citet{Shen_2006} is good enough for sheets and filaments.} Hence, we use the latter for all our calculations, given its simpler mathematical form and better behaviour in the low mass (low $\nu$) limit \sB{and use the \citet{Shen_2006} mass functions for sheets and filaments only}. The shape of the filament mass function (green line) is similar to the spherical collapse halo mass function by \citet{Press_1974} due to the small value of $\beta_\mathrm{f}$. They are however shifted with respect to each other along the $\nu$-axis, because for the former we have set $a=0.707$, while for the latter $a=1$ by construction.\footnote{Note, that equation~(\ref{eq:ing-mf-ps}) is equivalent to~(\ref{eq:ing-mf-st}) with $a=1$ and $p=0$.} We also show the \citet{Warren2006ApJ} fit to FoF haloes that we will use in section \ref{sec:1h}.

\subsection{Density profiles and window functions}
\label{sec:ing-dp}

This section is divided into two parts. In section \ref{sec:ing-dp-h}, we cover the standard case of virialised structures. By that, we mean haloes that have collapsed along three dimensions, are spherically symmetric, and have all their mass confined within their virial radius $r_\mathrm{v}$. Section \ref{sec:ing-dp-f} deals with the new part entering the web-halo model, deriving the density profiles and window functions of unvirialised structures such as sheets and filaments. These have not yet collapsed along all three dimensions, are not spherically symmetric, but instead follow cylindrical symmetry. They have mass outside of the virial radius $r_\mathrm{v}$, but within the radius $R$ of the initially collapsing sphere.

\subsubsection{Virialised structures (haloes)}
\label{sec:ing-dp-h}

The normalized Navarro-Frenk-White (NFW) profile \citet{NFW_1997} can be written as
\begin{align} \label{eq:ing-dpwf-NFW}
    \frac{\rho_\mathrm{h}(r, R)}{M(R)} = \left(\frac{c^3}{4\pi r_\mathrm{v}^3 f(c)}\right) \left(\frac{rc}{r_\mathrm{v}}\right)^{-1}\left(1+\frac{rc}{r_\mathrm{v}}\right)^{-2} \Theta(r_\mathrm{v}-r)~,
\end{align}
where $f(c) = \ln(1+c)-c/(1+c)$, $c(M)$ is the halo concentration \sB{(for conciseness, we may not always specify the mass dependence)}, and the halo mass $M(R)$ and halo virial radius $r_\mathrm{v}(R)$ are defined in equations \eqref{eq:whm-hm-peakheight} and \eqref{eq:whm-hm-virialoverdensity}, respectively. The Heaviside step function $\Theta(r_\mathrm{v}-r)$ and the normalisation ensure that all halo mass $M$ is within its virial radius $r_\mathrm{v}$, i.e., the volume integral of equation \eqref{eq:ing-dpwf-NFW} over the initially collapsing sphere with radius $R$ gives $1$. 

As a consequence, the normalized NFW halo Window function $W_\mathrm{h}(k,R)$ is obtained by plugging equation \eqref{eq:ing-dpwf-NFW} into \eqref{eq:whm-hm-window} yielding (for reference see \citealt{Cooray_2002,Mead_2015})
\begin{align} \label{eq:ing-dpwf-winNFW}
W_\mathrm{h}(k, R)f(c) = \left[\operatorname{Ci}\left(\frac{kr_\mathrm{v}(1 + c)}{c}\right) - \operatorname{Ci}\left(\frac{kr_\mathrm{v}}{c}\right)\right] \cos\left(\frac{kr_\mathrm{v}}{c}\right) \nonumber \\ + \left[\operatorname{Si}\left(\frac{kr_\mathrm{v}(1 + c)}{c}\right) - \operatorname{Si}\left(\frac{kr_\mathrm{v}}{c}\right)\right] \sin\left(\frac{kr_\mathrm{v}}{c}\right) - \frac{c\sin(kr_\mathrm{v})}{kr_\mathrm{v}(1 + c)}~,
\end{align}
where \(\operatorname{Ci}(x)\) and \(\operatorname{Si}(x)\) are the cosine and sine integral functions.

For the halo concentration-mass relation $c(M)$, we follow \citet{Mead_2021} and choose the fit from \citet{Bullock_2001} including the \citet{Dolag_2003} prescription for dark energy (DE) cosmologies. The latter depends on the ratio between linear growth rates $D(z)$ within a general DE model and its $\Lambda$ cold dark matter ($\Lambda$CDM) equivalent\footnote{As \citet{Mead_2021}, we do so by fixing $w_0 = -1$ and $w_a = 0$, enforcing flatness (fixing $\Omega_\Lambda= 1-\Omega_\mathrm{m}$) and converting neutrino mass into CDM mass.} at `infinite' redshift set to $z_\mathrm{inf}=10$  
\begin{align} \label{eq:ing-dpwf-concmass}
    c(M,z) = B \left[ \frac{1+z_\mathrm{form}(M,z)}{1+z} \right] \frac{D(z_\mathrm{inf})}{D_\Lambda(z_\mathrm{inf})} \frac{D_\Lambda(z)}{D(z)}~,
\end{align}
where the final fraction was added by \citet{Mead_2021} to suppress the DE dependence at high redshifts when the universe was still matter-dominated. The redshift of halo formation $z_\mathrm{form}$ corresponds to the redshift at which the fraction $\gamma=0.01$ of the halo mass $M$ crossed the linear collapse threshold. i.e., 
\begin{equation} \label{eq:ing-dpwf-zform}
    \frac{D(z_\mathrm{form})}{D(z)} = \frac{\delta_\mathrm{crit}(z)}{\sigma_\mathrm{cc}(\gamma M, z)}~.
\end{equation}
Here, we set the minimum concentration $B$ to the original \citet{Bullock_2001} value $B=4$, although it could be used to marginalise over baryonic feedback as in \citet{Mead_2015,Mead_2016a,Mead_2021}.

\subsubsection{Unvirialised structures (filaments and sheets)}
\label{sec:ing-dp-f}

To be able to analytically compute window functions for the intricate sheet-like and filamentary structure of the cosmic web, we make the following idealistic assumptions:
\begin{enumerate}
    \item[(i)] \textbf{Cylindrical Symmetry:} Starting from a spherical cloud with radius $R$ collapsing along the first and then the second dimension, we assume that the `still uncollapsed' dimensions retain the scale $R$, and collapse occurs homogeneously in each dimension. This results in sheets and then filaments with cylindrical symmetry.
    \item[(ii-a)]  \textbf{Constant density:} While haloes have attained a density profile with certain concentration after the virialisation process, for sheets and filaments we assume a constant density profile by default.
    \item[(ii-b)] \textbf{`Infinite' Density (thin web):} Alternatively, we assume sheets and filaments to be infinitely dense (and thin), but convolved with the window function in eq. \eqref{eq:ing-dpwf-winNFW} corresponding to halos of the same mass. This represents an upper bound for the sheet and filament concentration.
    \item[(iii)] \textbf{Dimensional scaling in overdensity:} Given that a spherical halo has undergone collapse along three dimensions to reach an overdensity of $\Delta_\mathrm{v}$, we assume that the gain in overdensity is equally distributed across each dimension. As a result, after collapse along one dimension the resulting sheet will have an overdensity of $\Delta_\mathrm{cyl,s} = \Delta_\mathrm{v}^{1/3}$ and after collapse along the second dimension into a filament it's overdensity will be $\Delta_\mathrm{cyl,f} = \Delta_\mathrm{v}^{2/3}$. This is in line with the reasoning presented in \citet{Shen_2006}.
\end{enumerate}

Thanks to assumptions (ii-a) and (iii), no additional parameters are needed for filaments and sheets; their profiles are purely determined by the virial overdensity $\Delta_\mathrm{v}$ of haloes with the same mass. 
However, we anticipate that this level of simplicity, especially assumption (ii-a), does not hold when compared to actual N-body simulation outputs. \sB{For example, \citet{Yang:2022ibs} analyse the profile of stacked filaments, finding indeed cylindrical symmetry (apart from environment effects at the edges that we ignore here), but a cored concentration of mass towards the central spine. \citet{Brinckmann:2014nia} argues sheet profiles are flat along the non-collapsed and Gaussian (cusp) along the collapsed dimension.}\footnote{\sB{\citet{Hertzsch:2025grv} model sheets as ellipsoids and hence non-uniform along the non-collapsed dimensions. We do not consider this here, as we model sheet substructure via the 1-filament and 1-halo terms.}} %In reality, we would expect sheets and filaments to present a certain concentration of mass towards their centres. 
To explore the potential impact on model predictions, we also consider as opposite extreme the `thin web' scenario (ii-b)\sB{. Here}, %where
sheets and filaments are as concentrated along their collapsed dimensions as haloes with the same mass\sB{, given that an even higher concentration than their child halo would be unphysical}. The difference between assuming homogeneity and mass profiles with the %same
\sB{(maximally allowed) same-mass halo} concentration %as haloes
for sheets and filaments can then be seen as a systematic uncertainty of the model.
We leave the prospect of improving the precision of the model by using actual filament and sheet profiles calibrated on simulations for future work.

\begin{figure}
    \centering
    \includegraphics[width=\columnwidth]{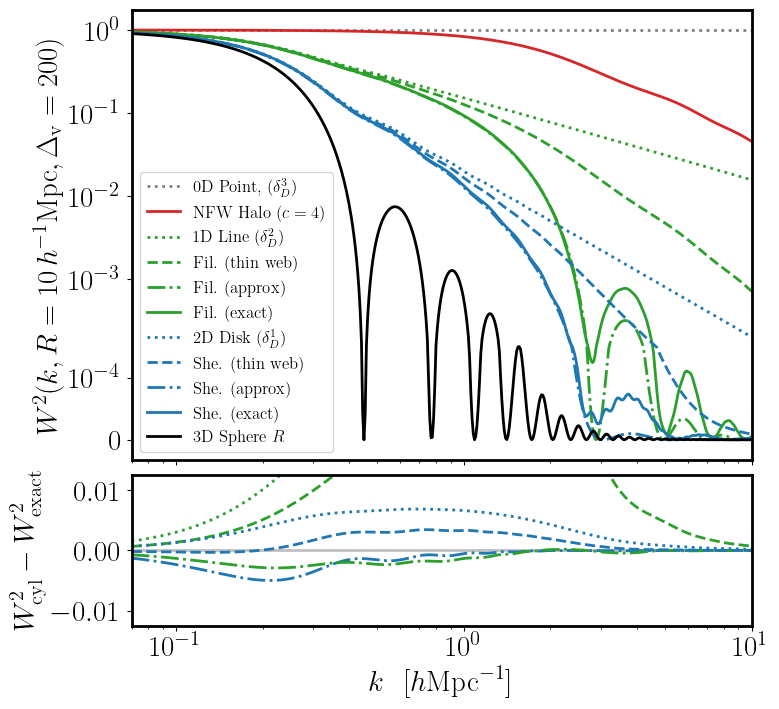}
    \caption{Shown are the squared, angle-averaged window functions for various objects that share the same mass, corresponding to a homogeneous sphere of radius $R=10\,h^{-1}\mathrm{Mpc}$ and constant density $\bar\rho$ (black \sB{solid}). First, it collapses into a sheet with overdensity $\Delta_\mathrm{v}^{1/3}$ (blue \sB{solid}), then into a filament with overdensity $\Delta_\mathrm{v}^{2/3}$ (green \sB{solid}), before finally creating a virialised \sB{NFW} halo with overdensity $\Delta_\mathrm{v}$ (red \sB{solid}). %The orange dotted curve shows a homogeneous halo for comparison.
    The blue and green dotted lines represent `infinitely thin' sheets and filaments, \sB{corresponding to 1D and 2D Dirac delta functions $\delta_\mathrm{D}$, i.e., the limit where collapsed dimension(s) have zero extent ($b=0$, $a=0$, respectively). T}heir multiplication with the \sB{squared NFW} halo \sB{window} %profile (dashed lines)
    corresponds to the \sB{`thin web' case introduced in the main text} %maximally allowed concentration of filaments and sheets. 
    %Finally, t
    \sB{T}he dash-dotted lines correspond to the approximations \eqref{eq:ing-dpwf-sheet} and \eqref{eq:ing-dpwf-filament} obtained from multiplying the `\sB{inf.} thin' limits with %the (slightly shifted to match the first trough) orange curve.
    \sB{a tophat accounting for the non-zero extent in the collapsed dimension(s).}
    \sB{As shown by the difference plot in the lower panel, they agree with the exact result within $0.005$, well below the difference towards the `thin web' case.}
    }
    \label{fig:ing-win}
\end{figure}

For a homogeneous cylinder with mass $M$, radius $a$ and height $b$, the normalized density profile in cylindrical coordinates $\mathbf{r}_\mathbf{cyl}=(r_\mathrm{cyl}, z, \phi)$ with axial radius $r_\mathrm{cyl}$, elevation $z$, and azimuth $\phi$ reads
\begin{equation} \label{eq:ing-dpwf-profilecyl}
    \frac{\rho_\mathbf{cyl}(\mathbf{r}_\mathbf{cyl}, a, b)}{M} = \frac{1}{\pi a^2b} \, \Theta\left(a-r_\mathrm{cyl}\right) \, \Theta\left(\frac{b}{2}-|z|\right) \ .
\end{equation}
To obtain the corresponding normalised window function, we evaluate the Fourier transform of \eqref{eq:ing-dpwf-profilecyl} in cylindrical wavenumber coordinates $\mathbf{k}_\mathbf{cyl}=(k_\mathrm{cyl}, k_z, \varphi)$
\begin{align} \label{eq:ing-dpwf-window-vartheta}
    &W_\mathrm{cyl}(\mathbf{k}_\mathbf{cyl}, a, b) = \int \!  \frac{\rho_\mathrm{cyl}(\mathbf{r}_\mathbf{cyl}, a, b)}{M} e^{i \mathbf{k}_\mathbf{cyl} \cdot \mathbf{r}_\mathbf{cyl}} \, \mathrm{d}^\mathrm{3} \mathbf{r}_\mathbf{cyl} \nonumber \\
     &= \frac{1}{\pi a^2b}
\int_0^a \! r_\mathrm{cyl} \int_0^{2\pi} \! e^{ik_\mathrm{cyl}r_\mathrm{cyl}\cos(\varphi-\phi)}\, \mathrm{d}\phi\mathrm{d}r_\mathrm{cyl} \int_{-b/2}^{b/2} \! e^{i k_z z}\,\mathrm{d}z  \nonumber \\
    &=  \frac{1}{\pi a^2b} \frac{2\pi a J_1(k_\mathrm{cyl}a)}{k_\mathrm{cyl}} \frac{\sin(k_z b/2)}{k_z/2}  = \frac{2 J_1(k_\mathrm{cyl}a)}{k_\mathrm{cyl}a} \frac{\sin(k_z b/2)}{k_z b/2} \nonumber \\
    &\Rightarrow W_\mathrm{cyl}(\mathbf{k},a,b) =  \frac{2 J_1(ka\sin(\vartheta))}{ka \sin(\vartheta)} \frac{\sin(kb\cos(\vartheta)/2)}{kb\cos(\vartheta)/2} ~,
\end{align}
where $J_1(x)$ is the first order Bessel function and in the last line we converted $k_\mathrm{cyl} = k\sin(\vartheta)$ and $k_z=k\cos(\vartheta)$ into spherical coordinates with standard wavenumber $k$ and zenith angle $\vartheta$, the latter representing the orientation of the standard wavevector $\mathbf{k}$ with respect to the cylinder. 
However, to evaluate the 1-sheet and 1-filament terms from equations \eqref{eq:whm-whm-1s} and \eqref{eq:whm-whm-1f}, we need to remove the dependence on orientation. For that, we compute the angle-average of the squared window function
\begin{align} \label{eq:ing-dpwf-window-average}
    W_\mathrm{cyl}^2(k,a,b) = \left\langle W_\mathrm{cyl}^2(\mathbf{k},a,b) \right\rangle_\vartheta \equiv \int_0^1 \! W_\mathrm{cyl}^2(\mathbf{k},a,b) \,\mathrm{d}\cos(\vartheta)\ .
\end{align}
Now we can use assumption (iii) and mass conservation of the initial sphere $M= \bar{\rho}\frac{4}{3}\pi R^3 =\Delta_\mathrm{cyl}\bar{\rho}V_\mathrm{cyl}$ after collapse into a cylinder with volume $V_\mathrm{cyl} = \pi a^2b $ to determine $a$ and $b$ as
\begin{equation}
\begin{aligned} \label{eq:ing-dpwf-numbers}
    %&&\mathrm{Sheets:}\quad  &a = R~, &&b = \frac{4}{3} \frac{R}{\Delta_\mathrm{v}^{1/3}}~, \\
    %&&\mathrm{Filaments:}\quad &a = \sqrt{\frac{2}{3}}\frac{R}{\Delta_\mathrm{v}^{1/3}}~,  &&b = 2R~.
    &&\mathrm{Sheets:}\quad  &a = R &&b = \frac{4}{3} \left(\frac{R}{\Delta_\mathrm{v}^{1/3}}\right) \\
    &&\mathrm{Filaments:}\quad &a = \sqrt{\frac{2}{3}}\left(\frac{R}{\Delta_\mathrm{v}^{1/3}}\right)  &&b = 2R~.    
\end{aligned}
\end{equation}
Plugging these values into equations \eqref{eq:ing-dpwf-window-vartheta} and \eqref{eq:ing-dpwf-window-average} we can compute $W_\mathrm{s}^2(k,R)$ and $W_\mathrm{f}^2(k,R)$ as functions of $R$ only (instead of $a$ and $b$). Their exact computation for $R=10\,h^{-1}\mathrm{Mpc}$ is shown as the blue and green solid curves in figure \ref{fig:ing-win}. 

In reality, we do not evaluate the exact integral in equation \eqref{eq:ing-dpwf-window-average}, as it does not have an analytic solution and %the numerical integration is computationally expensive. 
\sB{it appears within the $\nu$-integrals in equations \eqref{eq:whm-whm-1s}, so that even for fast FFTLog-based numerical integration methods the model evaluation would slow down significantly.}
Instead, we approximate the integral as follows. For sheets, we take the limit $b\rightarrow 0$, meaning that they become \sB{`2D Disks'} and only the first fraction with the Bessel function on the right-hand side of equation \eqref{eq:ing-dpwf-window-vartheta} remains. Conversely, for filaments we take the \sB{`1D Line'} limit $a\rightarrow 0$, such that the Bessel function term vanishes. In these cases, the integral in equation \eqref{eq:ing-dpwf-window-average} can be solved analytically and we account for the non-zero extent of sheets in $b$ and filaments in $a$ by convolving the result with spherical tophats $W_\mathrm{t}(x)$ (defined in equation \eqref{eq:whm-def-tophat}). We find
\begin{align} \label{eq:ing-dpwf-sheet}
    W_\mathrm{s}^2(k,R) &\approx \frac{2 \left[1- \,_0F_1\right(2,-k^2R^2\left)\right]}{k^2R^2}  W_\mathrm{t}^2\left(\sB{\frac{2\sqrt{5}}{5\sqrt{3}}}\frac{kR}{\Delta_\mathrm{v}^{1/3}}\right) \\
    W_\mathrm{f}^2(k,R) &\approx \frac{\cos(2kR) - 1+2kR \, \operatorname{Si}(2kR)}{2k^2R^2} W_\mathrm{t}^2\left(\sqrt{\frac{5}{6}}\frac{kR}{\Delta_\mathrm{v}^{1/3}}\right)~,
    \label{eq:ing-dpwf-filament}
\end{align}
where $_0F_1(a,x)$ is the confluent hypergeometric limit function. Note that the factors within the tophat arguments differ from $a$ and $b$ in equation \eqref{eq:ing-dpwf-numbers}. They were instead found by matching the \sB{leading order Taylor expansion of the cylinder to the tophat window function as explained in appendix \ref{app:window-numbers}}. 
\sB{In figure \ref{fig:ing-win}, we compare the exact result of the `Constant density' case (green and blue solid lines), i.e., the numerical integration of equations \eqref{eq:ing-dpwf-window-average} and \eqref{eq:ing-dpwf-numbers}, to the approximate analytical solution (green and blue dash-dotted lines) from equations \eqref{eq:ing-dpwf-sheet} and \eqref{eq:ing-dpwf-filament}. In the upper panel, we find excellent agreement on large scales and a small discrepancy on small scales. However, the residual plot in the lower panel reveals that the latter is subdominant due to the small absolute values of the window function in that regime. The analytic solution of the `1D line' and `2D disk', i.e., equations \eqref{eq:ing-dpwf-sheet} and \eqref{eq:ing-dpwf-filament} without the tophat window convolution is shown via the green and blue dotted lines (and there is also the `0D point' corresponding to a Dirac delta function in 3D space for comparison as a grey dotted line). These `helper' functions corresponding to sheets and filaments with infinite concentration are unrealistic, as they would exceed an NFW halo of the same mass (red solid line) at small scales. Therefore, we construct the `thin web' case by multiplying with the squared NFW window function, shown as blue and green dashed lines.} 
Their difference with respect to the \sB{`constant density case'} (solid and dashed-dotted blue and green lines) represents the maximum systematic uncertainty on the sheet and filament window functions, \sB{due to} the lack of knowledge of their concentration. Given the excellent agreement between the solid and dash-dotted blue and green lines, we use the approximate window functions (equations \ref{eq:ing-dpwf-sheet} and \ref{eq:ing-dpwf-filament}) for the web-halo model in equations (\ref{eq:whm-whm-powersum}-\ref{eq:whm-whm-1h}). 
\sB{Note that any residual impact of choosing the approximation rather than the exact result is included within the systematic uncertainty arising between the `Constant Density' (ii-a) and `thin web' (ii-b) cases.} 

One may ask why we perform the angle average of the squared window function in \eqref{eq:ing-dpwf-window-average} rather than the unsquared one from equation \eqref{eq:ing-dpwf-window-vartheta} itself. This is because for the WHM \eqref{eq:whm-whm-powersum} we need the window function to evaluate the 1-sheet and 1-filament terms that represent convolutions of collapsed objects with themselves. If we had to explicitly evaluate 2-sheet or 2-filament terms analogous to equation \eqref{eq:whm-hm-twohalo-full}, the average would need to be performed over the product of two individual window functions. However, in these cases, the correlation in orientation between adjacent sheets and filaments would not be negligible.
Here, we assume that such a correlation is automatically included at the perturbative level, and any residual impact on the 1-sheet and 1-filament terms is subdominant.

In appendix \ref{app:window}, we validate the derived window functions (and the window compensation) using synthetically produced sheets and filaments in a box, finding that equations \eqref{eq:ing-dpwf-sheet} and \eqref{eq:ing-dpwf-filament} indeed deliver good power spectrum predictions for particles distributed in randomly oriented sheets and filaments.

\subsubsection{Perturbation theory window functions} 
\label{sec:ing-dp-pt}

To finally evaluate equation \eqref{eq:whm-whm-powersum}, we also need the window functions $W_\mathrm{PT}^2(k)$ of perturbatively evolved structures before their collapse. In this work, we combine the web-halo model either with linear PT (`L') or 1$\ell$-LPT. For linear theory, we follow \citet{Valageas_2011a} and assume it is given by the homogeneous sphere with background density (equation \ref{eq:whm-hm-peakheight}). For 1$\ell$-LPT, we assume the advected particles follow sheet profiles up to the 1-loop cut-off scale $r_\mathrm{nl}$ (although simulation-based approaches such as \citealt{LamSheth2008MNRAS} exist.) Hence,
\begin{align}
    W_\mathrm{L}^2(k,R) &= W_\mathrm{t}^2(k,R) \label{eq:ing-dpwf-wlin} \\
    W_\mathrm{1\ell-LPT}^2(k,R) &= W_\mathrm{s}^2(k,R)\cdot W_\mathrm{t}^2(k,r_\mathrm{nl})~, \label{eq:ing-dpwf-wlpt} 
\end{align}
where $r_\mathrm{nl}$ is defined in equation \eqref{eq:ing-pt-rnl}. 

\sB{In summary, within the WHM the cosmology dependence of the non-linear correction enters via the same four channels as for the standard halo model: PT, the mass fluctuation amplitude $\sigma(M)$, the concentration-mass relation c(M), and the spherical collapse quantities $\delta_\mathrm{crit}$ and $\Delta_\mathrm{v}$.}

\section{Results} \label{sec:res}
Knowing its ingredients, we can now obtain web-halo model predictions of non-linear matter power spectra. We compare our results with other prediction schemes introduced in section \ref{sec:res-app}, first at fixed cosmology in section \ref{sec:res-fix} before moving on to varying cosmologies in section \ref{sec:res-var}. For all results, we use the emulator by \citet{Angulo_2021}  for the reference non-linear power spectra, but consider other emulator choices in section \ref{sec:res-emu}.  

\subsection{Other approaches} \label{sec:res-app}

A variety of prescriptions exist to predict non-linear real-space matter power spectra, and we are not able to compare our web-halo model against an exhaustive set of these. Instead, we focus on comparing against PT (section \ref{sec:res-app-pt}), the original halo model (section \ref{sec:res-app-hm}) and its enhanced \texttt{HMcode-2020} version (section \ref{sec:res-app-hm2020}).

\subsubsection{Perturbation Theory} \label{sec:res-app-pt}

As mentioned before, we use either linear or Lagrangian PT within our web-halo model. When comparing our results to PT, i.e., the 2-sheet term only, we use exactly the setup introduced in section \ref{sec:ing-pt}.

\subsubsection{Original Halo Model} \label{sec:res-app-hm}

We also compare our results to the plain, original halo model prediction using the baseline choices introduced by \citet{Mead_2021}. They are defined by the fourth column in Table 2 and \sB{motivated by figure 3} therein.

\subsubsection{\texttt{HMcode-2020}} \label{sec:res-app-hm2020}

Furthermore, we consider the enhanced form of the halo model presented as \texttt{HMcode-2020} in \citet{Mead_2021}. It builds on the plain halo model as a baseline, but adds twelve free parameters to fit the cosmology and redshift dependence of the \texttt{FrankenEmu} emulator representing an update of the extended \texttt{CosmicEmu} emulator by \citet{Heitmann_2014}.\footnote{\href{https://www.hep.anl.gov/cosmology/CosmicEmu/emu.html}{https://www.hep.anl.gov/cosmology/CosmicEmu/emu.html}} Note that this emulator is outdated and only represents a subset of the updated \texttt{MiraTitan} emulator by \citet{MoranMiraTitan-2023} we use for comparison in section \ref{sec:res-emu}.\footnote{\href{https://github.com/lanl/CosmicEmu}{https://github.com/lanl/CosmicEmu}} The twelve \texttt{HMcode-2020} fitting parameters are provided in the fifth column of Table 2 in \citet{Mead_2021}. To summarise, five parameters modify the 2-halo term, two parameters the 1-halo term cutoff at large scales due to halo exclusion, and another two the 2-halo to 1-halo transition smoothing defined in their equation (23).
The remaining three free parameters modify the 1-halo term at small scales: two for a mass-dependent change of virial radius called `halo bloating', and one for the overall concentration amplitude (parameter $B$ in equation \ref{eq:ing-dpwf-concmass}).

Compared to our web-halo model, we avoid all these free parameters by using $1\ell$-LPT instead of linear theory for an accurate prediction of the 2-halo (2-sheet) term, modelling halo exclusion explicitly via the window compensation approach of equation \eqref{eq:whm-s-phcomp}, and including sheets and filaments instead of employing an \textit{ad-hoc} 2-halo to 1-halo transition smoothing. With these changes, we do not see the need for an additional `halo bloating' as in \citet{Mead_2021} and delegate potential changes in 1-halo power to the concentration parameter, which we, however, keep fixed to its default value $B=4$ in this work.

\subsubsection{Emulators} \label{sec:res-app-emu}

Nowadays, the most accurate \sB{(and fast)} non-linear matter power spectrum prediction schemes rely on emulators. They are typically built from a suite of N-body simulations that cover a certain cosmological parameter range. 

By default, we compare our model predictions to the power spectra obtained from the BACCO emulator \texttt{baccoemu} \footnote{\href{https://baccoemu.readthedocs.io/en/latest/}{https://baccoemu.readthedocs.io/en/latest/}} (see \citet{Angulo_2021} and companion publications \citet{Arico1_2021,Arico2_2021,Zennaro_2023,MPI_2023}). It is built from three high resolution runs within the \texttt{BACCO} simulation suite of $N=4320^3$ particles in a $L=1.44\,h^{-1}\mathrm{Gpc}$ box for three cosmological models dubbed `\texttt{Vilya}', `\texttt{Nenya}', and `\texttt{Narya}' (see Table 1 therein), intentionally chosen to cover a broad parameter range following \citet{Contreras_2020} using the cosmology-rescaling technique by \citet{Angulo_2010}. From these additional 800 rescaled simulations (i.e., post-processed from the initial three simulations) they construct the power spectrum emulator using a neural network. For $\Lambda$CDM, the achieved accuracy is $1\%$, while for the full $w_0 w_a$CDM$+\sum m_\nu$ parameter space, the accuracy is $3\%$, where the most dominant uncertainty is due to the cosmology rescaling. This represents the maximum uncertainty measured on the smallest emulated scale of $k_\mathrm{max}^\mathtt{baccoemu} = 4.8\,h\mathrm{Mpc}^{-1}$. The uncertainty is expected to gradually decrease with decreasing $k$, but due to the lack of a $k$-dependent error envelope function, we always plot uncertainty floors constant with $k$. 

In section \ref{sec:res-emu}, we compare our web-halo model predictions with other widely used emulators. The \texttt{EuclidEmu2} by the Euclid collaboration \citet{Euclid:2020rfv} is built from $300$ simulations within a $w_0 w_a$CDM$+\sum m_\nu$ range, each with $N=3000^3$ particles in an $L=1.0\,h^{-1}\mathrm{Gpc}$ box with mass resolution of $3.3\times10^9\,h^{-1}M_\odot $ comparable to \texttt{baccoemu}. The quoted uncertainty is $1\%$ for $k<5\,h\mathrm{Mpc}^{-1}$ and 2\% for $5\,h\mathrm{Mpc}^{-1}<k<k_\mathrm{max}^\mathrm{EE2}=9.41\,h\mathrm{Mpc}^{-1}$, although these might be underestimated according to \citet{Angulo_2010}. Finally, we also compare to the 112 node cosmologies of the \texttt{MiraTitan} emulator by \citet{MoranMiraTitan-2023}. These cover a wider parameter space in $w_0 w_a$CDM$+\sum m_\nu$ than both \texttt{baccoemu} and \texttt{EuclidEmu2}, in particular for the $w_a$ parameter (for a direct comparison, see Table D1 by \citealt{Mead_2021}). The emulator is constructed from a set of one high resolution ($L=2.1\,\mathrm{Gpc}$, $N=3300^3$) and 16 low resolution ($L=1.3\,\mathrm{Gpc}$, $N=512^3$) simulations at each node cosmology. These are glued together and matched to the renormalization group time-flow PT approach by \citet{Lesgourgues_2009} on large scales as described in \citet{Heitmann_2016}. 

\subsection{Comparison for fixed cosmology} \label{sec:res-fix}

As a first step, we compare the web-halo model and other analytical prediction schemes mentioned in section \ref{sec:res-app} against the \texttt{baccoemu} prediction at fixed cosmology. To avoid the emulation error, we choose the \texttt{BACCO} input `\texttt{Narya}' cosmology (see details in Table 1 of \citealt{Angulo_2021}).\footnote{We find similar results for the `Vilya' and `Nenya' cosmologies.} 

\begin{figure*}
    \begin{center}
    \includegraphics[width=\textwidth]{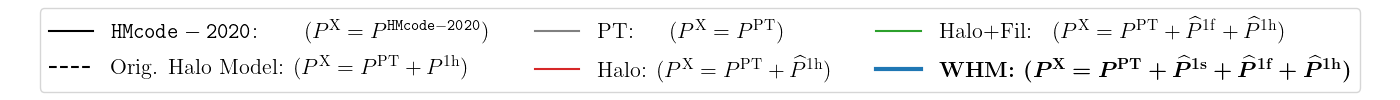}
    \includegraphics[width=\textwidth]{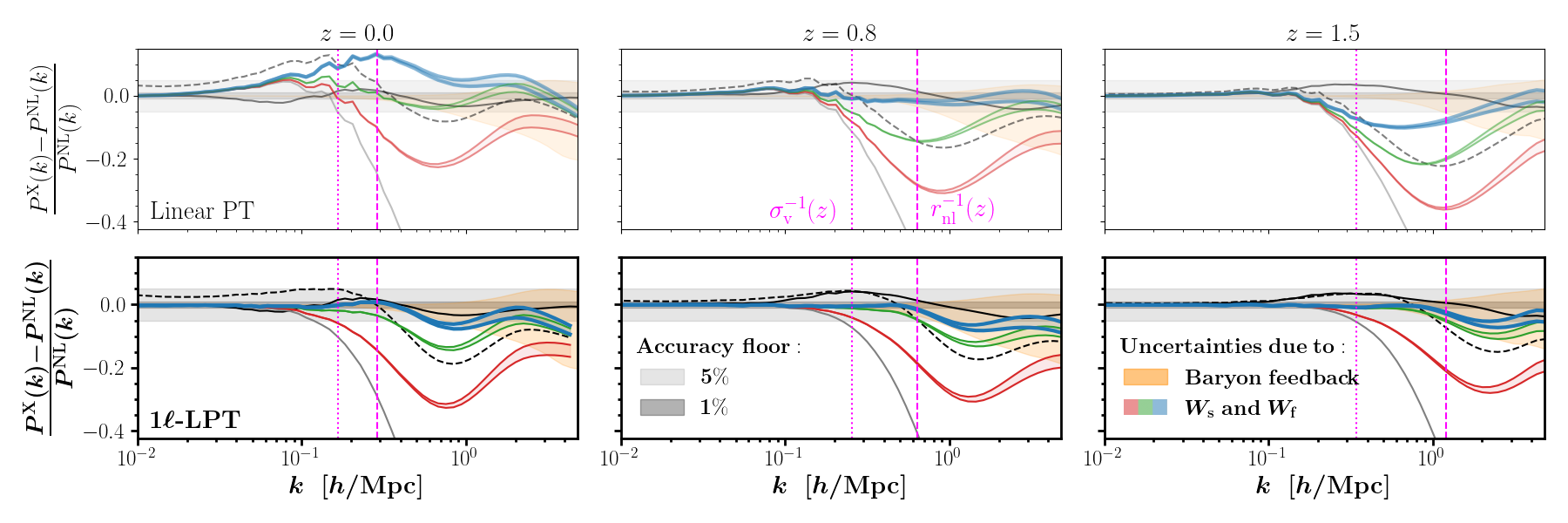}
    \end{center}
    \caption{We show the web-halo model (WHM) results breaking down the halo (red), halo+filament (green) and the full halo+filament+sheet (blue) contributions compared to PT only (grey), the original halo model without window compensation (dashed black), and the reference non-linear power spectrum obtained from \texttt{baccoemu} for the `\texttt{Narya}' cosmology. Each column shows the theory predictions at different redshifts. The top row uses linear PT, while the bottom row shows our baseline Lagrangian PT setup. The magenta dotted and dashed lines correspond to the inverse displacement field variance $\sigma_\mathrm{v}^{-1}$ and the inverse non-linear scale $r_\mathrm{nl}^{-1}$ defined in equations \eqref{eq:ing-pt-sigmav} and \eqref{eq:ing-pt-rnl}, respectively. Grey bands indicate the 1\% and 5\% accuracy regions, and orange bands the \texttt{FLAMINGO} baryonic feedback uncertainty. The coloured uncertainty bands within the WHM predictions cover the baseline assumption of either homogeneous or infinitely thin (but filtered by the same-mass halo profile) filaments and sheets, for details see section \ref{sec:ing-dp-f}.}
    \label{fig:res-fixed}
\end{figure*}

In Figure \ref{fig:res-fixed} we compare the non-linear matter power spectrum predictions relative to the \texttt{baccoemu} reference at three different redshifts $z=\left\lbrace 0.0, 0.8, 1.5\right\rbrace$ in the left, centre and right columns, respectively, covering the full \texttt{baccoemu} redshift range. Upper (lower) panels show predictions based on Linear (Lagrangian) PT. Dark and light grey bands represent the \texttt{BACCO} simulation accuracy of about 1\% and the $2\sigma$ \texttt{HMcode-2020} accuracy of 5\%, respectively. We also display as an orange band the uncertainty associated to baryonic feedback, obtained from the \texttt{FlamingoBaryonResponseEmulator}\footnote{\href{https://github.com/FLAMINGOSIM/FlamingoBaryonResponseEmulator.git}{https://github.com/FLAMINGOSIM/FlamingoBaryonResponseEmulator.git}} by \citet{Schaller2025MNRAS.539.1337S} for typical model variations within the \texttt{FLAMINGO} hydrodynamical simulation suite (\citealt{Schaye2023MNRAS.526.4978S,Kugel2023MNRAS.526.6103K}).\footnote{In particular, we allow for variations in gas fraction $f_\mathrm{gas}$ within $(-10\sigma,+4\sigma)$ and stellar mass functions $M_\star$ within $(-3.0\sigma,+2.0\sigma)$ as indicated in Figures 4 and 8 of \citet{Schaller2025MNRAS.539.1337S}, respectively.} 
Web-halo model predictions are shown via coloured lines, while other approaches are shown as black and grey lines. We now describe the behaviour of each individual model prediction. 

\paragraph*{Perturbation Theory (grey lines).} 
As expected, linear PT (upper panels) matches the references on the largest scales. Then, it overpredicts the power by ~5\% at the pre-virialisation scale $k\approx0.07\,h\mathrm{Mpc}^{-1}$ before severely underpredicting it at smaller scales. Also, it does not capture well the suppression of BAO wiggles, leading to the observed scatter at  $k\approx0.1\,h\mathrm{Mpc}^{-1}$. The Lagrangian PT prediction performs much better, as it accurately captures both the pre-virialisation dip and the BAO wiggle suppression. However, it starts underpredicting the power by more than $5\%$ at the scale of the linear RMS displacement field $k\approx\sigma_\mathrm{v}^{-1}(z)$ indicated by the vertical, dotted, magenta lines.

\paragraph*{Original Halo Model (black dashed lines).} 
Now we investigate the plain halo model calculated as described in section \ref{sec:res-app-hm}. On the largest scales, we see a small upturn due to neglecting halo exclusion, leading to an unphysical excess of power. The same effect gives rise to the 5\% (up to 10\%) excess using Lagrangian (Linear) PT as a baseline at $k\approx\sigma_\mathrm{v}^{-1}(z)$. In the 2-halo to 1-halo transition regime, roughly centred around the non-linear scale $k\approx r_\mathrm{nl}^{-1}(z)$ (see definition in equation \ref{eq:ing-pt-rnl}), indicated by vertical, dashed, magenta lines, it shows the well-known underprediction of 10\% to 20\%. For linear PT (upper panels), the underprediction worsens with increasing redshift. This might be related to the fact that the distance between $\sigma_\mathrm{v}^{-1}(z)$ and $r_\mathrm{nl}^{-1}(z)$ increases with redshift. Interestingly, this is not observed in the case of Lagrangian PT (lower panels), where it improves with increasing redshift. We checked, however, that this is only the case, if the $1\ell$-LPT evaluation is performed using $r_\mathrm{nl}$ as cutoff scale as described in section \ref{sec:ing-pt}. Finally, we note that in the small-scale limit, there is better agreement with \texttt{baccoemu} in the Linear PT case than in the Lagrangian PT case. This indicates that either the Lagrangian PT 2-halo term or the standard 1-halo term is underestimated. We further explore the latter option in section \ref{sec:1h}.

\paragraph*{\texttt{HMcode-2020} (black solid lines).} 
The main difference with respect to the plain halo model relies on the smoothed 2-halo to 1-halo transition, for details see section \ref{sec:res-app-hm2020}. As it always relies on linear PT (plus a range of fitted parameters), the curves in the upper and lower panels are identical. In general, the non-linear matter power spectrum predictions remain within the quoted $2\sigma$ uncertainty band of 5\%, but reaching exactly 5\% deviation at $k\approx\sigma_\mathrm{v}^{-1}(z)$. This might be due to this particular `Narya' cosmology being outside the \texttt{CosmicEmu} emulated range \texttt{HMcode-2020} was trained on. Wen compared to the \texttt{MiraTitan} emulator, representing a superset of \texttt{CosmicEmu},  \texttt{HMcode-2020} performs slightly better (see section \ref{sec:res-emu}).

\paragraph*{Web-Halo Model (coloured lines).} 
We now turn our attention to the web-halo model predictions, breaking down the different terms entering equation \eqref{eq:whm-whm-powersum}. Red lines indicate the addition of the \textit{compensated} 1-halo term (equation \eqref{eq:whm-whm-1h}) to the baseline PT predictions, in contrast to the \textit{uncompensated}, standard 1-halo term (black dashed lines, equation \eqref{eq:whm-hm-onehalo}. The former shows better agreement on large scales due to the removed sensitivity to halo exclusion effects, thanks to the compensation scheme. On the other hand the underprediction at small scales of up to 40\% is more severe than in the standard case. Adding the 1-filament term (green lines) ameliorates this underprediction, and for the full web-halo model expression including the 1-sheet term, agreement within 1\% with \texttt{baccoemu} is achieved up to $k\approx r_\mathrm{nl}^{-1}(z)$ and within 5\% on smaller scales. This is only the case with $1\ell$-LPT as baseline PT, though.\footnote{Also note the relative impact of the sheet term is less for $1\ell$-LPT than for linear PT. This is due to the different window compensation kernels (equations \ref{eq:ing-dpwf-wlin} and \ref{eq:ing-dpwf-wlpt}) used for each case.} With Linear PT, the web-halo model suffers in similar ways as the original halo model itself, i.e. failing to cover the redshift dependence of the 2-halo to 1-halo transition regime. Naively, one could have expected the lack of power in the transition region to be fixed by including filaments and sheets, given that they occupy more mass than haloes on larger redshifts. However, including more accurate terms (up to 1-loop) in the perturbative expansion with appropriate cutoff scale seems more important to capture the redshift dependence.  As in the original halo model case, we note that at the high-$k$ end, using Linear PT delivers results in better agreement with the reference than $1\ell$-LPT. Again, this raises the question whether the 1-halo term needs to be adapted (see section \ref{sec:1h}), or the $1\ell$-LPT 2-halo term cannot be trusted for $k>r_\mathrm{nl}^{-1}(z)$ (see appendix \ref{app:1l-LPT}). In any case, the systematic underprediction of 5\% at such small scales is rather mild once considering the uncertainties due to baryonic feedback. There is another (yet subdominant) source of uncertainty arising from the lack of knowledge of sheet and filament density profiles, displayed as coloured bands for each WHM curve. For the blue bands, the lower limits correspond to the default case of assuming constant density, whereas higher limits to the case of `infinitely thin' sheets and filmaments filtered on a halo with the same mass, for details see section \ref{sec:ing-dp-f} and figure \ref{fig:ing-win}. From the difference in window function between dashed and solid lines therein, one might expect a bigger effect than the $\sim 3\%$ effect indicated in figure \ref{fig:res-fixed}. However, it is important to note that the increased power due to more concentrated sheet and filament profiles also enters with a negative sign in the compensation terms, meaning that the net effect is less pronounced than one would naively expect. 

\paragraph*{Summary.}
We conclude that from all prediction schemes, the full web-halo model combined with $1\ell$-LPT performs best. It agrees with the \texttt{baccoemu} reference up to the non-linear scale $r_\mathrm{nl}^{-1}(z)$ at the 1\% level and at smaller $k$ within the combined uncertainty from halo profiles (due to baryonic feedback effects, yellow band) and sheet and filament profiles (red, green and blue bands). At the same time, there is still room for improvement for modelling small scales. The worse performance of \texttt{HMcode-2020} is due to the investigated `Narya' cosmology being outside the range of the emulator it was trained on. We also found that $1\ell$-LPT performs better than Linear PT in all cases. We therefore perform all remaining comparisons using $1\ell$-LPT and drop the Linear PT case in what follows.

\begin{figure*}
    \centering
    \includegraphics[width=0.9\textwidth]{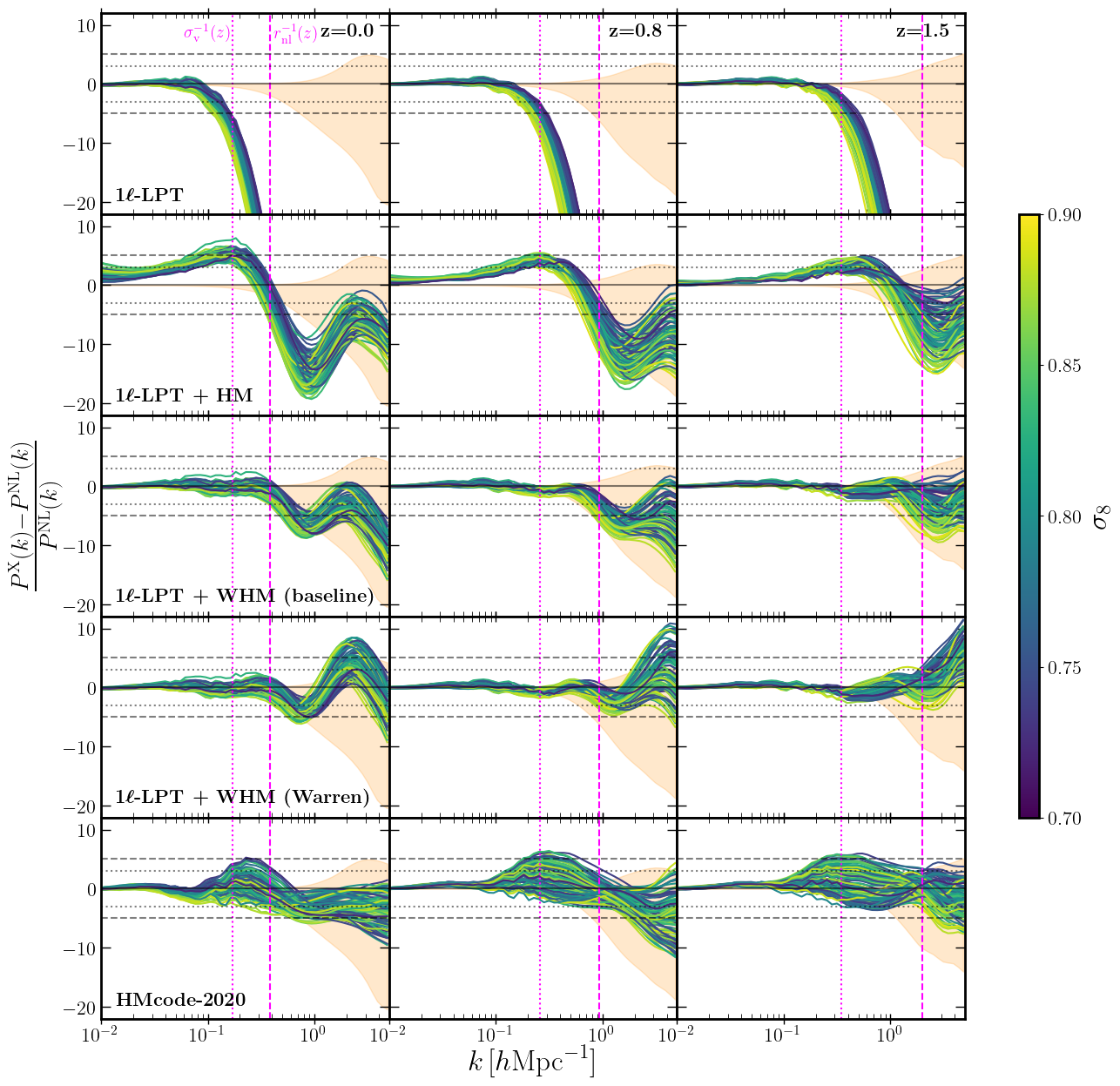}
    \caption{We compare our web-halo model (WHM) predictions using the baseline \citet{Sheth_2001} halo mass function  (third row) or the one from \citet{Warren2006ApJ} (fourth row) for 100 random cosmologies covering the full \texttt{baccoemu} range against Lagrangian PT only (first row), the original halo model (HM) without window compensation (second row), \texttt{HMcode-2020} (fifth row) and the reference non-linear power spectrum from \texttt{baccoemu}. Each column shows results at a different redshift and lines are colour-coded by $\sigma_8$. Again, magenta dotted and dashed lines correspond to the inverse displacement field variance $\sigma_\mathrm{v}^{-1}$ and the inverse non-linear scale $r_\mathrm{nl}^{-1}$, respectively. Orange bands indicate the \texttt{FLAMINGO} baryonic feedback uncertainty, horizontal dotted and dashed lines the maximum \texttt{baccoemu} uncertainty (3\%) and the $2\sigma$ \texttt{HMcode-2020} uncertainty (5\%), respectively.}
    \label{fig:res-varying}
\end{figure*}

\subsection{Comparison for varying cosmology} \label{sec:res-var}

\begin{figure*}
    \centering
    \includegraphics[width=\textwidth]{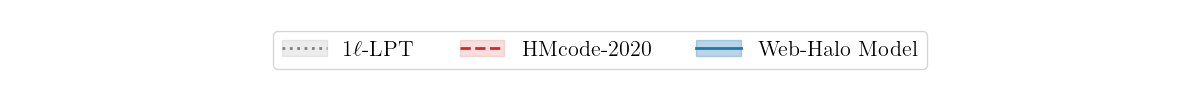}
    \includegraphics[width=0.9\textwidth]{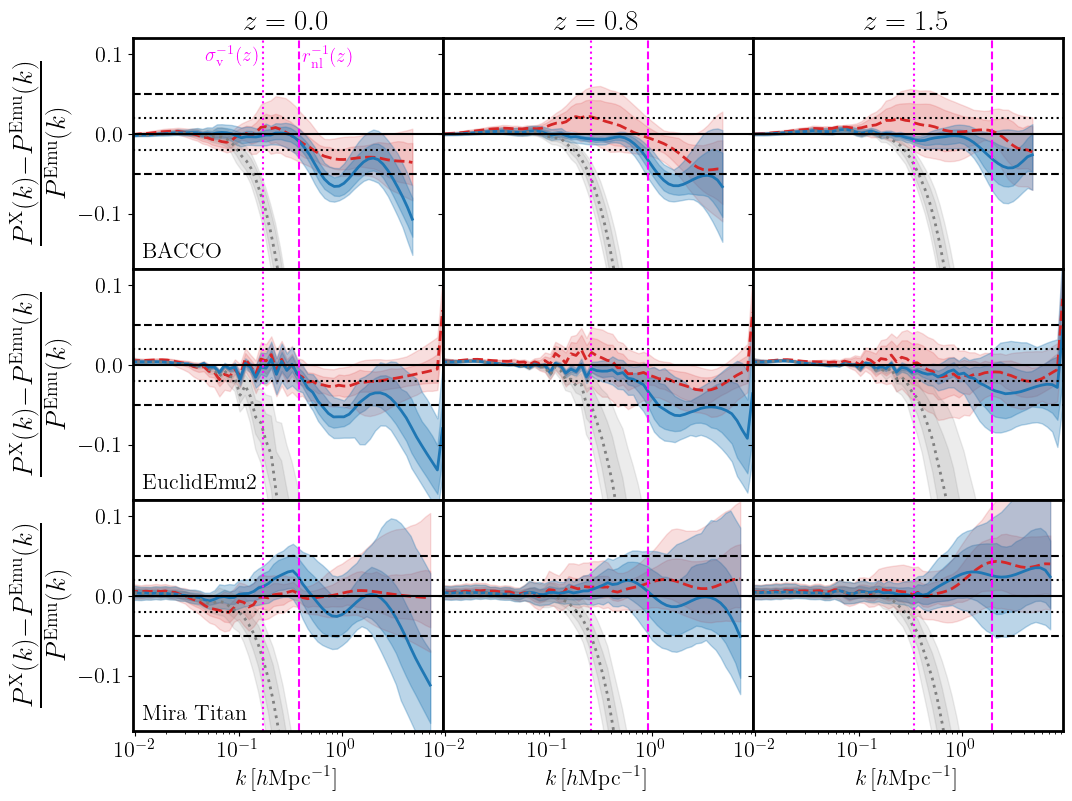}
    \caption{In this figure we compare the Lagrangian PT only (grey), \texttt{HMcode-2020} (red) and web-halo model (blue) precision with respect to the \texttt{baccoemu} (top row, $k_\mathrm{max}=4.8\,h\mathrm{Mpc}^{-1}$), \texttt{EuclidEmu2} (center row, $k_\mathrm{max}=9.4\,h\mathrm{Mpc}^{-1}$), and /texttt{MiraTitan} (bottom row, $k_\mathrm{max}=5\,\mathrm{Mpc}^{-1}$) emulators. To match the units, we converted the \texttt{MiraTitan} $x$-axis units to $h_\mathrm{fid}\mathrm{Mpc}^{-1}$ units using a fiducial value of $h^\mathrm{fid}=0.676$. The inner and outer coloured bands indicate the $1\sigma$ and $2\sigma$ fluctuations around the mean (central lines) obtained from 100 random cosmologies within the emulator ranges of \texttt{baccoemu} and \texttt{EuclidEmu2}, each, and the 112 node cosmologies of \texttt{MiraTitan}.  Again, we show predictions at a different redshift in each column, and magenta dotted and dashed lines represent the displacement field variance and non-linear scale. Dotted and dashed horizontal lines indicate the $2\%$ and $5\%$ precision levels, respectively.}
    \label{fig:res-emulators}
\end{figure*}

Having compared different models and contributions to the WHM in depth at a fixed cosmology, we now perform a similar comparison to the full range emulated by \texttt{baccoemu}. To that end, we generate 100 random cosmologies following a uniform distribution for each of the $w_0w_a$CDM$+\sum m_\nu$ parameters with ranges provided by equation (2) of \citet{Angulo_2021}. 

The resulting non-linear matter power spectra predictions for all 100 model parameter combinations colour-coded by $\sigma_8$ with respect to the \texttt{baccoemu} reference are shown in figure \ref{fig:res-varying}. As before, each column represents a different redshift, grey horizontal lines the \texttt{baccoemu} (dotted, 3\%)\footnote{This number increased from 1\% in the previous section to 3\% due to the use of $w_0w_a$CDM$+\sum m_\nu$ instead of $\Lambda$CDM} and \texttt{HMcode-2020} (dashed, 5\%) statistical error floors, and magenta vertical lines the characteristic scales $\sigma_\mathrm{v}^{-1}$ (dotted) and  $r_\mathrm{nl}^{-1}$ (dashed) for a fiducial Planck cosmology. From top to bottom, different rows show $1\ell$-LPT as baseline PT, plus the original halo model, the web-halo model using either the baseline \citet{Sheth_2001} halo mass function or the \citet{Warren2006ApJ} fit to FoF haloes (see section \ref{sec:1h}), and finally \texttt{HMcode-2020}.
Overall, for the bulk of cosmologies, we observe the same trends already discussed in section \ref{sec:res-fix} and not repeated here. Instead, we focus our discussion on the observed scatter across the residuals of the 100 selected random cosmologies. First, it is important to note that the usage of $1\ell$-LPT with appropriate cutoff scale is of major importance to reduce the scatter (also see appendix \ref{app:1l-LPT}) down to 1\% on scales $k<\sigma_\mathrm{v}^{-1}$. For example, linear PT (not shown here) severely fails in that regard, showing up to 10\% scatter for the same range of scales. Moving on to the web-halo model predictions, even at redshift $z=0$ all cosmologies remain within the \texttt{baccoemu} uncertainty band up to $k \approx r_\mathrm{nl}^{-1}$ and as expected the scatter decreases further with higher redshifts. Remarkably, in the region $\sigma_\mathrm{v}^{-1} < k < r_\mathrm{nl}^{-1}$, the WHM scatter remains as small as at larger scales, whereas both $1\ell$-LPT alone and its combination with the original halo model gradually increase scatter. Moreover, the WHM correctly cancels the small systematic trend with $\sigma_8$ observed for $1\ell$-LPT. This demonstrates the power of the ellipsoidal collapse formalism employed in the WHM to capture subtle non-linear effects at scales that used to be N-body simulation territory only. Even towards smaller scales $k > r_\mathrm{nl}^{-1}$, the WHM shows less scatter than the standard halo model. 

The predictive power of the WHM becomes even more striking once compared to \texttt{HMcode-2020}. The latter shows significantly more scatter of up to 5\% on a wide range of intermediate scales, becoming worse with increasing redshift, as opposed to the WHM. On smaller scales, however, \texttt{HMcode-2020} performs better at redshift $z=0$ capturing well the power spectrum shapes, while the WHM shows a systematic \sB{underprediction}. This effect is less pronounced for the larger redshifts, though, where both the scatter and mean become competitive with \texttt{HMcode-2020}. Also, we note that the choice of halo mass function plays a crucial role (up to $5\%$ difference) in the high-$k$ regime. We further investigate this issue in section \ref{sec:1h}. 

We reiterate that the WHM is parameter-free, whereas \texttt{HMcode-2020} consists of 12 fitting parameters. But as mentioned earlier, \texttt{HMcode-2020} uses linear PT as a baseline, while for the WHM we take advantage of $1\ell$-LPT predictions. This choice alone would make 5 of the 12 \texttt{HMcode-2020} parameters redundant. Also, 3 of the 12 \texttt{HMcode-2020} parameters are fit to dark matter only simulations, although strictly speaking, they capture baryonic feedback effects, a strategy we choose to avoid in this work. Finally, it is important to repeat that \texttt{HMcode-2020} was fitted to an independent emulator whose parameter space does not completely overlap with \texttt{baccoemu}. As a next step, we therefore consider different reference emulators, providing a tighter performance comparison between \texttt{HMcode-2020} and WHM. 

\subsection{Comparison with other emulators} \label{sec:res-emu}

In figure \ref{fig:res-emulators} we compare $1\ell$-LPT (grey), \texttt{HMcode-2020} (red), and the full baseline web-halo model (blue) to three different emulators, from top to bottom \texttt{baccoemu} \citet{Angulo_2021}, \texttt{EuclidEmu2} \citet{Euclid:2020rfv}, and \texttt{MiraTitan} \citet{MoranMiraTitan-2023}.  Again, each column represents a different redshift. Black horizontal lines indicate the approximate uncertainty floors due to the emulators (on average roughly 2\%, dotted) and \texttt{HMcode-2020} (5\%, dashed). The dotted and dashed magenta vertical lines indicate the displacement field variance $\sigma_\mathrm{v}^{-1}$ and non-linear scale $r_\mathrm{nl}^{-1}$, respectively. 

In the first row (\texttt{baccoemu} reference), there is no extra information compared to figure \ref{fig:res-varying}. Basically, instead of plotting individual model curves, the mean (central lines) and the $1\sigma$+$2\sigma$ scatter (less and more transparent coloured bands) across the 100 random cosmologies is shown. Furthermore, the three models are shown within the same panels, allowing a more direct, by-eye comparison. This reinforces the impressive WHM precision compared to \texttt{HMcode} at all scales and redshift and its percent-level accuracy up to $k\approx r_\mathrm{nl}^{-1}$.

The situation is very similar for the \texttt{EuclidEmu2} reference (middle rows). Again, the mean and scatter are obtained from 100 random cosmological models uniformly covering the emulated $w_0w_a$CDM$+\sum m_\nu$ parameter space provided in Table 2 of \citet{Euclid:2020rfv}. Compared to \texttt{baccoemu}, it covers a wider range in the time-dependent dark energy equation of state ($-0.7<w_a<0.7$ versus $-0.3<w_a<0.3$), but is constrained to smaller neutrino mass sums ($\sum m_\nu<0.15\,\mathrm{eV}$ versus $\sum m_\nu<0.4\,\mathrm{eV}$). Another important difference is the sampling in primordial fluctuation amplitude $A_\mathrm{s}$ rather than late-time fluctuation amplitude $\sigma_8$ used in \texttt{baccoemu}, leading to a fairly complementary emulated parameter space. It is therefore encouraging that the WHM also, in this case, provides predictions as accurate as the emulator itself up to $k\approx r_\mathrm{nl}^{-1}$. All models show noisy behaviour at the BAO wiggle regime, which is not present for \texttt{baccoemu}. This is already visible for $1\ell$-LPT, which also shows larger scatter across cosmological models than for \texttt{baccoemu}, propagating into the WHM predictions as well. For \texttt{HMcode-2020}, the bias and scatter are similar in magnitude and shape as for \texttt{baccoemu}. 

Finally, we use the \texttt{MiraTitan} emulator as a reference in the lower panels; in particular, we use all their 112 node cosmologies to determine the mean and scatter bands. Note that the emulator operates in $\mathrm{Mpc}^{-1}$ units, which we rescale to $h_\mathrm{fid}\mathrm{Mpc}^{-1}$ units using a fiducial value of $h^\mathrm{fid}=0.676$ to approximately match the x-axes of the upper panels. We do not use the `true '$h \mathrm{Mpc}^{-1}$ units corresponding to each cosmology, as this would introduce varying $k_\mathrm{max}$ and hence complicate the mean and scatter determination. \texttt{MiraTitan} also covers a $w_0w_a$CDM$+\sum m_\nu$ parameter space, but wider than \texttt{baccoemu} and \texttt{EuclidEmu2} especially for $h$, $n_s$ and time-varying dark energy ($-1.73\leq w_a \leq 1.28$), see equations (1a) to (1h) in \citet{MoranMiraTitan-2023} for completeness. Interestingly, it samples physical density fractions $\omega_i$ rather than absolute fractions $\Omega_i$, hence providing another complementary parameter space compared to \texttt{baccoemu} and \texttt{EuclidEmu2}. For a direct quantitative comparison of the different parameter ranges, we again refer to Table D1 by \citet{Mead_2021}. As expected, \texttt{HMcode-2020} agrees better with \texttt{MiraTitan} than other emulators, as its 12 parameters were fitted against a subset of the former. In particular, the mean \texttt{HMcode-2020} curve (red dashed line) shows less bias on intermediate scales, apart from a 2\% bias at $k\approx0.1 \, h_\mathrm{fid}\mathrm{Mpc}^{-1}$ at $z=0$. Still, it shows $2\sigma$ scatter up to 4\% even on intermediate scales $k\approx \sigma_\mathrm{v}^{-1}$, where the WHM performs similarly (at $z=0$) or even better (with only 2\% scatter at $z\geq0.8$). So also in this case, we observe the influence of the underlying PT resulting in \texttt{HMcode-2020} struggling at higher redshifts (as the Linear PT + original halo model case), while the WHM performs better with increasing redshift thanks to $1\ell$-LPT. At $z=0$, the WHM $1\sigma$ band still agrees within $5\%$ at intermediate scales, but drops towards 15\% at $k_\mathrm{max}$. At $z\geq 0.8$, both \texttt{HMcode-2020} and the WHM achieve similar accuracy and precision at the smallest scales. For the WHM, there is still a systematic downwards shift in power at the non-linear scale.

In summary, we have very good agreement between the WHM and the three independent and highly synergetic emulators studied here, especially given the lack of fitted parameters in the WHM, contrary to \texttt{HMcode-2020}. The WHM agreement and scatter with respect to all emulators are comparable, considering the differences in parameter spaces among these. While in all cases the WHM agreement is within the intrinsic emulator errors up to $k< r_\mathrm{nl}^{-1}$, there is a systematic downward shift in power just at $k\approx r_\mathrm{nl}^{-1}$. This consistently leads to an underprediction of $\approx 5\%$ at small scales in the \texttt{baccoemu} and \texttt{EuclidEmu2} cases, and less obvious in the \texttt{MiraTitan} case, depending on redshift. We investigate this issue further in the next section.

\section{Improved 1-halo term from simulation} \label{sec:1h}

To explore if the $5\%$ power deficit observed earlier is due to inaccuracies in modelling the 1-halo term, we will now utilise N-body simulations. The simulations are described in section \ref{sec:1h-nbody}, the measurement of the 1-halo term in \ref{sec:1h-con} and the results and conclusions in section \ref{sec:1h-mod}.

\subsection{N-body simulation} \label{sec:1h-nbody}

The simulation used is a companion simulation of the \texttt{BACCO} suite described in \cite{Angulo_2021} used to test the scaling technique \citep{Angulo_2010} in a variety of scenarios (see e.g. \citealt{Contreras_2020}). This gravity only simulation follows the evolution of $1536^3$ particles in a periodic box of volume $512^3[h^{-1}\rm{Mpc}]^3$, with a particle mass resolution of $m_p=3.18\times 10^9 h^{-1}\rm{M}_\odot$, equal to the \texttt{BACCO} suite. The cosmology of the simulation is a $\Lambda$CDM Planck-like model with parameters $\Omega_m=0.30964$, $h=0.6766$, $\Omega_b=0.04897$, $\sigma_8=0.8102$, $n_s=0.9665$, $\tau=0.0561$, and $\Omega_\nu=0$. The gravitational evolution is computed using L-Gadget3 \citep{Springel_2005,Angulo_2012,Angulo_2021}, a variant of the Gadget code, with a Plummer-equivalent softening length of $\epsilon=5h^{-1}\rm{kpc}$. The numerical parameters of the suite, including force and mass resolution, were chosen to achieve convergence at the 1\% level in the non-linear power spectrum at $k=10h^{-1}\rm{Mpc}$. The initial conditions were generated at $z=49$ using second-order Lagrangian perturbation theory (2LPT). To minimise cosmic variance, the amplitudes of Fourier modes were fixed to the ensemble average of the linear power spectrum. The particle catalogue was downsampled by a factor of $4^3$ for computational and memory efficiency.

\subsection{Measuring the 1-halo term} \label{sec:1h-con}

We measure the 1-halo term from the simulation described above by first identifying haloes using the FoF algorithm \citep{Davis_1985}. Haloes are defined as groups containing at least 20 matter particles, with a linking length of 0.2. Once identified, the data–data pairs, $DD(r)$, are measured as a function of separation using the public code \texttt{corrfunc} \citep{manodeep_sinha_2016,Sinha_2020}. Since this is an additive quantity, we compute it for each halo individually and then sum over all haloes. As the simulation is performed in a periodic box, the random–random pairs, $RR(r)$, correspond to the mean number within a shell at separation $r$, which sets the normalisation of the correlation function: $\xi^{1h}(r) = DD^{1h}(r)/RR(r)$.\footnote{We checked that this FoF-based measurement of $\xi^{1h}(r)$ converges towards $\xi^{NL}(r)$ at scales of $r\leq0.1\,\mathrm{Mpc}/h$ at the 0.5\% level.} To obtain the power spectrum, we Fourier transform this quantity, which in our case reduces to a one-dimensional integral.

\begin{figure}
    \centering
    \includegraphics[width=\columnwidth]{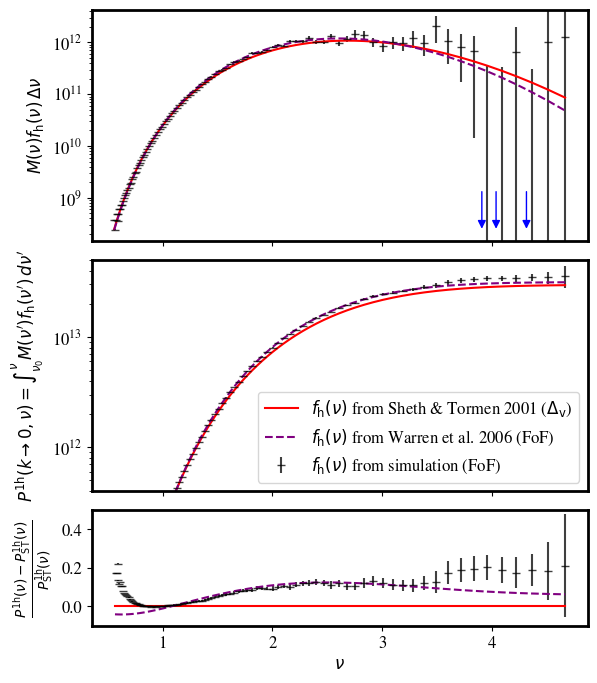}
    \caption{We show the 1-halo term integrand (top) of equation \eqref{eq:whm-hm-onehalo} in the large-scale limit ($W_\mathrm{h}(k\rightarrow0)=1$), the cumulative integrand (middle), and its residuals (bottom) with respect to the \citet{Sheth_2001} prediction (red, solid) as a function of peak-height $\nu$. Black datapoints indicate the simulation halo mass function measurement with error bars corresponding to the Poisson noise associated to the expected halo number counts in each bin given the \citet{Warren2006ApJ} prediction (purple, dashed). This avoids divergences in case of zero counts (blue arrows, top). Cumulative errors (middle and bottom) are obtained by adding all errors from lower $\nu$-bins in quadrature, i.e., assuming no correlation across bins.
    }
    \label{fig:1h-fnu}
\end{figure}

To quantify the (maximum) statistical uncertainty of our measurement we show in figure \ref{fig:1h-fnu} the 1-halo term integrand (top panel) and the cumulative integrand (middle and bottom panels) in the low-$k$ limit, where the halo window function in equation \eqref{eq:whm-hm-onehalo} becomes unity, as a function of peak-height $\nu$. The black data points represent the result using the measured halo mass function $f_\mathrm{h}(\nu)$ from the FoF haloes described above. We see that, naturally, the \sB{phenomenological} \citet{Warren2006ApJ} fit to FoF haloes represents a better fit to the measurement than the baseline, virial overdensity halo mass function by \citet{Sheth_2001}. The deviation on high $\nu$ (high masses) is attributed to the increased level of noise, as halo number counts within those bins approach order unity. This motivates our test of employing the \citet{Warren2006ApJ} halo mass function in section \ref{sec:res-var}.

Note that, since the simulation is part of the \texttt{BACCO} suite, the 1-halo term computed here shares the same parameter choices as the simulations used to train \texttt{baccoemu}. Given their larger volume, the impact of noise at the high mass end is expected to be at the sub-percent level, in contrast to the $\sim 20\%$ effect visible in figure \ref{fig:1h-fnu}.

\subsection{Resulting model predictions} \label{sec:1h-mod}

\begin{figure*}
    \centering
    \includegraphics[width=\textwidth]{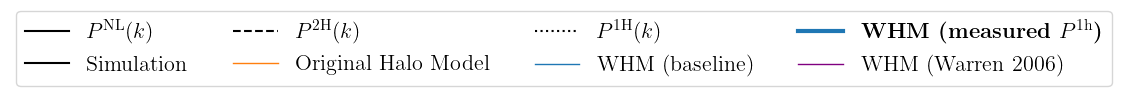}
    \includegraphics[width=\textwidth]{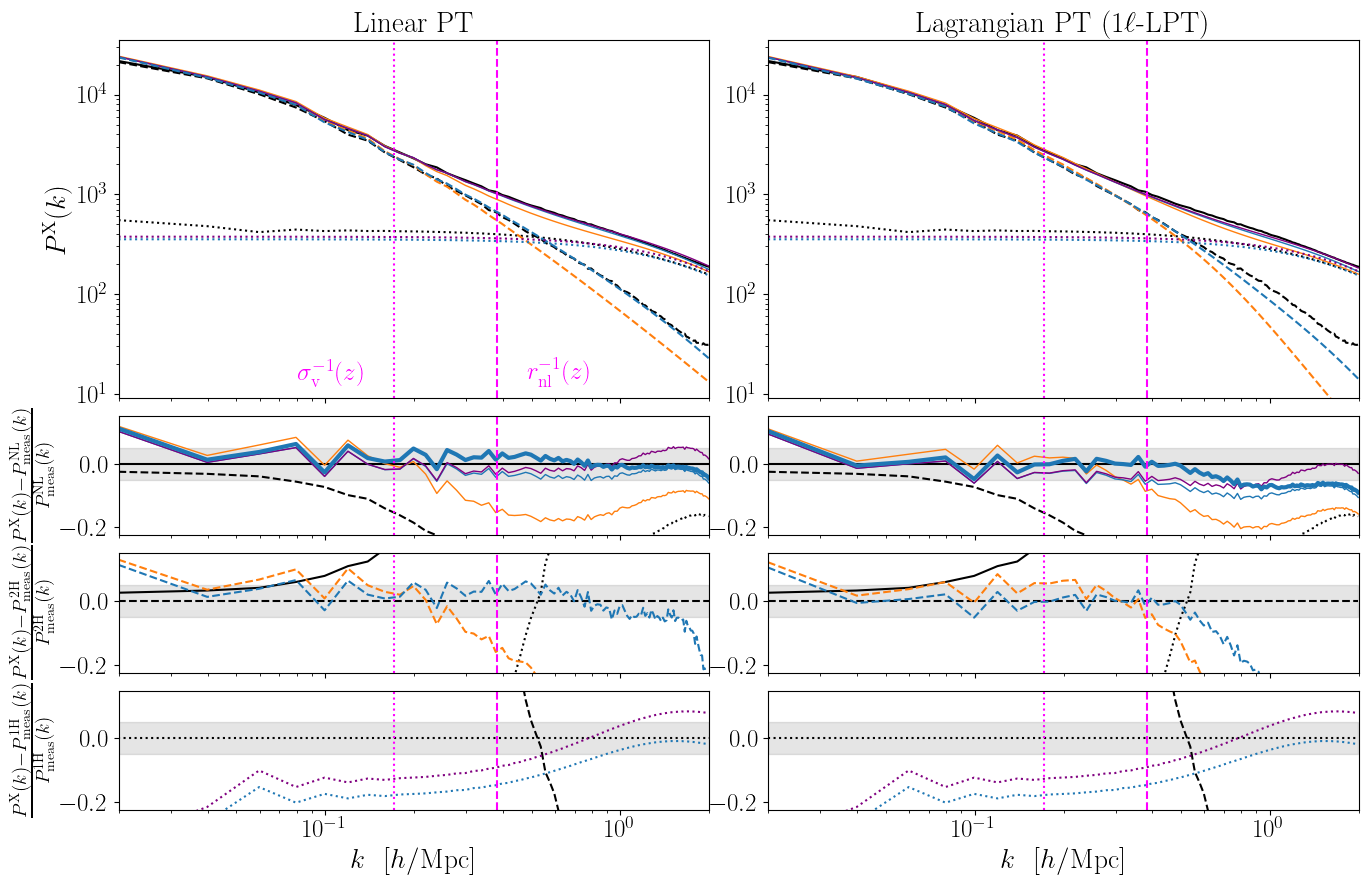}
    \caption{We compare the original halo model (orange) and WHM (blue) predictions against the simulations (black) for the 2-halo term (dashed), 1-halo term (dotted), and their sum, the full non-linear matter power spectrum (solid) at $z=0$. We use either Linear PT (left column) or 1-$\ell$LPT (right column) as baseline PT. The top row shows all power spectra, and centre and bottom rows their residuals with respect to the simulation $P^\mathrm{NL}(k)$ and $P^\mathrm{2h}(k)$ measurements, respectively. The thick blue solid line shows the impact of replacing the theoretical (uncompensated) 1-halo term by the measurement with nearly 20\% larger amplitude.}
    \label{fig:1h-sim}
\end{figure*}

Having measured (see sections \ref{sec:1h-nbody} and \ref{sec:1h-con}) the simulation non-linear matter power spectrum $P^\mathrm{NL}(k)$ and 1-halo term $P^\mathrm{1h}(k)$, we can infer the simulation 2-halo term as
\begin{align} \label{eq:1h-NL-2h}
P^\mathrm{2h}(k) = P^\mathrm{NL}(k) - P^\mathrm{1h}(k)~. 
\end{align}
In figure \ref{fig:1h-sim} we compare these simulation power spectra (black) to the original halo model (orange) and WHM (blue) predictions using either Linear PT (left column) or $1\ell$-LPT (right column) as baseline PT. In both cases, we see that the WHM theory prediction for $P^\mathrm{NL}(k)$ (center row) matches the simulation measurements quite well (within $5\%$) up to the non-linear scale, but even further for Linear PT.\footnote{The good agreement for Linear PT is just a coincidence of the Planck cosmology at $z=0$ and does not hold for general cosmological models and larger redshift ranges, as demonstrated in figure \ref{fig:res-fixed}.} The situation is similar for the 2-halo term (lower row). However, as can be seen from the dotted lines in the upper row, the theoretical 1-halo term (identical for the original halo model and WHM) is about 20\% lower in amplitude than the simulation measurement in the large-scale limit.

As a next step, we investigate the impact of replacing the theoretical 1-halo term given by eq. \eqref{eq:whm-hm-onehalo} by the measured 1-halo term.\footnote{For the WHM, we hence replace the first term in eq. \eqref{eq:whm-whm-1h} by the measurement, including the compensation term proportional to $W_f^2(k,\nu)$ in the model. That term is hence interpreted as a contribution to the theoretical two-halo term, following eq. \eqref{eq:1h-NL-2h}.} The resulting non-linear power spectrum prediction is shown as the thick blue lines in the centre row. This procedure improves the agreement with the simulation significantly, for both Linear PT and $1\ell$-LPT. For the latter, the regime of 5\% precision now extends well beyond the non-linear scale. At scales around $1\,h\mathrm{Mpc}^{-1}$, the non-linear power is damped due to the sudden drop in 2-halo power (dashed blue line in lower right panel), although the 1-halo term starts dominating in this regime. To conclude, we find that
\begin{enumerate}
    \item Replacing the theoretical 1-halo term by the one measured from the simulation improves the agreement in $P^\mathrm{NL}(k)$ between WHM and simulation and doubles the $k$-range within the 5\% accuracy band from the non-linear scale $k\approx0.4\,h\mathrm{Mpc}^{-1}$ towards $k\approx0.8\,h\mathrm{Mpc}^{-1}$. \sB{This is true even when comparing the measurement to the \citet{Warren2006ApJ} model, even though it is found to deliver a good fit to the measured halo mass function in figure \ref{fig:1h-fnu}. However, the scatter in the measurement at high masses dominates, leading to the difference in the 1-halo term.}
    \item Perfect agreement across the full 2-halo to 1-halo transition regime up to $k\approx2\,h\mathrm{Mpc}^{-1}$ requires accurate 2-halo term predictions up to $k\approx0.7\,h\mathrm{Mpc}^{-1}$. The WHM formalism provides a faithful 2-halo term prediction up to $k\approx0.5\,h\mathrm{Mpc}^{-1}$, but not beyond. 
\end{enumerate}

\section{Conclusions} \label{sec:concl}

In this work, we presented and validated a novel analytic approach to faithfully predict the non-linear matter power spectrum, the \textit{web-halo model}. We showed that, without any free fitting parameters, it is able to fit all the investigated emulators within their intrinsic uncertainty up to the non-linear scale in the full $w_0 w_a$CDM$+\sum m_\nu$ cosmological parameter space. To our knowledge, this is the first model of its kind to achieve this, even outperforming the \texttt{HMcode-2020} 12-parameter fit for the emulator it has been fit to. 

The web-halo model physically extends the basic halo model calculation by incorporating the ellipsoidal collapse excursion set framework of \citet{Shen_2006}. The general idea, based on the work by \citet{Sheth_1999} and~\citet{Sheth_2001}, is already widely used to predict halo mass functions. The additional step in this work was to make use of the sheet and filament mass functions, which necessarily arise as a result of anisotropic collapse. When paired with 1-loop LPT ($1\ell$-LPT), we showed that the additional sheet and filament terms contribute exactly the missing power to fit the 2-halo to 1-halo transition regime.

On scales smaller than the non-linear scale, we observed a systematic underprediction of $3\%-5\%$, depending on the scenario. When using linear PT, the small-scale limit was recovered well, indicating that the $1\ell$-LPT based 2-halo (or 2-sheet) term lacks some power on the smallest scales. This would not come as a surprise, given that the non-linear scale is used as a cutoff scale for the $1\ell$-LPT kernels. On the other hand, we showed in section \ref{sec:1h} that the small-scale limit may be partially recovered when measuring the 1-halo term directly from the simulation, suggesting that there is an issue with the theoretical 1-halo term, rather than the 2-halo (or 2-sheet) term. Given that there is not much cosmological information on these small scales due to degeneracies with baryonic feedback processes, etc, we leave the resolution of this issue to future work.

The WHM is available as an extension of the \texttt{HMcode-2020} package by \citet{Mead_2021} and takes advantage of all the benefits associated with the halo model calculations. It is fast, baryonic feedback models can be easily incorporated \sB{using the same implementation as \citet{Mead_2020}}, and it provides a natural framework extending the \texttt{ReAct} formalism \citet{Cataneo_2019a}. 

We recommend using the WHM in tandem with $1\ell$-LPT. The latter represents a bottleneck,\footnote{To evaluate a Planck $\Lambda$CDM cosmology, on a 8 core machine, running \texttt{CAMB+WHMcode} takes about $2$s (only $0.02s$ more than \texttt{CAMB+HMcode}), while \texttt{velocileptors} takes $8$s to evaluate 100 $k$-bins between $0.02 < k\,[h/\mathrm{Mpc}]<2$ at each redshift.} and it would be interesting to investigate potential speed-ups (like constructing a $1\ell$-LPT, or even a $1\ell$-LPT+WHM emulator using software such as \texttt{COSMOPOWER}\footnote{\href{https://alessiospuriomancini.github.io/cosmopower/}{https://alessiospuriomancini.github.io/cosmopower/}} by \citealt{SpurioMancini2022MNRAS.511.1771S}, \texttt{Effort}\footnote{\href{https://github.com/CosmologicalEmulators/Effort.jl}{https://github.com/CosmologicalEmulators/Effort.jl}} 
 by \citealt{2025JCAP...09..044B}, or similar). The $1\ell$-LPT+WHM combination also comes with the prospect that it would allow for the first time a proper combination of galaxy clustering and weak lensing model predictions. In future work, we plan to investigate generalising the WHM towards a similar prescription as \citet{Sheth_2003}. We foresee that this could pave the way for a more accurate bias expansion beyond PT based on the WHM. This would have a long-lasting impact on the cosmological analyses of large-scale structure datasets. 

\section*{Acknowledgements}
\sB{We thank the referee, Christos Georgiou, for useful comments that helped us improve the clarity of our work, in particular concerning figure \ref{fig:ing-win} and the addition of appendix \ref{app:window-numbers}.}
We thank Benjamin Bose, Yan-Chuan Cai, Pedro Carrilho, Joanne Cohn, Héctor Gil-Marín, Oliver Hahn, Raúl Jimenez, Julien Lesgourgues, Farnik Nikakhtar, John Peacock, Ravi Sheth, Licia Verde, and Martin White for interesting discussions and Catherine Heymans, Alexander Mead, Farnik Nikakhtar, John Peacock, and Ravi Sheth for valuable comments on an advanced version of the manuscript. We especially thank Ravi Sheth for suggesting the additional test of filament substructure in appendix \ref{app:window}. We are indebted to Jens Stücker for publishing his 2D simulation code. \sB{We thank Alexander Tipp for pointing us to a typo in equation \eqref{eq:ing-mf-numbers}.} 
SB and FB acknowledge support from the European Research Council (ERC) under the European Union’s Horizon 2020 research and innovation program (FutureLSS, grant agreement 853291). FB is a University Research Fellow. 

%%%%%%%%%%%%%%%%%%%%%%%%%%%%%%%%%%%%%%%%%%%%%%%%%%
\section*{Data Availability}

\sB{We release the code associated with the web-halo model as an additional option for calculating non-linear corrections in the Boltzmann-codes \texttt{CAMB} and \texttt{CLASS} within their respective \texttt{HMcode} modules, as well as notebook to generate the figures of this work. The materials are available at \href{https://github.com/SamuelBrieden/WHM}{https://github.com/SamuelBrieden/WHM}.}

%%%%%%%%%%%%%%%%%%%% REFERENCES %%%%%%%%%%%%%%%%%%

% The best way to enter references is to use BibTeX:

\bibliographystyle{mnras}
\bibliography{example} % if your bibtex file is called example.bib

\begin{thebibliography}{}
\makeatletter
\relax
\def\mn@urlcharsother{\let\do\@makeother \do\$\do\&\do\#\do\^\do\_\do\%\do\~}
\def\mn@doi{\begingroup\mn@urlcharsother \@ifnextchar [ {\mn@doi@}
  {\mn@doi@[]}}
\def\mn@doi@[#1]#2{\def\@tempa{#1}\ifx\@tempa\@empty \href
  {http://dx.doi.org/#2} {doi:#2}\else \href {http://dx.doi.org/#2} {#1}\fi
  \endgroup}
\def\mn@eprint#1#2{\mn@eprint@#1:#2::\@nil}
\def\mn@eprint@arXiv#1{\href {http://arxiv.org/abs/#1} {{\tt arXiv:#1}}}
\def\mn@eprint@dblp#1{\href {http://dblp.uni-trier.de/rec/bibtex/#1.xml}
  {dblp:#1}}
\def\mn@eprint@#1:#2:#3:#4\@nil{\def\@tempa {#1}\def\@tempb {#2}\def\@tempc
  {#3}\ifx \@tempc \@empty \let \@tempc \@tempb \let \@tempb \@tempa \fi \ifx
  \@tempb \@empty \def\@tempb {arXiv}\fi \@ifundefined
  {mn@eprint@\@tempb}{\@tempb:\@tempc}{\expandafter \expandafter \csname
  mn@eprint@\@tempb\endcsname \expandafter{\@tempc}}}

\bibitem[\protect\citeauthoryear{{Abbott} et~al.,}{{Abbott}
  et~al.}{2025}]{DES:2025xii}
{Abbott} T.~M.~C.,  et~al., 2025, \mn@doi [\prd] {10.1103/3dzh-d8f5}, \href
  {https://ui.adsabs.harvard.edu/abs/2025PhRvD.112h3535A} {112, 083535}

\bibitem[\protect\citeauthoryear{{Abdul Karim} et~al.,}{{Abdul Karim}
  et~al.}{2025a}]{DESI:2025zpo}
{Abdul Karim} M.,  et~al., 2025a, \mn@doi [\prd] {10.1103/2wwn-xjm5}, \href
  {https://ui.adsabs.harvard.edu/abs/2025PhRvD.112h3514A} {112, 083514}

\bibitem[\protect\citeauthoryear{{Abdul Karim} et~al.,}{{Abdul Karim}
  et~al.}{2025b}]{desidr2bao2025arXiv250314738D}
{Abdul Karim} M.,  et~al., 2025b, \mn@doi [\prd] {10.1103/tr6y-kpc6}, \href
  {https://ui.adsabs.harvard.edu/abs/2025PhRvD.112h3515A} {112, 083515}

\bibitem[\protect\citeauthoryear{{Adame} et~al.,}{{Adame}
  et~al.}{2025a}]{desidr1fscosmo2024arXiv241112022D}
{Adame} A.~G.,  et~al., 2025a, \mn@doi [\jcap] {10.1088/1475-7516/2025/07/028},
  \href {https://ui.adsabs.harvard.edu/abs/2025JCAP...07..028A} {2025, 028}

\bibitem[\protect\citeauthoryear{{Adame} et~al.,}{{Adame}
  et~al.}{2025b}]{desidr1fs2024arXiv241112021D}
{Adame} A.~G.,  et~al., 2025b, \mn@doi [\jcap] {10.1088/1475-7516/2025/09/008},
  \href {https://ui.adsabs.harvard.edu/abs/2025JCAP...09..008A} {2025, 008}

\bibitem[\protect\citeauthoryear{Alonso, Eardley  \& Peacock}{Alonso
  et~al.}{2015}]{Alonso:2014zfa}
Alonso D.,  Eardley E.,   Peacock J.~A.,  2015, \mn@doi [MNRAS]
  {10.1093/mnras/stu2632}, 447, 2683

\bibitem[\protect\citeauthoryear{{Angulo} \& {White}}{{Angulo} \&
  {White}}{2010}]{Angulo_2010}
{Angulo} R.~E.,  {White} S.~D.~M.,  2010, \mn@doi [\mnras]
  {10.1111/j.1365-2966.2010.16459.x}, \href
  {https://ui.adsabs.harvard.edu/abs/2010MNRAS.405..143A} {405, 143}

\bibitem[\protect\citeauthoryear{{Angulo}, {Springel}, {White}, {Jenkins},
  {Baugh}  \& {Frenk}}{{Angulo} et~al.}{2012}]{Angulo_2012}
{Angulo} R.~E.,  {Springel} V.,  {White} S.~D.~M.,  {Jenkins} A.,  {Baugh}
  C.~M.,   {Frenk} C.~S.,  2012, \mn@doi [\mnras]
  {10.1111/j.1365-2966.2012.21830.x}, \href
  {https://ui.adsabs.harvard.edu/abs/2012MNRAS.426.2046A} {426, 2046}

\bibitem[\protect\citeauthoryear{{Angulo}, {Zennaro}, {Contreras}, {Aric{\`o}},
  {Pellejero-Iba{\~n}ez}  \& {St{\"u}cker}}{{Angulo}
  et~al.}{2021}]{Angulo_2021}
{Angulo} R.~E.,  {Zennaro} M.,  {Contreras} S.,  {Aric{\`o}} G.,
  {Pellejero-Iba{\~n}ez} M.,   {St{\"u}cker} J.,  2021, \mn@doi [\mnras]
  {10.1093/mnras/stab2018}, \href
  {https://ui.adsabs.harvard.edu/abs/2021MNRAS.507.5869A} {507, 5869}

\bibitem[\protect\citeauthoryear{{Aric{\`o}}, {Angulo}, {Contreras},
  {Ondaro-Mallea}, {Pellejero-Iba{\~n}ez}  \& {Zennaro}}{{Aric{\`o}}
  et~al.}{2021}]{Arico2_2021}
{Aric{\`o}} G.,  {Angulo} R.~E.,  {Contreras} S.,  {Ondaro-Mallea} L.,
  {Pellejero-Iba{\~n}ez} M.,   {Zennaro} M.,  2021, \mn@doi [\mnras]
  {10.1093/mnras/stab1911}, \href
  {https://ui.adsabs.harvard.edu/abs/2021MNRAS.506.4070A} {506, 4070}

\bibitem[\protect\citeauthoryear{{Aric{\`o}}, {Angulo}  \&
  {Zennaro}}{{Aric{\`o}} et~al.}{2022}]{Arico1_2021}
{Aric{\`o}} G.,  {Angulo} R.~E.,   {Zennaro} M.,  2022, \mn@doi [Open Res
  Europe] {10.12688/openreseurope.14310.2}, \href
  {https://ui.adsabs.harvard.edu/abs/2021arXiv210414568A} {1, 152}

\bibitem[\protect\citeauthoryear{{Asgari}, {Mead}  \& {Heymans}}{{Asgari}
  et~al.}{2023}]{Asgari_2023}
{Asgari} M.,  {Mead} A.~J.,   {Heymans} C.,  2023, \mn@doi [The Open Journal of
  Astrophysics] {10.21105/astro.2303.08752}, \href
  {https://ui.adsabs.harvard.edu/abs/2023OJAp....6E..39A} {6, 39}

\bibitem[\protect\citeauthoryear{Baldauf, Seljak, Smith, Hamaus  \&
  Desjacques}{Baldauf et~al.}{2013}]{Baldauf_2013}
Baldauf T.,  Seljak U. c.~v.,  Smith R.~E.,  Hamaus N.,   Desjacques V.,  2013,
  \mn@doi [Phys. Rev. D] {10.1103/PhysRevD.88.083507}, 88, 083507

\bibitem[\protect\citeauthoryear{{Baldauf}, {Schaan}  \&
  {Zaldarriaga}}{{Baldauf} et~al.}{2016}]{Baldauf_2016}
{Baldauf} T.,  {Schaan} E.,   {Zaldarriaga} M.,  2016, \mn@doi [\jcap]
  {10.1088/1475-7516/2016/03/007}, \href
  {https://ui.adsabs.harvard.edu/abs/2016JCAP...03..007B} {2016, 007}

\bibitem[\protect\citeauthoryear{{Baumann}, {Nicolis}, {Senatore}  \&
  {Zaldarriaga}}{{Baumann} et~al.}{2012}]{Baumann2012JCAP...07..051B}
{Baumann} D.,  {Nicolis} A.,  {Senatore} L.,   {Zaldarriaga} M.,  2012, \mn@doi
  [\jcap] {10.1088/1475-7516/2012/07/051}, \href
  {https://ui.adsabs.harvard.edu/abs/2012JCAP...07..051B} {2012, 051}

\bibitem[\protect\citeauthoryear{Bernardeau, Colombi, Gaztañaga  \&
  Scoccimarro}{Bernardeau et~al.}{2002}]{BERNARDEAU_2001}
Bernardeau F.,  Colombi S.,  Gaztañaga E.,   Scoccimarro R.,  2002, \mn@doi
  [Physics Reports] {https://doi.org/10.1016/S0370-1573(02)00135-7}, 367, 1

\bibitem[\protect\citeauthoryear{Bharadwaj}{Bharadwaj}{1996}]{Bharadwaj_1996}
Bharadwaj S.,  1996, \mn@doi [The Astrophysical Journal] {10.1086/178036}, 472,
  1

\bibitem[\protect\citeauthoryear{{Bianco} et~al.,}{{Bianco}
  et~al.}{2022}]{lsst2022ApJS..258....1B}
{Bianco} F.~B.,  et~al., 2022, \mn@doi [\apjs] {10.3847/1538-4365/ac3e72},
  \href {https://ui.adsabs.harvard.edu/abs/2022ApJS..258....1B} {258, 1}

\bibitem[\protect\citeauthoryear{Bird, Viel  \& Haehnelt}{Bird
  et~al.}{2012}]{Bird_2014}
Bird S.,  Viel M.,   Haehnelt M.~G.,  2012, \mn@doi [\mnras]
  {10.1111/j.1365-2966.2011.20222.x}, 420, 2551

\bibitem[\protect\citeauthoryear{Blas, Lesgourgues  \& Tram}{Blas
  et~al.}{2011}]{Blas_2011}
Blas D.,  Lesgourgues J.,   Tram T.,  2011, \mn@doi [\jcap]
  {10.1088/1475-7516/2011/07/034}, 2011, 034

\bibitem[\protect\citeauthoryear{{Bond} \& {Myers}}{{Bond} \&
  {Myers}}{1996}]{Bond_1996}
{Bond} J.~R.,  {Myers} S.~T.,  1996, \mn@doi [\apjs] {10.1086/192267}, \href
  {https://ui.adsabs.harvard.edu/abs/1996ApJS..103....1B} {103, 1}

\bibitem[\protect\citeauthoryear{{Bond}, {Cole}, {Efstathiou}  \&
  {Kaiser}}{{Bond} et~al.}{1991}]{Bond_1991}
{Bond} J.~R.,  {Cole} S.,  {Efstathiou} G.,   {Kaiser} N.,  1991, \mn@doi
  [\apj] {10.1086/170520}, \href
  {https://ui.adsabs.harvard.edu/abs/1991ApJ...379..440B} {379, 440}

\bibitem[\protect\citeauthoryear{{Bond}, {Kofman}  \& {Pogosyan}}{{Bond}
  et~al.}{1996}]{Bond_1996CosmicWeb}
{Bond} J.~R.,  {Kofman} L.,   {Pogosyan} D.,  1996, \mn@doi [\nat]
  {10.1038/380603a0}, \href
  {https://ui.adsabs.harvard.edu/abs/1996Natur.380..603B} {380, 603}

\bibitem[\protect\citeauthoryear{{Bonici}, {D'Amico}, {Bel}  \&
  {Carbone}}{{Bonici} et~al.}{2025}]{2025JCAP...09..044B}
{Bonici} M.,  {D'Amico} G.,  {Bel} J.,   {Carbone} C.,  2025, \mn@doi [\jcap]
  {10.1088/1475-7516/2025/09/044}, \href
  {https://ui.adsabs.harvard.edu/abs/2025JCAP...09..044B} {2025, 044}

\bibitem[\protect\citeauthoryear{Bose, Cataneo, Tröster, Xia, Heymans  \&
  Lombriser}{Bose et~al.}{2020}]{Bose_2020}
Bose B.,  Cataneo M.,  Tröster T.,  Xia Q.,  Heymans C.,   Lombriser L.,
  2020, \mn@doi [\mnras] {10.1093/mnras/staa2696}, 498, 4650

\bibitem[\protect\citeauthoryear{Bose et~al.,}{Bose et~al.}{2021}]{Bose_2021}
Bose B.,  et~al., 2021, \mn@doi [\mnras] {10.1093/mnras/stab2731}, 508, 2479

\bibitem[\protect\citeauthoryear{Brinckmann, Lindholmer, Hansen  \&
  Falco}{Brinckmann et~al.}{2016}]{Brinckmann:2014nia}
Brinckmann T.,  Lindholmer M.,  Hansen S.~H.,   Falco M.,  2016, \mn@doi [JCAP]
  {10.1088/1475-7516/2016/04/007}, 04, 007

\bibitem[\protect\citeauthoryear{{Bullock}, {Kolatt}, {Sigad}, {Somerville},
  {Kravtsov}, {Klypin}, {Primack}  \& {Dekel}}{{Bullock}
  et~al.}{2001}]{Bullock_2001}
{Bullock} J.~S.,  {Kolatt} T.~S.,  {Sigad} Y.,  {Somerville} R.~S.,  {Kravtsov}
  A.~V.,  {Klypin} A.~A.,  {Primack} J.~R.,   {Dekel} A.,  2001, \mn@doi
  [\mnras] {10.1046/j.1365-8711.2001.04068.x}, \href
  {https://ui.adsabs.harvard.edu/abs/2001MNRAS.321..559B} {321, 559}

\bibitem[\protect\citeauthoryear{Carrilho, Moretti  \& Pourtsidou}{Carrilho
  et~al.}{2023}]{Carrilho_2022_priors}
Carrilho P.,  Moretti C.,   Pourtsidou A.,  2023, \mn@doi [JCAP]
  {10.1088/1475-7516/2023/01/028}, 01, 028

\bibitem[\protect\citeauthoryear{{Castorina}, {Sefusatti}, {Sheth},
  {Villaescusa-Navarro}  \& {Viel}}{{Castorina}
  et~al.}{2014}]{Castorina_2014JCAP}
{Castorina} E.,  {Sefusatti} E.,  {Sheth} R.~K.,  {Villaescusa-Navarro} F.,
  {Viel} M.,  2014, \mn@doi [\jcap] {10.1088/1475-7516/2014/02/049}, \href
  {https://ui.adsabs.harvard.edu/abs/2014JCAP...02..049C} {2014, 049}

\bibitem[\protect\citeauthoryear{Cataneo, Lombriser, Heymans, Mead, Barreira,
  Bose  \& Li}{Cataneo et~al.}{2019a}]{Cataneo_2019a}
Cataneo M.,  Lombriser L.,  Heymans C.,  Mead A.~J.,  Barreira A.,  Bose S.,
  Li B.,  2019a, \mn@doi [\mnras] {10.1093/mnras/stz1836}, 488, 2121

\bibitem[\protect\citeauthoryear{Cataneo, Emberson, Inman, Harnois-Déraps  \&
  Heymans}{Cataneo et~al.}{2019b}]{Cataneo_2019b}
Cataneo M.,  Emberson J.~D.,  Inman D.,  Harnois-Déraps J.,   Heymans C.,
  2019b, \mn@doi [\mnras] {10.1093/mnras/stz3189}, 491, 3101

\bibitem[\protect\citeauthoryear{{Chen} \& {Afshordi}}{{Chen} \&
  {Afshordi}}{2020}]{Chen_2020}
{Chen} A.~Y.,  {Afshordi} N.,  2020, \mn@doi [\prd]
  {10.1103/PhysRevD.101.103522}, \href
  {https://ui.adsabs.harvard.edu/abs/2020PhRvD.101j3522C} {101, 103522}

\bibitem[\protect\citeauthoryear{{Chen}, {Vlah}  \& {White}}{{Chen}
  et~al.}{2020}]{Chen_LPT_2020}
{Chen} S.-F.,  {Vlah} Z.,   {White} M.,  2020, \mn@doi [\jcap]
  {10.1088/1475-7516/2020/07/062}, \href
  {https://ui.adsabs.harvard.edu/abs/2020JCAP...07..062C} {2020, 062}

\bibitem[\protect\citeauthoryear{{Chen}, {Vlah}, {Castorina}  \&
  {White}}{{Chen} et~al.}{2021}]{Chen_velo_2021}
{Chen} S.-F.,  {Vlah} Z.,  {Castorina} E.,   {White} M.,  2021, \mn@doi [\jcap]
  {10.1088/1475-7516/2021/03/100}, \href
  {https://ui.adsabs.harvard.edu/abs/2021JCAP...03..100C} {2021, 100}

\bibitem[\protect\citeauthoryear{{Contreras}, {Angulo}, {Zennaro}, {Aric{\`o}}
  \& {Pellejero-Iba{\~n}ez}}{{Contreras} et~al.}{2020}]{Contreras_2020}
{Contreras} S.,  {Angulo} R.~E.,  {Zennaro} M.,  {Aric{\`o}} G.,
  {Pellejero-Iba{\~n}ez} M.,  2020, \mn@doi [\mnras] {10.1093/mnras/staa3117},
  \href {https://ui.adsabs.harvard.edu/abs/2020MNRAS.499.4905C} {499, 4905}

\bibitem[\protect\citeauthoryear{Cooray \& Sheth}{Cooray \&
  Sheth}{2002}]{Cooray_2002}
Cooray A.,  Sheth R.,  2002, \mn@doi [Physics Reports]
  {https://doi.org/10.1016/S0370-1573(02)00276-4}, 372, 1

\bibitem[\protect\citeauthoryear{{DESI Collaboration} et~al.,}{{DESI
  Collaboration} et~al.}{2025a}]{DESI:2024lzq}
{DESI Collaboration} et~al., 2025a, \mn@doi [\jcap]
  {10.1088/1475-7516/2025/01/124}, \href
  {https://ui.adsabs.harvard.edu/abs/2025JCAP...01..124A} {2025, 124}

\bibitem[\protect\citeauthoryear{{DESI Collaboration} et~al.,}{{DESI
  Collaboration} et~al.}{2025b}]{DESIDR1cosmo2025JCAP...02..021A}
{DESI Collaboration} et~al., 2025b, \mn@doi [\jcap]
  {10.1088/1475-7516/2025/02/021}, \href
  {https://ui.adsabs.harvard.edu/abs/2025JCAP...02..021A} {2025, 021}

\bibitem[\protect\citeauthoryear{{DESI Collaboration} et~al.,}{{DESI
  Collaboration} et~al.}{2025c}]{desidr1bao2025JCAP...04..012A}
{DESI Collaboration} et~al., 2025c, \mn@doi [\jcap]
  {10.1088/1475-7516/2025/04/012}, \href
  {https://ui.adsabs.harvard.edu/abs/2025JCAP...04..012A} {2025, 012}

\bibitem[\protect\citeauthoryear{{Dalal} et~al.,}{{Dalal}
  et~al.}{2023}]{Dalal:2023olq}
{Dalal} R.,  et~al., 2023, \mn@doi [\prd] {10.1103/PhysRevD.108.123519}, \href
  {https://ui.adsabs.harvard.edu/abs/2023PhRvD.108l3519D} {108, 123519}

\bibitem[\protect\citeauthoryear{{Davis}, {Efstathiou}, {Frenk}  \&
  {White}}{{Davis} et~al.}{1985}]{Davis_1985}
{Davis} M.,  {Efstathiou} G.,  {Frenk} C.~S.,   {White} S.~D.~M.,  1985,
  \mn@doi [\apj] {10.1086/163168}, \href
  {https://ui.adsabs.harvard.edu/abs/1985ApJ...292..371D} {292, 371}

\bibitem[\protect\citeauthoryear{Desjacques, Jeong  \& Schmidt}{Desjacques
  et~al.}{2018}]{Desjacques:2016bnm}
Desjacques V.,  Jeong D.,   Schmidt F.,  2018, \mn@doi [Phys. Rept.]
  {10.1016/j.physrep.2017.12.002}, 733, 1

\bibitem[\protect\citeauthoryear{Dolag, Bartelmann, Perrotta, Baccigalupi,
  Moscardini, Meneghetti  \& Tormen}{Dolag et~al.}{2004}]{Dolag_2003}
Dolag K.,  Bartelmann M.,  Perrotta F.,  Baccigalupi C.,  Moscardini L.,
  Meneghetti M.,   Tormen G.,  2004, \mn@doi [Astron. Astrophys.]
  {10.1051/0004-6361:20031757}, 416, 853

\bibitem[\protect\citeauthoryear{{Euclid Collaboration} et~al.,}{{Euclid
  Collaboration} et~al.}{2025a}]{euclid2025arXiv250315302E}
{Euclid Collaboration} et~al., 2025a, \mn@doi [arXiv e-prints]
  {10.48550/arXiv.2503.15302}, \href
  {https://ui.adsabs.harvard.edu/abs/2025arXiv250315302E} {p. arXiv:2503.15302}

\bibitem[\protect\citeauthoryear{{Euclid Collaboration} et~al.,}{{Euclid
  Collaboration} et~al.}{2025b}]{euclid2025A&A...697A...1E}
{Euclid Collaboration} et~al., 2025b, \mn@doi [\aap]
  {10.1051/0004-6361/202450810}, \href
  {https://ui.adsabs.harvard.edu/abs/2025A&A...697A...1E} {697, A1}

\bibitem[\protect\citeauthoryear{{Gil-Mar{\'\i}n}, {Jimenez}  \&
  {Verde}}{{Gil-Mar{\'\i}n} et~al.}{2011}]{GilMarin2011MNRAS}
{Gil-Mar{\'\i}n} H.,  {Jimenez} R.,   {Verde} L.,  2011, \mn@doi [\mnras]
  {10.1111/j.1365-2966.2011.18456.x}, \href
  {https://ui.adsabs.harvard.edu/abs/2011MNRAS.414.1207G} {414, 1207}

\bibitem[\protect\citeauthoryear{{Gunn} \& {Gott}}{{Gunn} \&
  {Gott}}{1972}]{Gunn_1972}
{Gunn} J.~E.,  {Gott} III J.~R.,  1972, \mn@doi [\apj] {10.1086/151605}, \href
  {https://ui.adsabs.harvard.edu/abs/1972ApJ...176....1G} {176, 1}

\bibitem[\protect\citeauthoryear{Heitmann, Lawrence, Kwan, Habib  \&
  Higdon}{Heitmann et~al.}{2014}]{Heitmann_2014}
Heitmann K.,  Lawrence E.,  Kwan J.,  Habib S.,   Higdon D.,  2014, \mn@doi
  [The Astrophysical Journal] {10.1088/0004-637X/780/1/111}, 780, 111

\bibitem[\protect\citeauthoryear{Heitmann et~al.,}{Heitmann
  et~al.}{2016}]{Heitmann_2016}
Heitmann K.,  et~al., 2016, \mn@doi [The Astrophysical Journal]
  {10.3847/0004-637X/820/2/108}, 820, 108

\bibitem[\protect\citeauthoryear{{Hertzsch}, {Feldbrugge}, {Rodriguez}  \& {van
  de Weygaert}}{{Hertzsch} et~al.}{2025}]{Hertzsch:2025grv}
{Hertzsch} B.,  {Feldbrugge} J.,  {Rodriguez} M.,   {van de Weygaert} R.,
  2025, \mn@doi [arXiv e-prints] {10.48550/arXiv.2510.02419}, \href
  {https://ui.adsabs.harvard.edu/abs/2025arXiv251002419H} {p. arXiv:2510.02419}

\bibitem[\protect\citeauthoryear{{Icke} \& {van de Weygaert}}{{Icke} \& {van de
  Weygaert}}{1987}]{IckevdW1987}
{Icke} V.,  {van de Weygaert} R.,  1987, \aap, \href
  {https://ui.adsabs.harvard.edu/abs/1987A&A...184...16I} {184, 16}

\bibitem[\protect\citeauthoryear{{Ivanov}, {Simonovi{\'c}}  \&
  {Zaldarriaga}}{{Ivanov} et~al.}{2020}]{Ivanov2020JCAP...05..042I}
{Ivanov} M.~M.,  {Simonovi{\'c}} M.,   {Zaldarriaga} M.,  2020, \mn@doi [\jcap]
  {10.1088/1475-7516/2020/05/042}, \href
  {https://ui.adsabs.harvard.edu/abs/2020JCAP...05..042I} {2020, 042}

\bibitem[\protect\citeauthoryear{Knabenhans et~al.}{Knabenhans
  et~al.}{2021}]{Euclid:2020rfv}
Knabenhans M.,  et~al., 2021, \mn@doi [Mon. Not. Roy. Astron. Soc.]
  {10.1093/mnras/stab1366}, 505, 2840

\bibitem[\protect\citeauthoryear{{Knebe} et~al.,}{{Knebe}
  et~al.}{2011}]{Knebe_2011}
{Knebe} A.,  et~al., 2011, \mn@doi [\mnras] {10.1111/j.1365-2966.2011.18858.x},
  \href {https://ui.adsabs.harvard.edu/abs/2011MNRAS.415.2293K} {415, 2293}

\bibitem[\protect\citeauthoryear{{Kugel} et~al.,}{{Kugel}
  et~al.}{2023}]{Kugel2023MNRAS.526.6103K}
{Kugel} R.,  et~al., 2023, \mn@doi [\mnras] {10.1093/mnras/stad2540}, \href
  {https://ui.adsabs.harvard.edu/abs/2023MNRAS.526.6103K} {526, 6103}

\bibitem[\protect\citeauthoryear{{Lam} \& {Sheth}}{{Lam} \&
  {Sheth}}{2008}]{LamSheth2008MNRAS}
{Lam} T.~Y.,  {Sheth} R.~K.,  2008, \mn@doi [\mnras]
  {10.1111/j.1365-2966.2008.13038.x}, \href
  {https://ui.adsabs.harvard.edu/abs/2008MNRAS.386..407L} {386, 407}

\bibitem[\protect\citeauthoryear{Lesgourgues, Matarrese, Pietroni  \&
  Riotto}{Lesgourgues et~al.}{2009}]{Lesgourgues_2009}
Lesgourgues J.,  Matarrese S.,  Pietroni M.,   Riotto A.,  2009, \mn@doi
  [Journal of Cosmology and Astroparticle Physics]
  {10.1088/1475-7516/2009/06/017}, 2009, 017

\bibitem[\protect\citeauthoryear{{Libeskind} et~al.,}{{Libeskind}
  et~al.}{2018}]{Libeskind_2018}
{Libeskind} N.~I.,  et~al., 2018, \mn@doi [\mnras] {10.1093/mnras/stx1976},
  \href {https://ui.adsabs.harvard.edu/abs/2018MNRAS.473.1195L} {473, 1195}

\bibitem[\protect\citeauthoryear{Ma \& Fry}{Ma \& Fry}{2000}]{Ma_2000}
Ma C.-P.,  Fry J.~N.,  2000, \mn@doi [\apj] {10.1086/317146}, 543, 503

\bibitem[\protect\citeauthoryear{Massara, Villaescusa-Navarro  \& Viel}{Massara
  et~al.}{2014}]{Massara_2014}
Massara E.,  Villaescusa-Navarro F.,   Viel M.,  2014, \mn@doi [\jcap]
  {10.1088/1475-7516/2014/12/053}, 2014, 053

\bibitem[\protect\citeauthoryear{{McQuinn} \& {White}}{{McQuinn} \&
  {White}}{2016}]{McQuinn_2016}
{McQuinn} M.,  {White} M.,  2016, \mn@doi [\jcap]
  {10.1088/1475-7516/2016/01/043}, \href
  {https://ui.adsabs.harvard.edu/abs/2016JCAP...01..043M} {2016, 043}

\bibitem[\protect\citeauthoryear{Mead}{Mead}{2016}]{Mead_2016b}
Mead A.~J.,  2016, \mn@doi [\mnras] {10.1093/mnras/stw2312}, 464, 1282

\bibitem[\protect\citeauthoryear{Mead \& Verde}{Mead \&
  Verde}{2021}]{MeadVerde_2021}
Mead A.~J.,  Verde L.,  2021, \mn@doi [\mnras] {10.1093/mnras/stab748}, 503,
  3095

\bibitem[\protect\citeauthoryear{{Mead, A. J.}, {Tr\"oster, T.}, {Heymans, C.},
  {Van Waerbeke, L.}  \& {McCarthy, I. G.}}{{Mead, A. J.}
  et~al.}{2020}]{Mead_2020}
{Mead, A. J.} {Tr\"oster, T.} {Heymans, C.} {Van Waerbeke, L.}  {McCarthy, I.
  G.} 2020, \mn@doi [A\&A] {10.1051/0004-6361/202038308}, 641, A130

\bibitem[\protect\citeauthoryear{Mead, Peacock, Heymans, Joudaki  \&
  Heavens}{Mead et~al.}{2015}]{Mead_2015}
Mead A.~J.,  Peacock J.~A.,  Heymans C.,  Joudaki S.,   Heavens A.~F.,  2015,
  \mn@doi [\mnras] {10.1093/mnras/stv2036}, 454, 1958

\bibitem[\protect\citeauthoryear{Mead, Heymans, Lombriser, Peacock, Steele  \&
  Winther}{Mead et~al.}{2016}]{Mead_2016a}
Mead A.~J.,  Heymans C.,  Lombriser L.,  Peacock J.~A.,  Steele O.~I.,
  Winther H.~A.,  2016, \mn@doi [\mnras] {10.1093/mnras/stw681}, 459, 1468

\bibitem[\protect\citeauthoryear{Mead, Brieden, Tröster  \& Heymans}{Mead
  et~al.}{2021}]{Mead_2021}
Mead A.~J.,  Brieden S.,  Tröster T.,   Heymans C.,  2021, \mn@doi [\mnras]
  {10.1093/mnras/stab082}, 502, 1401

\bibitem[\protect\citeauthoryear{Mohammed \& Seljak}{Mohammed \&
  Seljak}{2014}]{Mohammed_2014}
Mohammed I.,  Seljak U.,  2014, \mn@doi [\mnras] {10.1093/mnras/stu1972}, 445,
  3382

\bibitem[\protect\citeauthoryear{{Moran} et~al.,}{{Moran}
  et~al.}{2023}]{MoranMiraTitan-2023}
{Moran} K.~R.,  et~al., 2023, \mn@doi [\mnras] {10.1093/mnras/stac3452}, \href
  {https://ui.adsabs.harvard.edu/abs/2023MNRAS.520.3443M} {520, 3443}

\bibitem[\protect\citeauthoryear{{Navarro}, {Frenk}  \& {White}}{{Navarro}
  et~al.}{1997}]{NFW_1997}
{Navarro} J.~F.,  {Frenk} C.~S.,   {White} S. D.~M.,  1997, \mn@doi [\apj]
  {10.1086/304888}, \href
  {https://ui.adsabs.harvard.edu/abs/1997ApJ...490..493N} {490, 493}

\bibitem[\protect\citeauthoryear{Peacock \& Smith}{Peacock \&
  Smith}{2000}]{Peacock_2000}
Peacock J.~A.,  Smith R.~E.,  2000, \mn@doi [\mnras]
  {10.1046/j.1365-8711.2000.03779.x}, 318, 1144

\bibitem[\protect\citeauthoryear{{Pellejero Iba{\~n}ez}, {Angulo}, {Zennaro},
  {St{\"u}cker}, {Contreras}, {Aric{\`o}}  \& {Maion}}{{Pellejero Iba{\~n}ez}
  et~al.}{2023}]{MPI_2023}
{Pellejero Iba{\~n}ez} M.,  {Angulo} R.~E.,  {Zennaro} M.,  {St{\"u}cker} J.,
  {Contreras} S.,  {Aric{\`o}} G.,   {Maion} F.,  2023, \mn@doi [\mnras]
  {10.1093/mnras/stad368}, \href
  {https://ui.adsabs.harvard.edu/abs/2023MNRAS.520.3725P} {520, 3725}

\bibitem[\protect\citeauthoryear{Philcox, Spergel  \&
  Villaescusa-Navarro}{Philcox et~al.}{2020}]{Philcox_2020}
Philcox O. H.~E.,  Spergel D.~N.,   Villaescusa-Navarro F.,  2020, \mn@doi
  [Phys. Rev. D] {10.1103/PhysRevD.101.123520}, 101, 123520

\bibitem[\protect\citeauthoryear{{Press} \& {Schechter}}{{Press} \&
  {Schechter}}{1974}]{Press_1974}
{Press} W.~H.,  {Schechter} P.,  1974, \mn@doi [\apj] {10.1086/152650}, \href
  {https://ui.adsabs.harvard.edu/abs/1974ApJ...187..425P} {187, 425}

\bibitem[\protect\citeauthoryear{{Pritchard} \& {Loeb}}{{Pritchard} \&
  {Loeb}}{2012}]{pritchard2012RPPh...75h6901P}
{Pritchard} J.~R.,  {Loeb} A.,  2012, \mn@doi [Reports on Progress in Physics]
  {10.1088/0034-4885/75/8/086901}, \href
  {https://ui.adsabs.harvard.edu/abs/2012RPPh...75h6901P} {75, 086901}

\bibitem[\protect\citeauthoryear{{Ravoux} et~al.,}{{Ravoux}
  et~al.}{2025}]{Ravoux:2025uik}
{Ravoux} C.,  et~al., 2025, \mn@doi [\jcap] {10.1088/1475-7516/2025/11/079},
  \href {https://ui.adsabs.harvard.edu/abs/2025JCAP...11..079R} {2025, 079}

\bibitem[\protect\citeauthoryear{{Reichardt} et~al.,}{{Reichardt}
  et~al.}{2021}]{2021ApJ...908..199R}
{Reichardt} C.~L.,  et~al., 2021, \mn@doi [\apj] {10.3847/1538-4357/abd407},
  \href {https://ui.adsabs.harvard.edu/abs/2021ApJ...908..199R} {908, 199}

\bibitem[\protect\citeauthoryear{{Schaller}, {Schaye}, {Kugel}, {Broxterman}
  \& {van Daalen}}{{Schaller} et~al.}{2025}]{Schaller2025MNRAS.539.1337S}
{Schaller} M.,  {Schaye} J.,  {Kugel} R.,  {Broxterman} J.~C.,   {van Daalen}
  M.~P.,  2025, \mn@doi [\mnras] {10.1093/mnras/staf569}, \href
  {https://ui.adsabs.harvard.edu/abs/2025MNRAS.539.1337S} {539, 1337}

\bibitem[\protect\citeauthoryear{{Schaye} et~al.,}{{Schaye}
  et~al.}{2023}]{Schaye2023MNRAS.526.4978S}
{Schaye} J.,  et~al., 2023, \mn@doi [\mnras] {10.1093/mnras/stad2419}, \href
  {https://ui.adsabs.harvard.edu/abs/2023MNRAS.526.4978S} {526, 4978}

\bibitem[\protect\citeauthoryear{{Scoccimarro}, {Sheth}, {Hui}  \&
  {Jain}}{{Scoccimarro} et~al.}{2001}]{Scocc2001ApJ}
{Scoccimarro} R.,  {Sheth} R.~K.,  {Hui} L.,   {Jain} B.,  2001, \mn@doi [\apj]
  {10.1086/318261}, \href
  {https://ui.adsabs.harvard.edu/abs/2001ApJ...546...20S} {546, 20}

\bibitem[\protect\citeauthoryear{Seljak}{Seljak}{2000}]{Seljak_2000}
Seljak U.,  2000, \mn@doi [\mnras] {10.1046/j.1365-8711.2000.03715.x}, 318, 203

\bibitem[\protect\citeauthoryear{Seljak \& Vlah}{Seljak \&
  Vlah}{2015}]{Seljak_2015}
Seljak U. c.~v.,  Vlah Z.,  2015, \mn@doi [Phys. Rev. D]
  {10.1103/PhysRevD.91.123516}, 91, 123516

\bibitem[\protect\citeauthoryear{Shen, Abel, Mo  \& Sheth}{Shen
  et~al.}{2006}]{Shen_2006}
Shen J.,  Abel T.,  Mo H.~J.,   Sheth R.~K.,  2006, \mn@doi [\apj]
  {10.1086/504513}, 645, 783

\bibitem[\protect\citeauthoryear{{Sheth} \& {Jain}}{{Sheth} \&
  {Jain}}{2003}]{Sheth_2003}
{Sheth} R.~K.,  {Jain} B.,  2003, \mn@doi [\mnras]
  {10.1046/j.1365-8711.2003.06974.x}, \href
  {https://ui.adsabs.harvard.edu/abs/2003MNRAS.345..529S} {345, 529}

\bibitem[\protect\citeauthoryear{Sheth \& Lemson}{Sheth \&
  Lemson}{1999}]{ShethLemson_1999}
Sheth R.~K.,  Lemson G.,  1999, \mn@doi [MNRAS]
  {10.1046/j.1365-8711.1999.02378.x}, 304, 767

\bibitem[\protect\citeauthoryear{{Sheth} \& {Tormen}}{{Sheth} \&
  {Tormen}}{1999}]{Sheth_1999}
{Sheth} R.~K.,  {Tormen} G.,  1999, \mn@doi [\mnras]
  {10.1046/j.1365-8711.1999.02692.x}, \href
  {https://ui.adsabs.harvard.edu/abs/1999MNRAS.308..119S} {308, 119}

\bibitem[\protect\citeauthoryear{Sheth, Mo  \& Tormen}{Sheth
  et~al.}{2001}]{Sheth_2001}
Sheth R.~K.,  Mo H.~J.,   Tormen G.,  2001, \mn@doi [\mnras]
  {10.1046/j.1365-8711.2001.04006.x}, 323, 1

\bibitem[\protect\citeauthoryear{Sinha}{Sinha}{2016}]{manodeep_sinha_2016}
Sinha M.,  2016, Corrfunc: Corrfunc-1.1.0, \mn@doi{10.5281/zenodo.55161}, \url
  {https://doi.org/10.5281/zenodo.55161}

\bibitem[\protect\citeauthoryear{{Sinha} \& {Garrison}}{{Sinha} \&
  {Garrison}}{2020}]{Sinha_2020}
{Sinha} M.,  {Garrison} L.~H.,  2020, \mn@doi [\mnras] {10.1093/mnras/stz3157},
  \href {https://ui.adsabs.harvard.edu/abs/2020MNRAS.491.3022S} {491, 3022}

\bibitem[\protect\citeauthoryear{Smith \& Angulo}{Smith \&
  Angulo}{2019}]{Smith_2019}
Smith R.~E.,  Angulo R.~E.,  2019, \mn@doi [\mnras] {10.1093/mnras/stz890},
  486, 1448

\bibitem[\protect\citeauthoryear{Smith et~al.,}{Smith
  et~al.}{2003}]{Smith_2003}
Smith R.~E.,  et~al., 2003, \mn@doi [\mnras]
  {10.1046/j.1365-8711.2003.06503.x}, 341, 1311

\bibitem[\protect\citeauthoryear{Smith, Scoccimarro  \& Sheth}{Smith
  et~al.}{2007}]{Smith_2007}
Smith R.~E.,  Scoccimarro R.,   Sheth R.~K.,  2007, \mn@doi [Phys. Rev. D]
  {10.1103/PhysRevD.75.063512}, 75, 063512

\bibitem[\protect\citeauthoryear{{Springel}}{{Springel}}{2005}]{Springel_2005}
{Springel} V.,  2005, \mn@doi [\mnras] {10.1111/j.1365-2966.2005.09655.x},
  \href {https://ui.adsabs.harvard.edu/abs/2005MNRAS.364.1105S} {364, 1105}

\bibitem[\protect\citeauthoryear{{Spurio Mancini}, {Piras}, {Alsing},
  {Joachimi}  \& {Hobson}}{{Spurio Mancini}
  et~al.}{2022}]{SpurioMancini2022MNRAS.511.1771S}
{Spurio Mancini} A.,  {Piras} D.,  {Alsing} J.,  {Joachimi} B.,   {Hobson}
  M.~P.,  2022, \mn@doi [\mnras] {10.1093/mnras/stac064}, \href
  {https://ui.adsabs.harvard.edu/abs/2022MNRAS.511.1771S} {511, 1771}

\bibitem[\protect\citeauthoryear{Sullivan, Seljak  \& Singh}{Sullivan
  et~al.}{2021}]{Sullivan_2021}
Sullivan J.~M.,  Seljak U.,   Singh S.,  2021, \mn@doi [\jcap]
  {10.1088/1475-7516/2021/11/026}, 2021, 026

\bibitem[\protect\citeauthoryear{{Sunyaev} \& {Zeldovich}}{{Sunyaev} \&
  {Zeldovich}}{1972}]{1972CoASP...4..173S}
{Sunyaev} R.~A.,  {Zeldovich} Y.~B.,  1972, Comments on Astrophysics and Space
  Physics, \href {https://ui.adsabs.harvard.edu/abs/1972CoASP...4..173S} {4,
  173}

\bibitem[\protect\citeauthoryear{Takahashi, Sato, Nishimichi, Taruya  \&
  Oguri}{Takahashi et~al.}{2012}]{Takahashi_2012}
Takahashi R.,  Sato M.,  Nishimichi T.,  Taruya A.,   Oguri M.,  2012, \mn@doi
  [\apj] {10.1088/0004-637X/761/2/152}, 761, 152

\bibitem[\protect\citeauthoryear{{Valageas} \& {Nishimichi}}{{Valageas} \&
  {Nishimichi}}{2011}]{Valageas_2011a}
{Valageas} P.,  {Nishimichi} T.,  2011, \mn@doi [\aap]
  {10.1051/0004-6361/201015685}, \href
  {https://ui.adsabs.harvard.edu/abs/2011A&A...527A..87V} {527, A87}

\bibitem[\protect\citeauthoryear{{Voivodic}, {Rubira}  \& {Lima}}{{Voivodic}
  et~al.}{2020}]{Voivodic2020JCAP}
{Voivodic} R.,  {Rubira} H.,   {Lima} M.,  2020, \mn@doi [\jcap]
  {10.1088/1475-7516/2020/10/033}, \href
  {https://ui.adsabs.harvard.edu/abs/2020JCAP...10..033V} {2020, 033}

\bibitem[\protect\citeauthoryear{Wang, Beutler  \& Sugiyama}{Wang
  et~al.}{2023}]{Wang2023}
Wang M.~S.,  Beutler F.,   Sugiyama N.~S.,  2023, \mn@doi [Journal of Open
  Source Software] {10.21105/joss.05571}, 8, 5571

\bibitem[\protect\citeauthoryear{{Warren}, {Abazajian}, {Holz}  \&
  {Teodoro}}{{Warren} et~al.}{2006}]{Warren2006ApJ}
{Warren} M.~S.,  {Abazajian} K.,  {Holz} D.~E.,   {Teodoro} L.,  2006, \mn@doi
  [\apj] {10.1086/504962}, \href
  {https://ui.adsabs.harvard.edu/abs/2006ApJ...646..881W} {646, 881}

\bibitem[\protect\citeauthoryear{{Wright} et~al.,}{{Wright}
  et~al.}{2025}]{Wright_2025}
{Wright} A.~H.,  et~al., 2025, \mn@doi [\aap] {10.1051/0004-6361/202554908},
  \href {https://ui.adsabs.harvard.edu/abs/2025A&A...703A.158W} {703, A158}

\bibitem[\protect\citeauthoryear{Yang, Hudson  \& Afshordi}{Yang
  et~al.}{2022}]{Yang:2022ibs}
Yang T.,  Hudson M.~J.,   Afshordi N.,  2022, \mn@doi [Mon. Not. Roy. Astron.
  Soc.] {10.1093/mnras/stac2564}, 516, 6041

\bibitem[\protect\citeauthoryear{{Zel'dovich}}{{Zel'dovich}}{1970}]{1970A&A.....5...84Z}
{Zel'dovich} Y.~B.,  1970, \aap, \href
  {https://ui.adsabs.harvard.edu/abs/1970A&A.....5...84Z} {5, 84}

\bibitem[\protect\citeauthoryear{{Zennaro}, {Angulo},
  {Pellejero-Ib{\'a}{\~n}ez}, {St{\"u}cker}, {Contreras}  \&
  {Aric{\`o}}}{{Zennaro} et~al.}{2023}]{Zennaro_2023}
{Zennaro} M.,  {Angulo} R.~E.,  {Pellejero-Ib{\'a}{\~n}ez} M.,  {St{\"u}cker}
  J.,  {Contreras} S.,   {Aric{\`o}} G.,  2023, \mn@doi [\mnras]
  {10.1093/mnras/stad2008}, \href
  {https://ui.adsabs.harvard.edu/abs/2023MNRAS.524.2407Z} {524, 2407}

\bibitem[\protect\citeauthoryear{{d'Amico}, {Gleyzes}, {Kokron}, {Markovic},
  {Senatore}, {Zhang}, {Beutler}  \& {Gil-Mar{\'\i}n}}{{d'Amico}
  et~al.}{2020}]{Senatore2020JCAP...05..005D}
{d'Amico} G.,  {Gleyzes} J.,  {Kokron} N.,  {Markovic} K.,  {Senatore} L.,
  {Zhang} P.,  {Beutler} F.,   {Gil-Mar{\'\i}n} H.,  2020, \mn@doi [\jcap]
  {10.1088/1475-7516/2020/05/005}, \href
  {https://ui.adsabs.harvard.edu/abs/2020JCAP...05..005D} {2020, 005}

\makeatother
\end{thebibliography}

% Alternatively you could enter them by hand, like this:
% This method is tedious and prone to error if you have lots of references
%\begin{thebibliography}{99}
%\bibitem[\protect\citeauthoryear{Author}{2012}]{Author2012}
%Author A.~N., 2013, Journal of Improbable Astronomy, 1, 1
%\bibitem[\protect\citeauthoryear{Others}{2013}]{Others2013}
%Others S., 2012, Journal of Interesting Stuff, 17, 198
%\end{thebibliography}

%%%%%%%%%%%%%%%%%%%%%%%%%%%%%%%%%%%%%%%%%%%%%%%%%%

%%%%%%%%%%%%%%%%% APPENDICES %%%%%%%%%%%%%%%%%%%%%

\appendix

\section{LPT validation} \label{app:1l-LPT}

Here, we describe our practical implementation of $1\ell$-LPT using the \texttt{`lpt\_rsd\_fftw'} module within the \texttt{velocileptors}. We obtain the real-space matter power spectrum $P_\mathrm{}^\mathrm{LPT}(k)$ by passing the CAMB linear power spectrum $P_\mathrm{}^\mathrm{L}(k)$ and associated wavevector $k$ to the \texttt{LPT\_RSD()} class. On that class we then call the \texttt{make\_pell\_fixedbias()} function, passing as arguments the growth rate and all Lagrangian bias terms set to zero, since we are interested in matter clustering in real space. 

As anticipated in section \ref{sec:ing-pt-lag}, we need to validate the LPT power spectrum output in the deeply non-linear regime, where \texttt{velocileptors} is not designed to provide accurate predictions. 

\begin{figure}
    \centering
    \includegraphics[width=\columnwidth]{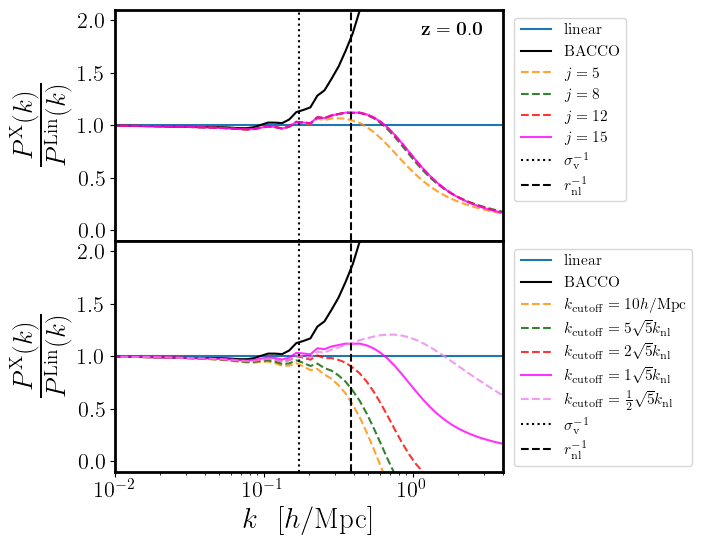}
    \caption{We explore varying different \texttt{velocileptors} settings from our baseline choice (solid magenta) for the LPT power spectrum and the \texttt{baccoemu} prediction (solid black), normalised by the linear power spectrum (solid blue). Top panel: We vary the Bessel series truncation order $j$ from the default value $j=5$ (dashed orange) until it converges well before our baseline choice $j=15$. Bottom panel: We show the impact of changing the cutoff and IR wavenumbers around our baseline choice, setting it to the inverse non-linear scale transformed to the tophat filter.}
    \label{fig:app-lpt-comp1}
\end{figure}

For example, the calculation involves a Bessel series expansion truncated at order $j=5$ by default, sufficient for sub-percent level precision up to $k=0.25\,h\mathrm{Mpc}^{-1}$ according to appendix C.6 of \citet{Chen_velo_2021}. But since we use LPT  for the 2-halo (or 2-sheet) term, we need to make sure that the series is converged up to scales that still contribute to the total power, i.e., up to  $k=2\,h\mathrm{Mpc}^{-1}$. Hence, we test the impact of increasing the \texttt{cutoff\_jn} parameter within the \texttt{LPT\_RSD()} class. This test is shown in the upper panel of figure \ref{fig:app-lpt-comp1}, showing the LPT model predictions for $j\in\left\lbrace 5,8,12,15 \right\rbrace$ along with the non-linear $P_\mathrm{}^\mathrm{NL}(k)$ from \texttt{baccoemu} \citet{Angulo_2021} normalised by the linear power spectrum. While all 1$\ell$-LPT curves coincide up to $k=0.25\,h\mathrm{Mpc}^{-1}$, we note that for higher wavenumbers the default choice $j=5$ is discrepant from the fully converged choice $j>12$ by $\approx10\%$. Hence, for our baseline, we conservatively set $j=15$ (solid magenta line).

The other decision to make relates to the choice of loop integral cutoff scale  $k_\mathrm{cutoff}$ and infrared resummation scale $k_\mathrm{IR}$. The latter has been introduced in section 4.3 of \citet{Chen_LPT_2020} and serves as a `dial' between pure LPT in case all modes are resummed ($k_\mathrm{IR}=\infty$), and pure EPT in case no modes are resummed ($k_\mathrm{IR}=0$). By default, it is set to $k_\mathrm{IR}=0.2\,h\mathrm{Mpc}^{-1}$, as in section 5.1 of \citet{Chen_velo_2021}, this was found to best reproduce the redshift space galaxy hexadecapole. The cutoff scale is set arbitrarily to $k_\mathrm{cutoff}=10\,h^{-1}\mathrm{Mpc}$ by default. 

For the application to galaxy clustering, these choices barely impact the analysis, as the differences are absorbed by so-called counterterms whose amplitude is marginalised over. In our case, where we replace the counterterms with unpredictable amplitude by the deterministic web-halo model, we need to carefully select sensible values for $k_\mathrm{cutoff}$ and $k_\mathrm{IR}$. To reduce the dimensionality of the problem, we set $k_\mathrm{cutoff} = k_\mathrm{IR}$, making sure that all modes entering the loop integrals are also resummed, and hence preserving the spirit of LPT. Then, we aim to capture the cosmology dependence of the cutoff scale, demanding it should be a function of the non-linear scale $r_\mathrm{nl}$ defined in equation \eqref{eq:ing-pt-rnl} at which spherical collapse typically occurs. 

The lower panel of figure \ref{fig:app-lpt-comp1} shows the impact of varying $k_\mathrm{cutoff}$ as a function of $r_\mathrm{nl}=2.7\,h^{-1}\mathrm{Mpc}$ assuming a Planck cosmology at $z=0$. Naively, we would expect the cutoff wavenumber $k_\mathrm{cutoff} = \sqrt{5}\,r_\mathrm{nl}^{-1}$ to be the inverse of the non-linear scale multiplied by $\sqrt{5}$ to map the Gaussian filter used internally within \texttt{velocileptors} to the tophat filter used here, as explained in section \ref{sec:ing-pt-lag}. As we can see from the figure, varying this choice within $\sim1$ order of magnitude significantly impacts the predictions on small scales, with larger values of $k_\mathrm{cutoff}$ showing more suppressed power, eventually crossing zero, and smaller values approaching the \texttt{baccoemu} prediction. 

\begin{figure}
    \centering
    \includegraphics[width=\columnwidth]{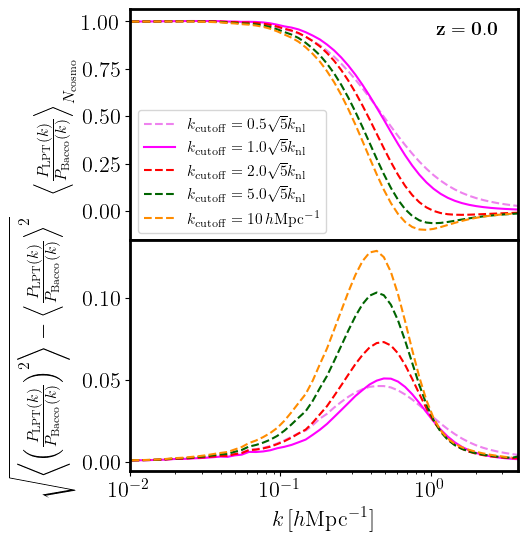}
    \caption{Again, we show the impact of changing the cutoff and IR wavenumbers around the baseline (solid magenta), but normalised to the \texttt{baccoemu} prediction for its mean (top) and standard deviation (bottom).}
    \label{fig:app-lpt-comp2}
\end{figure}

To make a sensible choice, we perform a more systematic test. We run \texttt{velocileptors} for the five different choices of cutoff scale presented in figure \ref{fig:app-lpt-comp1} for $N_\mathrm{cosmo}=100$ randomly selected cosmological models within the \texttt{baccoemu} range. We compute the mean ratio of LPT with respect to \texttt{baccoemu} across cosmologies for these choices (upper panel of figure \ref{fig:app-lpt-comp2}) and its respective standard deviation (lower panel). We observe that our naive choice of $k_\mathrm{cutoff} = \sqrt{5}\,r_\mathrm{nl}^{-1}$ (orange curve) represents a `sweet spot' minimising the scatter of LPT with respect to the \texttt{baccoemu} predictions at the relevant, perturbative scales. Therefore, in this work, we stick to this physically motivated choice.

\section{Window function validation} \label{app:window}

\begin{figure}
    \centering
    \includegraphics[width=0.8\columnwidth]{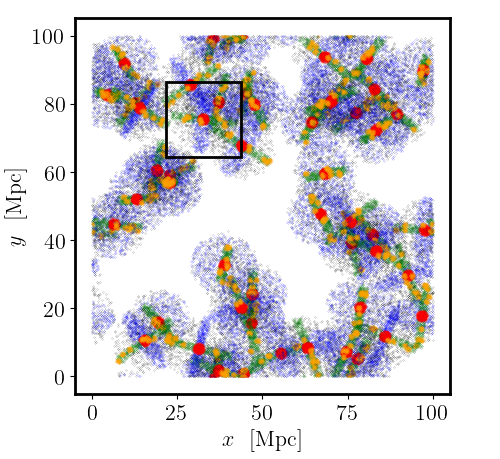}
    \includegraphics[width=0.8\columnwidth]{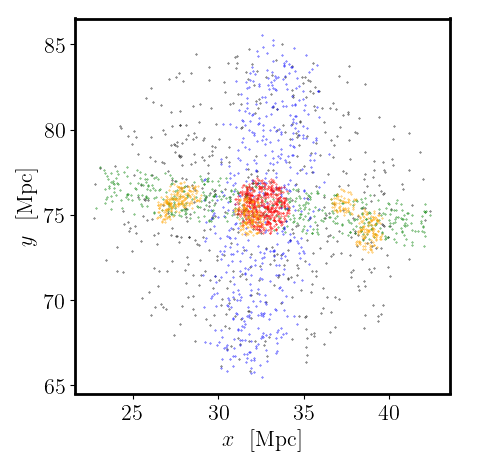}
    \caption{Top panel: We show the $x-y$ plane of the $L=100\,h^{-1}\mathrm{Mpc}$ sidelength box including the various objects around 50 randomly subsampled halo-centres. Bottom panel: This is a zoom into one of them. Black, blue, green, and red dots represent the homogeneous $R=10\,\mathrm{Mpc}$ sphere, $(a,b)=(R,(4/3)R/\Delta_\mathrm{v}^{1/3})$ sheet, $(a,b)=(\sqrt{2/3}R/\Delta_\mathrm{v}^{1/3}, 2R)$ filament, and $r_\mathrm{v}=R/\Delta_\mathrm{v}^{1/3}$ halo, with virial overdensity $\Delta_\mathrm{v}=200$ and $(a,b)$ representing the radius and height of randomly oriented cylinders centered around the haloes.}
    \label{fig:app-win-field}
\end{figure}

To validate the cylindrical window functions derived in section \ref{sec:ing-dp-f}, we perform a numerical experiment. We populate an $L=100\,\mathrm{Mpc}$ periodic box with $N_\mathrm{h} = 64^3$ randomly (Poisson) distributed spheres with radius $R=10\,\mathrm{Mpc}$, consisting of $N_\mathrm{DM} / N_\mathrm{h} =512$ randomly distributed DM particles each. Figure \ref{fig:app-win-field} shows the DM particles (black dots) within a subsample of 50 spheres for the full volume in the top panel, and a zoom into one of the spheres in the bottom panel. Next, we randomly select an angle for each sphere, along which we randomly place additional 512 DM particles representing filaments (green dots) and in the orthogonal plane sheets (blue dots). The radius $a$ and height $b$ of the respective cylinders follow equation \eqref{eq:ing-dpwf-numbers}, ensuring that the density is $\Delta_\mathrm{v}^{2/3}$ and $\Delta_\mathrm{v}^{1/3}$ times the density of each sphere, respectively, assuming $\Delta_\mathrm{v}=200$. Then, we randomly place additional 512 DM particles (red dots) within $r_\mathrm{v}=R/\Delta_\mathrm{v}^{1/3}$ from the sphere centres, representing homogeneous haloes with $\Delta_\mathrm{v}$ times the sphere density. Finally, we also consider the case that the mass is split into 8 homogeneous haloes with virial radii $r_{\mathrm{v},1/8} = r_\mathrm{v}/\sqrt[3]{8}$ whose centers are randomly distributed within the ``parent'' filament (orange dots). This allows us to validate our window compensation scheme from equation \eqref{eq:whm-s-compensation} and compare its performance with the more general substructure prescription of \citet{Sheth_2003}.

\begin{figure}
    \centering
    \includegraphics[width=\columnwidth]{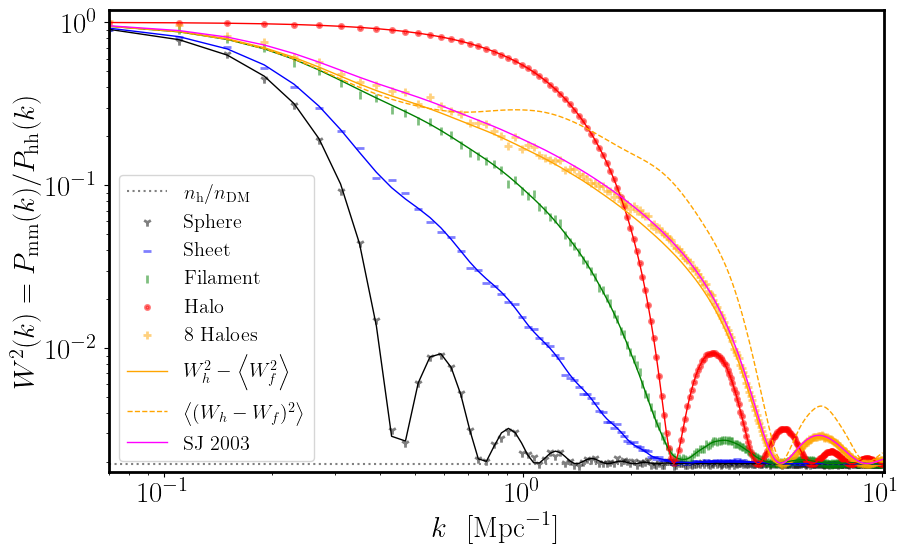}
    \caption{Here, we show the measured squared window functions (coloured datapoints) compared to their theoretical predictions (coloured lines)}
    \label{fig:app-win-fit}
\end{figure}

Now, we compute the power spectra of each DM sample corresponding to the different objects (tophats, sheets, filaments, and haloes) using the \texttt{Triumvirate}\footnote{\href{https://github.com/MikeSWang/Triumvirate}{https://github.com/MikeSWang/Triumvirate}} code by \citet{Wang2023}. Conveniently, since all objects have the same mass and (within each sample) radii, dividing the spectra by the halo power spectrum $P_\mathrm{hh} \approx n_\mathrm{h}^{-1} = \frac{L^3}{N_\mathrm{h}} $, i.e., the correlation between the centers of the collapsed objects only, we directly measure the squared window function of each subsample shown in figure \ref{fig:app-win-fit}. As expected, the measurements converge to $W^2(k\rightarrow0)=1$ on large scales, and to $W_\mathrm{u}^2(k\rightarrow \infty)=n_\mathrm{h}/n_\mathrm{DM}$, i.e., the shot-noise limit given the DM particles per collapsed object, on small scales. 

We compare the window function measurements in figure \ref{fig:app-win-fit} to the theoretical predictions (coloured lines), i.e. the tophat window (equation \eqref{eq:whm-def-tophat}) in case of the spheres (black and red) and the angle-averaged cylinder window (equations \eqref{eq:ing-dpwf-window-vartheta}-\eqref{eq:ing-dpwf-numbers}) in case of filaments (green) and sheets (blue). These predictions provide an excellent fit to the measurement, hence validating the window functions used within the web-halo model.

Next, we consider the case of the 8 orange subhaloes randomly placed within each filament. Following our default prescription from equation \eqref{eq:whm-whm-1h}, we can write (note that the halo and filament mass functions collapse to a delta-function in our case) their window function as
\begin{equation} \label{eq:app-h8}
    W_\mathrm{h,1/8}^2(k,R) = W_\mathrm{f}^2(k,R) + \left( W_\mathrm{h}^2(k,R/\sqrt[3]{8}) - W_\mathrm{f}^2(k,R/\sqrt[3]{8}) \right)/8~,
\end{equation}
where $W_\mathrm{f}^2(k,R)$ takes the role of the ``two-halo'' term and the division by $8$ arises from the mass within the integrand being $1/8$ of the nominal mass. This model, shown as a solid orange line in figure \ref{fig:app-win-fit}, provides a good fit to the measured data. In particular, it provides a notably better fit than performing the window compensation before squaring (dashed orange line) justifying our choice in equation \eqref{eq:whm-s-compensation}. We also compare the measurement to prediction from \citet{Sheth_2003}, in particular equation (21) therein, which in our toy model --under the approximation that the subhaloes are significantly denser than the parent filament-- reduces to
\begin{equation} \label{eq:app-h8sj}
\left(W_\mathrm{h,1/8}^{SJ}\right)^2(k,R) = W_\mathrm{f}^2(k,R) + \left( W_\mathrm{h}^2(k,R/\sqrt[3]{8}) - W_\mathrm{f}^2(k,R) \right)/8~.
\end{equation}
From figure \ref{fig:app-win-fit} we see that this prediction (purple solid line) provides an even better fit to the data. This is because equation \eqref{eq:app-h8sj} correctly compensates each subhalo for $1/8$ of the mass of the ``parent'' filament, whereas in \eqref{eq:app-h8} we compensate the subhaloes by ``imaginary'', randomly oriented filaments of the same mass as the subhaloes with radius $a$ and height $b$ following equation \eqref{eq:ing-dpwf-numbers} with Radius $R/\sqrt[3]{8}$. Clearly, this is just an approximation, but a very good one, that allows us to perform the WHM integrals using general mass functions only, i.e., without making assumptions of the substructure using conditional mass functions entering the model as a nested hierarchy as in \citet{Sheth_2003}.

\section{\sB{Cylinder Window Approximation}} \label{app:window-numbers}

\sB{In section \ref{sec:ing-dp-f} we describe how we obtain the analytical approximations \eqref{eq:ing-dpwf-sheet} and \eqref{eq:ing-dpwf-filament} for the volume-averaged sheet and filament window functions. They depend on two ingredients, (i) the analytical solution of \eqref{eq:ing-dpwf-window-average} in the limits $(a = R, ~b\rightarrow 0)$ for sheets and $a\rightarrow 0,~ b=2R$ for filaments, (ii) the tophat window function accounting for the non-zero extent of the collapsed dimension $b = \frac{4}{3} \left(\frac{R}{\Delta_\mathrm{v}^{1/3}}\right)$ for sheets and $a = \sqrt{\frac{2}{3}}\left(\frac{R}{\Delta_\mathrm{v}^{1/3}}\right)$ for filaments, also see equation \eqref{eq:ing-dpwf-numbers}. However, using these latter values as arguments in the tophat filter does not yield a good fit to the exact numerical result. This is expected, since we are trying to fit a sphere to objects with cylindrical symmetry. }

\sB{We hence employ the following trick. We write down the leading order Taylor expansion of all the window functions we are using
\begin{align}
    W_G(x) &\approx 1 - x^2 + \mathcal{O}(x^4) \label{eq:app-cyl-gauss} \\
    W_t^2(x) &\approx 1 - \frac{x^2}{5} + \mathcal{O}(x^4) \label{eq:app-cyl-t}\\
    W_\mathrm{Disk}^2(x) &\approx 1 - \frac{x^2}{4} + \mathcal{O}(x^4) \label{eq:app-cyl-disk} \\
    W_\mathrm{Line}^2(x) &\approx 1 - \frac{x^2}{3} + \mathcal{O}(x^4)~, \label{eq:app-cyl-line}
\end{align}
where $W_\mathrm{Disk}$ corresponds to the 2D tophat and $W_\mathrm{Line}$ to the 1D tophat, i.e., the $2J_1(x)/x$ and the $\sin(x)/x$ term in equation \eqref{eq:ing-dpwf-window-vartheta}, respectively. 
}

\sB{To calculate the approximate sheet window function we want to match the 3D tophat variance \eqref{eq:app-cyl-t} to the geometry of a sheet with small but non-zero height, which is given by $W_\mathrm{Line}(b/2)$ in equation \eqref{eq:app-cyl-line}. Hence, we set
\begin{equation}
    \frac{1}{5} R_\mathrm{sphere}^2 = \frac{1}{3} \left(\frac{b}{2}\right)^2 \longrightarrow R_\mathrm{sphere} = \sqrt{\frac{5}{12}}b
\end{equation}
Next, we use equation \eqref{eq:ing-dpwf-numbers} to substitute the height $b$ as
\begin{equation}
    R_\mathrm{sphere} = \sqrt{\frac{5}{12}} \frac{4}{3} \left(\frac{R}{\Delta_\mathrm{v}^{1/3}}\right) = \frac{2\sqrt{5}}{3\sqrt{3}} \left(\frac{R}{\Delta_\mathrm{v}^{1/3}}\right)~.
\end{equation}
}
\sB{Similarly, to approximate the filament window function we scale the 3D tophat radius to match the non-zero extent of the filament described by $W_\mathrm{Disk}(a)$ in equation \eqref{eq:app-cyl-disk}
\begin{equation} \label{eq:app-cyl-sheet}
    \frac{1}{5} R_\mathrm{sphere}^2 = \frac{1}{4} a^2 \longrightarrow R_\mathrm{sphere} = \sqrt{\frac{5}{4}}a~,
\end{equation}
finding, again using \eqref{eq:ing-dpwf-numbers},
\begin{equation} \label{eq:app-cyl-fil}
    R_\mathrm{sphere} = \sqrt{\frac{5}{4}} \cdot \sqrt{\frac{2}{3}}\left(\frac{R}{\Delta_\mathrm{v}^{1/3}}\right) = \sqrt{\frac{5}{6}} \left(\frac{R}{\Delta_\mathrm{v}^{1/3}}\right)~.
\end{equation}
}
\sB{The right hand sides of equations \eqref{eq:app-cyl-sheet} and \eqref{eq:app-cyl-fil} correspond exactly to the scales used within the 3D tophat in equations \eqref{eq:ing-dpwf-sheet} and \eqref{eq:ing-dpwf-filament}.}

\sB{One may ask why we need to go through the process of matching the Taylor expansion of $W_\mathrm{t}$ to the cylindrical geometry instead of using $W_\mathrm{Line}$ and $W_\mathrm{Disk}$ themselves. This is because we are computing an approximation for the angular average, such that the spherical geometry of the rescaled 3D tophat is better suited.}

\sB{Finally, it is worth noting that the factor $\sqrt{5}$ introduced in equation \eqref{eq:ing-pt-rnl} immediately follows from equations \eqref{eq:app-cyl-gauss} and \eqref{eq:app-cyl-t}}.

%%%%%%%%%%%%%%%%%%%%%%%%%%%%%%%%%%%%%%%%%%%%%%%%%%

% Don't change these lines
\bsp	% typesetting comment
\label{lastpage}
\end{document}